\documentclass[orivec]{article}

\usepackage{authblk}
\usepackage[ruled,lined,vlined]{algorithm2e}
\usepackage{amsfonts}
\usepackage{amssymb}
\usepackage{amsmath}
\usepackage{amsthm}
\usepackage[titletoc]{appendix}
\usepackage[english]{babel}
\usepackage{booktabs}
\usepackage{caption}
\usepackage{cite}
\usepackage{xcolor}
\usepackage{colortbl}
\usepackage{comment}
\usepackage[bottom]{footmisc}
\usepackage{graphicx}
\usepackage{hyperref}
\usepackage[utf8]{inputenc}
\usepackage{listings}
\usepackage{pgfplots}
\usepackage[author={svzelst}]{pdfcomment}
\usepackage{subcaption}
\usepackage{tikz}
\usepackage{wrapfig}

\usetikzlibrary{arrows,decorations.pathmorphing,backgrounds,positioning,fit,petri}
\usetikzlibrary{patterns}
\usepgfplotslibrary{groupplots}

\AtBeginDocument{%
	\mathchardef\mathcomma\mathcode`\,
	\mathcode`\,="8000
}
{\catcode`,=\active
	\gdef,{\mathcomma\discretionary{}{}{}}
}
\mathchardef\breakingcomma\mathcode`\,
{\catcode`,=\active
	\gdef,{\breakingcomma\discretionary{}{}{}}
}

\numberwithin{equation}{section}

\newtheorem{definition}{Definition}
\newtheorem{lemma}{Lemma}
\newtheorem{theorem}{Theorem}

\newcommand{\xrightarrowdbl}[2][]{%
	\xrightarrow[#1]{#2}\mathrel{\mkern-14mu}\rightarrow
}

\newcommand{\activity}{a}
\newcommand{\activityUniverse}{\mathcal{A}}
\newcommand{\activities}{A}
\newcommand{\bag}{B}
\newcommand{\allBags}{\mathcal{B}}
\newcommand{\causalOracle}{\gamma_c}
\newcommand{\element}{e}
\newcommand{\eventLog}{L}
\newcommand{\flowRelation}{F}
\newcommand{\labelFunc}{\lambda}
\newcommand{\labels}{\Lambda}
\newcommand{\lang}{\mathcal{L}}
\newcommand{\marking}{M}
\newcommand{\naturals}{\mathbb{N}}
\newcommand{\parikh}[2]{\#_{#2}(#1)}
\newcommand{\petriNet}{N}
\newcommand{\place}{p}
\newcommand{\places}{P}
\newcommand{\powerSet}{\mathcal{P}}
\newcommand{\region}{r}
\newcommand{\regions}{R}
\newcommand{\silent}{\tau}
\newcommand{\sequence}{\sigma}
\newcommand{\setGeneric}{X}
\newcommand{\transition}{t}
\newcommand{\transitions}{T}
\newcommand{\useProj}{\pi}
\newcommand{\workflowNet}{W}

\begin{document}
	
	\title{Discovering Relaxed Sound Workflow Nets using Integer Linear Programming}
			
	\author{S.J. van Zelst\thanks{Corresponding Author: \texttt{s.j.v.zelst@tue.nl}}}
	\author{B.F. van Dongen}
	\author{W.M.P. van der Aalst}
	\author{H.M.W. Verbeek}
	\affil{Department of Mathematics and Computer Science\\ 
			Eindhoven University of Technology\\ 
			P.O. Box 513, 5600 MB Eindhoven, The Netherlands\\ }

	\date{\today}

	\maketitle
	
	\begin{abstract}
		Process mining is concerned with the analysis, understanding and improvement of business processes. 
		Process discovery, i.e. discovering a process model based on an event log, is considered the most challenging process mining task.
		State-of-the-art process discovery algorithms only discover local control-flow patterns and are unable to discover complex, non-local patterns.
		Region theory based techniques, i.e. an established class of process discovery techniques, do allow for discovering such patterns.
		However, applying region theory directly results in complex, over-fitting models, which is less desirable.
		Moreover, region theory does not cope with guarantees provided by state-of-the-art process discovery algorithms, both w.r.t. structural and behavioural properties of the discovered process models.
		In this paper we present an ILP-based process discovery approach, based on region theory, that guarantees to discover relaxed sound workflow nets.
		Moreover, we devise a filtering algorithm, based on the internal working of the ILP-formulation, that is able to cope with the presence of infrequent behaviour.
		We have extensively evaluated the technique using different event logs with different levels of exceptional behaviour.
		Our experiments show that the presented approach allow us to leverage the inherent shortcomings of existing region-based approaches.
		The techniques presented are implemented and readily available in the HybridILPMiner package in the open-source process mining tool-kits ProM and RapidProM.\footnote{\url{http://promtools.org}; \url{http://rapidprom.org}}
	\end{abstract}

	\section{Introduction}
	\label{sec:introduction}
	The execution of business processes within a company generates traces of event data in its supporting information system. 
	The goal of \textit{process mining}~\cite{DBLP:books/sp/Aalst16} is to turn this data, recorded in \textit{event logs}, into actionable knowledge. 
	Three core branches form the basis of process mining: \textit{process discovery}, \textit{conformance checking} and \textit{process enhancement}. 
	In process discovery, this paper's focus, the goal is to construct a process model based on an event log. 
	In conformance checking the goal is to assess whether a given process model and event log conform with respect to each other in terms of described behaviour.
	In process enhancement the goal is to improve processes models, primarily, though not exhaustively, using the two aforementioned fields.
	
	Several \textit{different process models} exist that (largely) describe the behaviour in an event log.
	Hence, we need means to rank and compare these different process models.
	In process mining we typically judge the quality of process models based on four essential quality dimensions: \textit{replay-fitness}, \textit{precision}, \textit{generalization} and \textit{simplicity}~\cite{DBLP:books/sp/Aalst16, DBLP:journals/widm/AalstAD12, DBLP:conf/otm/BuijsDA12}. 
	Replay-fitness describes the fraction of behaviour in the event log that is also described by the model.
	Precision describes the fraction of behaviour described by the model that is also present in the event log.
	Generalization indicates a model's ability to account for behaviour not part of the event log, e.g. in case of parallelism, it is often impossible to observe all behaviour in the event log.
	Simplicity refers to a model's interpretability by a human analyst.
	A process discovery result ideally strikes and adequate balance between these four quality dimensions.

	A field closely related to process discovery is \textit{Petri net synthesis}~\cite{DBLP:series/txtcs/BadouelBD15}.
	Here the problem is to, given a \textit{behavioural system description}, decide whether there exists a Petri net\cite{murata_1989_petri_nets} that allows for all behaviour of the system description.
	Moreover, it needs to minimize additional behaviour.
	Most Petri net synthesis approaches use \textit{region theory}~\cite{DBLP:conf/ac/BadouelD96} which comes in two forms: \textit{state-based region theory}~\cite{DBLP:journals/acta/EhrenfeuchtR89,DBLP:journals/acta/EhrenfeuchtR89a, DBLP:conf/apn/Bernardinello93} (using \textit{transition systems}), and \textit{language-based region theory}~\cite{DBLP:conf/tapsoft/BadouelBD95,DBLP:conf/concur/Darondeau98} (using \textit{languages}).
	Applying classical region theory using an event log as a system description results in Petri nets with \textit{maximal} replay-fitness.
	Moreover, precision is \textit{maximized}.
	An implicit consequence is \textit{poor generalization} and \textit{poor simplicity}.
	Using these techniques directly on real event logs therefore results in process models that are not an adequate representation of the event log and do not allow us to reach the global goal of process mining, i.e. turning data into actionable knowledge. 	
	
	In~\cite{DBLP:journals/fuin/derWerfDHS09} a process discovery algorithm is proposed on top of language-based region theory.
	The core of the algorithm is an Integer Linear Programming (ILP)-formulation that is solved multiple times using slight variations.
	The main contribution is a relaxation of the precision maximization property of language-based region theory. 
	The algorithm still guarantees that the resulting Process model is able to replay all behaviour in the event log.
	Opposed to state-of-the-art process discovery algorithms, the algorithm provides limited guarantees w.r.t. structural and behavioural properties of the resulting process models.
	Moreover, the algorithm only works well under the assumption that the event log only holds frequent behaviour that fits nicely into some underlying process model.

	Real event logs typically include low-frequent exceptional behaviour, e.g. caused by people deviating from the normative process, cases that require special treatment, employees solving unexpected issues in an ad-hoc fashion etc.
	Considering all irregularities together with ``normal behaviour'' yields incomprehensible models, both in classical region-based synthesis and region-based process discovery techniques.
	In this paper we tackle these problems by extending and improving existing, region theory based, algorithms~\cite{DBLP:journals/fuin/derWerfDHS09,DBLP:conf/bpm/ZelstDA15a,DBLP:conf/apn/ZelstDA15}.
	This paper's contributions are summarized as follows:
	\begin{enumerate}
		\item We show that our approach is able discover \textit{relaxed sound workflow nets}.
		\item We present an effective, integrated, filtering algorithm that results in process models that abstract from infrequent and/or exceptional behaviour.
	\end{enumerate}	
	
	The proposed algorithm is implemented in the process mining framework \texttt{ProM}~\cite{DBLP:conf/caise/VerbeekBDA10} (\textit{HybridILPMiner} package) and is available in \texttt{RapidProM}~\cite{DBLP:journals/corr/AalstBZ2017,DBLP:journals/sttt/BoltLA16}.
	We have compared our technique with two state-of-the-art filtering techniques~\cite{DBLP:conf/bpm/LeemansFA13,DBLP:journals/tkde/ConfortiRH17}.
	We additionally validated the applicability of our approach on two real life event logs~\cite{https://doi.org/10.4121/uuid:270fd440-1057-4fb9-89a9-b699b47990f5,https://doi.org/10.4121/uuid:915d2bfb-7e84-49ad-a286-dc35f063a460}.
	Our experiments confirm the effectiveness of the proposed approach, both in terms of resulting model quality and computational complexity.
	
	The remainder of this paper is organized as follows.
	In \autoref{sec:motivation} we motivate the need to further develop ILP-based process discovery.
	In \autoref{sec:related_work} we discuss related work.
	In \autoref{sec:preliminaries} we present background related to event logs, Petri nets and region theory.
	In \autoref{sec:regions} we show how to incorporate regions within process discovery.
	In \autoref{sec:disc_wf_nets} we show that we are able to guarantee discovery of relax sound workflow nets.
	In \autoref{sec:eliminating_replay_fitness} we present an integrated effective algorithm to eliminate infrequent exceptional behaviour.
	In \autoref{sec:evaluation} we present an evaluation of the proposed approach.
	\autoref{sec:conclusion} concludes the paper.
	
	\section{Motivation}
	\label{sec:motivation}
	A multitude of process discovery algorithms exists~\cite{DBLP:books/sp/Aalst16, DBLP:journals/topnoc/DongenMW09, DBLP:journals/is/WeerdtBVB12}.
	Some notable algorithms concern the $\alpha$-Miner Family~\cite{DBLP:journals/tkde/AalstWM04,DBLP:conf/caise/MedeirosDAW04,DBLP:journals/datamine/WenAWS07,DBLP:journals/dke/WenWAHS10}, the Heuristics Miner~\cite{DBLP:journals/icae/WeijtersA03,DBLP:conf/cidm/WeijtersR11}, the Evolutionary Tree Miner (ETM)~\cite{DBLP:conf/cec/BuijsDA12} and the Inductive Miner~\cite{DBLP:conf/apn/LeemansFA13,DBLP:conf/bpm/LeemansFA13}.
	However, there are good reasons to study and develop region-based techniques.
	Most prominently because they allow us to discover \textit{complex non-local control-flow patterns}.

\begin{figure}[b]
		\centering
		\includegraphics[width=0.9\textwidth]{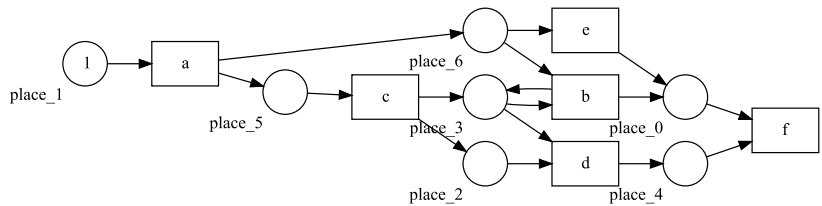}
		\caption{Result of ILP Based Discovery~\cite{DBLP:journals/fuin/derWerfDHS09}, containing a milestone pattern.}
		\label{fig:milestone_ilp}
	\end{figure}

	Consider the following set of sequences of executed business process activities: $\langle a,c,d,e,f \rangle$, $\langle a, c, b, d, f \rangle$, $\langle a,c,e,d,f \rangle$ and $\langle a,e,c,d,f \rangle$.
	If we apply ILP-Based process discovery~\cite{DBLP:journals/fuin/derWerfDHS09}, i.e. using region theory, we obtain the process model in \autoref{fig:milestone_ilp}.

	The model describes that activity $a$ is always executed first.
	After activity $a$ we are able to execute activity $c$ and $e$ in any order, i.e. they are in a parallel construct.
	However, after executing activity $c$ we are able to perform activity $b$ instead of $e$.
	However, this is only possible as long as we do not execute activity $d$, which we are able to execute after we have executed $c$.
	Finally we always execute activity $f$.	
	In the model, the choice of executing activity $b$ instead of $e$ is influenced by the global state of the system.
	Such pattern is called a \textit{milestone pattern}~\cite{DBLP:journals/dpd/AalstHKB03}. 
	
\begin{figure}[tb]
		\centering
		\begin{subfigure}[b]{\textwidth}
			\centering
			\includegraphics[width=0.9\textwidth]{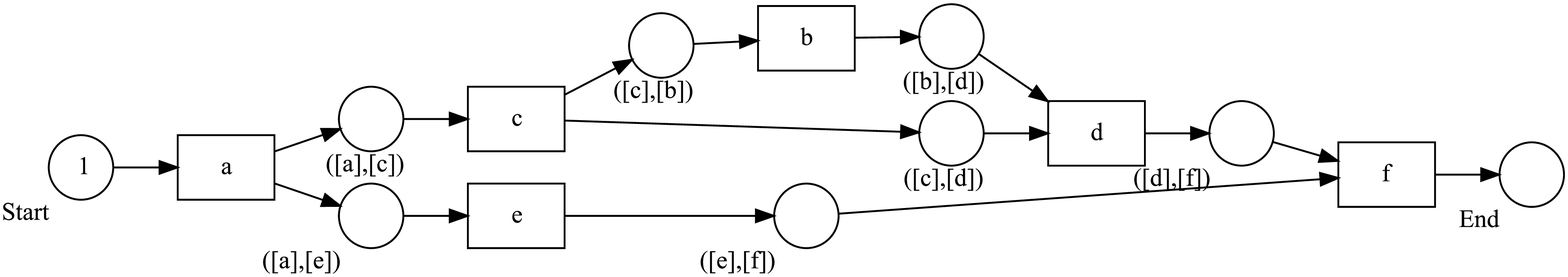}
			\caption{$\alpha$-, $\alpha^{+}$-, $\alpha^{++}$-Miner~\cite{DBLP:journals/tkde/AalstWM04,DBLP:conf/caise/MedeirosDAW04,DBLP:journals/datamine/WenAWS07}.}
			\label{fig:milestone_alpha}
		\end{subfigure}
		\begin{subfigure}[b]{\textwidth}
			\centering
			\includegraphics[width=0.9\textwidth]{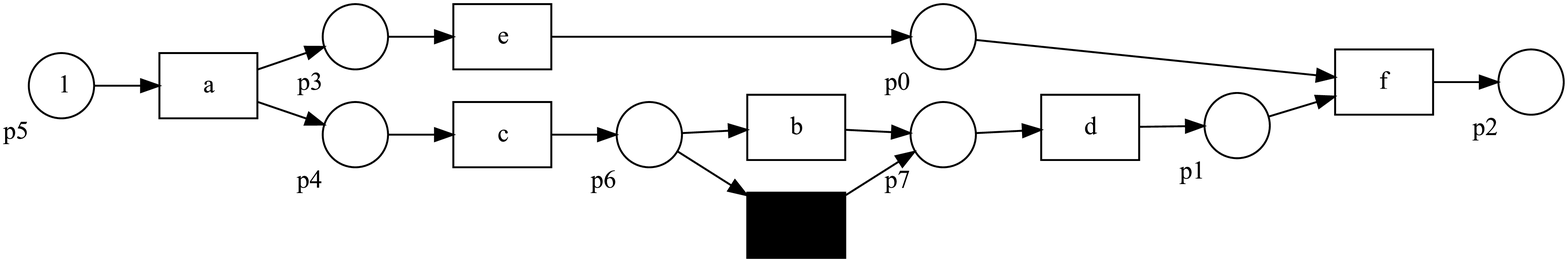}
			\caption{$\alpha^{\sharp}$-Miner~\cite{DBLP:journals/dke/WenWAHS10}.}
			\label{fig:milestone_alpha_sharp}
		\end{subfigure}
		\begin{subfigure}[b]{\textwidth}
			\centering
			\includegraphics[width=0.9\textwidth]{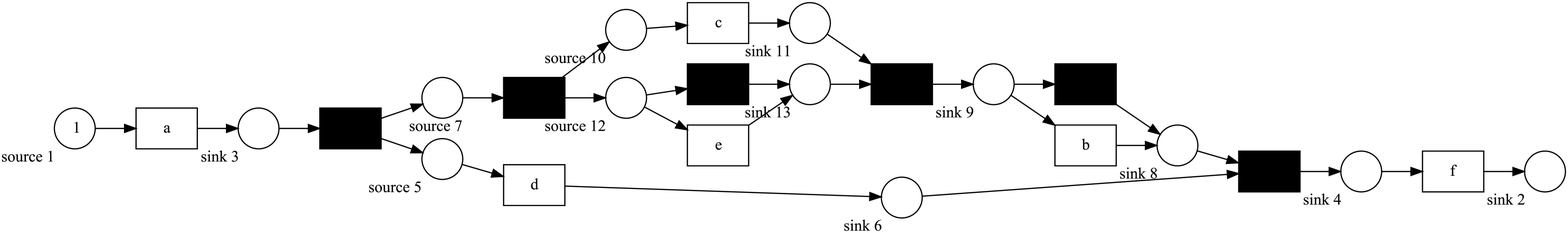}
			\caption{Inductive-Miner~\cite{DBLP:conf/apn/LeemansFA13}.}
			\label{fig:milestone_inductive}
		\end{subfigure}
		\begin{subfigure}[b]{\textwidth}
			\centering
			\includegraphics[width=0.9\textwidth]{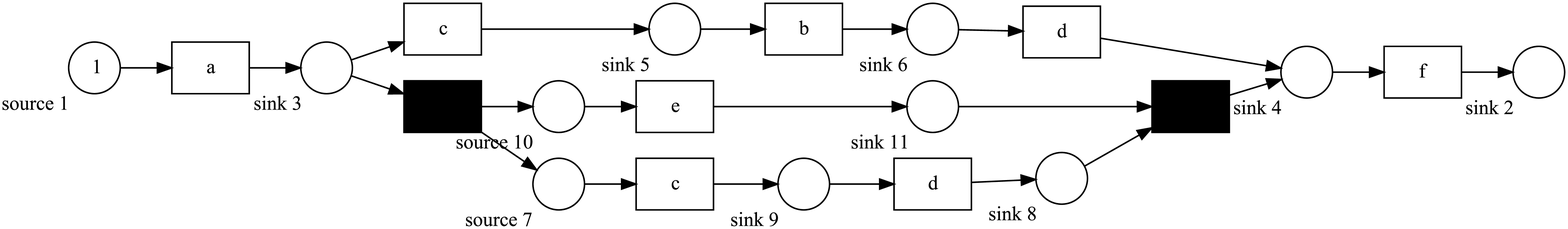}
			\caption{Evolutionary Tree Miner~\cite{DBLP:conf/cec/BuijsDA12}.}
			\label{fig:milestone_etm}
		\end{subfigure}
		\begin{subfigure}[b]{\textwidth}
			\centering
			\includegraphics[width=0.9\textwidth]{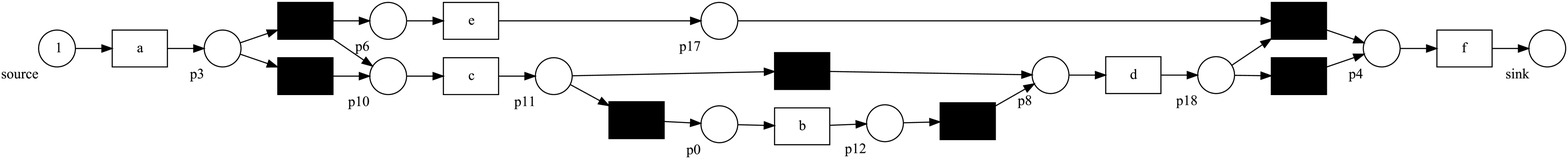}
			\caption{Heuristics Miner~\cite{DBLP:journals/icae/WeijtersA03,DBLP:conf/cidm/WeijtersR11}.}
			\label{fig:milestone_heuristics}
		\end{subfigure}
		\caption{Results of several state-of-the-art discovery algorithms in the ProM Framework~\cite{DBLP:conf/caise/VerbeekBDA10}, when given sequences containing a milestone pattern.}
		\label{fig:milestone_disc}
	\end{figure}

	If we apply the aforementioned state-of-the-art discovery algorithms on these same data, we obtain the models depicted in \autoref{fig:milestone_disc}.
	None of the models adequately describes the milestone pattern.
	Some models do not even guarantee perfect replay-fitness, i.e. \autoref{fig:milestone_alpha}.
	Other models, such as the model in \autoref{fig:milestone_inductive} have very low precision.
	The only model that actually describes the same behaviour is the model in \autoref{fig:milestone_etm}.
	However, the model does not capture the milestone pattern, i.e. we need to analyse the behaviour of the model to derive the conclusion that it describes the behaviour of a milestone pattern.	
	
\begin{figure}[tb]
		\centering
		\includegraphics[width=0.9\textwidth]{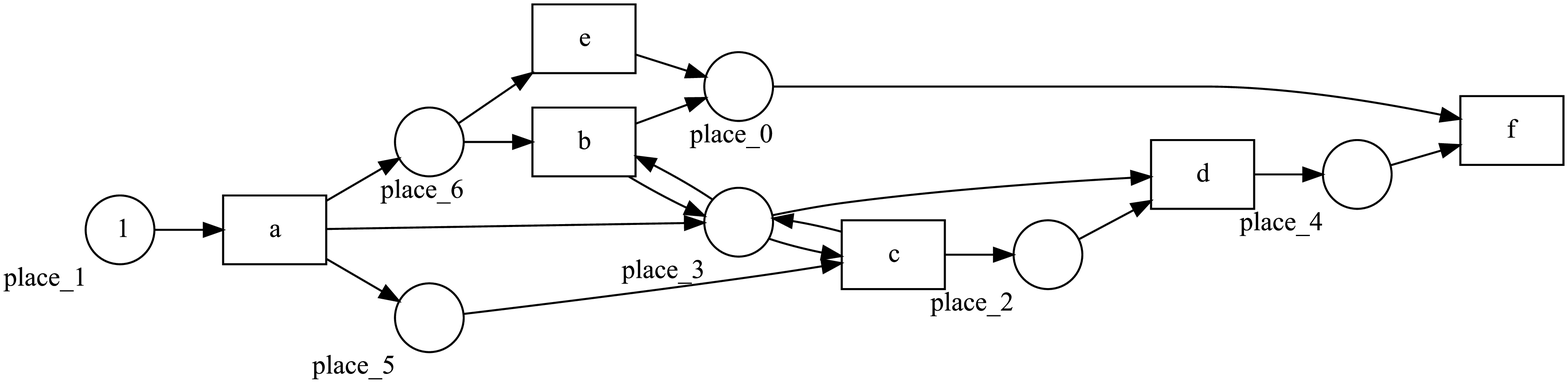}
		\caption{Result of applying ILP Based Discovery~\cite{DBLP:journals/fuin/derWerfDHS09} on an event log containing a milestone pattern, and, infrequent (faulty) behaviour.}
		\label{fig:milestone_ilp_noise}
	\end{figure}

	As the example shows there is a clear incentive for process discovery algorithms based on (language-based) region theory.
	However, the state-of-the-art technique based on language-based region theory~\cite{DBLP:journals/fuin/derWerfDHS09} has a number of deficiencies.
	It is not able to guarantee that the resulting Petri net is a workflow net, i.e. a Petri net with favourable graph-theoretical properties.
	For example, the Petri net in \autoref{fig:milestone_ilp} does not have a unique sink place.
	The Inductive Miner for example does guarantee to return (sound) workflow nets.
	Moreover, ILP-based process discovery greatly suffers from the presence of infrequent and/or exceptional behaviour.
	Assume we have an event log containing thousands of repetitions of the aforementioned sequences $\langle a,c,d,e,f \rangle$, $\langle a, c, b, d, f \rangle$, $\langle a,c,e,d,f \rangle$ and $\langle a,e,c,d,f \rangle$.
	If we inject just one exceptional sequence, e.g. $\langle a,b,c,d,f \rangle$, and apply the current state-of-art region-based discovery algorithm, we obtain the model depicted in \autoref{fig:milestone_ilp_noise}.
	Within the model, activity $b$ is now able to occur in parallel with activity $c$.	
	Thus, by adding one infrequent, exceptional sequence, the algorithm is no longer able to detect the milestone pattern.	
	
	In this paper we solve the two aforementioned issues.
	Firstly, we present an approach that guarantees to find relaxed sound workflow nets.
	Secondly, we present an effective integrated filtering technique that identifies and ignores infrequent and/or exceptional behaviour.

	\section{Related Work}
	\label{sec:related_work}
	We predominantly focus on related work in the area of region theory and its application to process discovery.
	We also focus on filtering techniques for process discovery.
	For a detailed overview of process discovery algorithms we refer to~\cite{DBLP:books/sp/Aalst16, DBLP:journals/topnoc/DongenMW09, DBLP:journals/is/WeerdtBVB12}.

	\paragraph{Region Theory and Petri Net Synthesis}
	\label{subsec:sec:preliminaries_subsec:petri_net_synthesis}
	Region theory is a solution to the Petri net synthesis problem~\cite{DBLP:journals/topnoc/Reisig13}.
	The two terms are therefore often used interchangeably.
	The synthesis problem is to, given a behavioural system description, decide whether there exists a Petri net that allows for all behaviour described by the system description, and, at the same time minimizes additional behaviour. 
	Initial work focused on solving the synthesis problem using \textit{transition systems} as a system description~\cite{DBLP:journals/acta/EhrenfeuchtR89,DBLP:journals/acta/EhrenfeuchtR89a, DBLP:conf/apn/Bernardinello93}, i.e. \textit{state-based region theory}.
	A set of states within the transition system forms a \textit{region}, which defines a place in the resulting Petri net.
	Region theory has also been applied using a \textit{prefix-closed language} as a system description~\cite{DBLP:conf/tapsoft/BadouelBD95, DBLP:conf/concur/Darondeau98}, i.e. \textit{language-based region theory}.
	Here, a region is an assignment of decision variables over the language's alphabet, again defining a place in the resulting Petri net.
	Finally, language-based region theory has also been extended for labelled partial orders~\cite{DBLP:conf/apn/LorenzJ06,DBLP:conf/wsc/LorenzMJ07,DBLP:journals/fuin/BergenthumDLM08}.
	
	\paragraph{Process Discovery}
	In contrast to Petri net synthesis, process discovery aims at extracting a \textit{generalizing process model} from an \textit{incomplete} behavioural system description, i.e. an event log.
	Additionally, we typically need to abstract from infrequent behaviour in order to focus on the mainstream behaviour in the event log.
	
	In~\cite{DBLP:journals/sosym/AalstRVDKG10} a process discovery approach is presented that transforms an event log into a transition system, after which state-based region theory is applied.
	Constructing the transition system is strongly parametrized, i.e. using different parameters yields different process discovery results.
	In~\cite{DBLP:conf/apn/SoleC10} a similar approach is presented. The main contribution is a complexity reduction w.r.t. conventional region-based techniques.
	
	In~\cite{DBLP:conf/bpm/BergenthumDLM07} a process discovery approach is presented based on language-based region theory.
	The method finds a minimal linear basis of a polyhedral cone of integer points, based on the event log.
	It guarantees perfect replay-fitness, whereas it does not maximize precision.
	The worst-case time complexity of the approach is exponential in the size of the event log.
	In~\cite{DBLP:journals/tkde/CarmonaC14} a process discovery algorithm is proposed based on the concept of numerical abstract domains. 
	Based on the event log's prefix-closure a convex polyhedron is approximated by means of calculating a convex hull. 
	The convex hull is used to compute causalities within the input log by deducing a set of linear inequalities which represent places. 
	In~\cite{DBLP:journals/fuin/derWerfDHS09} a first design of a process discovery ILP-formulation is presented.
	An objective function is presented, which is generalized in~\cite{DBLP:conf/apn/ZelstDA15}, that allows for expressing a preference for finding certain Petri net places.
	The work also presents means to formulate ILP-constraints that help finding more advanced Petri net-types, e.g. Petri nets with reset- and inhibitor arcs.
	
	All aforementioned techniques leverage the strict implications of region theory w.r.t. process discovery, i.e. precision maximization, poor generalization and poor simplicity, to some extend.
	However, the techniques still perform suboptimal.
	Since the techniques guarantee perfect replay-fitness, they tend to fail if exceptional behaviour is present in the event log, i.e. they produce models that are incorporating infrequent behaviour (outliers).
	
	\paragraph{Filtering Infrequent Behaviour}
	Little work has been done regarding filtering of infrequent behaviour in context of process mining.
	The majority of work concerns unpublished/undocumented ad-hoc filtering implementations in the ProM framework~\cite{DBLP:conf/caise/VerbeekBDA10}.	
	
	In~\cite{DBLP:journals/tkde/ConfortiRH17} an event log filtering technique is presented that filters on \textit{event level}.
	Events within the event log are removed in case they do not fit an underlying, event log based, automaton.
	The technique can be used as a \textit{pre-processing} step prior to invoking a discovery algorithm.
	
	In~\cite{DBLP:conf/bpm/LeemansFA13} Leemans et al. show how to extend the Inductive Miner~\cite{DBLP:conf/apn/LeemansFA13} with filtering capabilities to handle infrequent behaviour.
	The technique is tailored towards the internal working of the Inductive Miner algorithm and considers three different types of filters.
	Moreover, the technique exploits the inductive nature of the underlying algorithm, i.e. filters are applied on multiple levels.
	
	\section{Background}
	\label{sec:preliminaries}
	In this section we present basic notational conventions, event logs and workflow nets.
	\subsection{Bags, Sequences and Vectors}
	\label{subsec:sec:preliminaries_subsec:bags}
	$\setGeneric = \{e_1,e_2, ..., e_n\}$ denotes a set. 
	$\powerSet(\setGeneric)$ denotes the power set of $\setGeneric$.
	$\mathbb{N}$ denotes the set of positive integers \textit{including} $0$ whereas $\mathbb{N}^+$ \textit{excludes} $0$. 
	$\mathbb{R}$ denotes the set of real numbers.
	A bag (bag) over $\setGeneric$ is a function $\bag : \setGeneric \rightarrow \naturals$ which we write as $[e_1^{v_1}, e_2^{v_2}, ..., e_n^{v_n}]$, where for $1 \leq i \leq n$ we have $e_i \in \setGeneric$, $v_i \in \mathbb{N}^+$ and $e_i^{v_i} \equiv B(e_i) = v_i$. 
	If for some element $e$, $B(e) = 1$, we omit its superscript. 
	An empty bag is denoted as $\emptyset$. 
	Element inclusion applies to bags: if $e \in \setGeneric$ and $\bag(e) > 0$ then also $e \in B$.
	Set operations, i.e. $\uplus$, $\setminus$, $\cap$, extend to bags.
	The set of all bags over $\setGeneric$ is denoted $\allBags(\setGeneric)$.
	
	A \textit{sequence} $\sigma$ of length $k$ relates positions to elements $\element \in \setGeneric$, i.e. $\sigma : \{1,2,...,k\} \rightarrow \setGeneric$. 
	An empty sequence is denoted as $\epsilon$. 
	We write every non-empty sequence as $\langle e_1, e_2, ..., e_k \rangle$.
	The set of all possible sequences over a set $\setGeneric$ is denoted as $\setGeneric^*$. We define \textit{concatenation} of sequences $\sigma_1$ and $\sigma_2$ as $\sigma_1 \cdot \sigma_2$, e.g., $\langle a , b \rangle \cdot \langle c, d \rangle = \langle a,b,c,d \rangle$. 
	Let $\setGeneric' \subseteq \setGeneric$, we define $\downarrow_{\setGeneric'} \colon \setGeneric^* \to \setGeneric'^*$ recursively with $\downarrow_{\setGeneric'}(\epsilon) = \epsilon$ and $\downarrow_{\setGeneric'}(\langle x \rangle \cdot \sequence) = \langle x \rangle \cdot \downarrow_{\setGeneric'}(\sequence)$ if $x \in \setGeneric'$ and $\downarrow_{\setGeneric'}(\sequence)$ otherwise.
	We write $\sequence_{\downarrow_{X'}}$ for $\downarrow_{\setGeneric'}(\sequence)$.
	
	Given $\setGeneric' \subseteq \setGeneric^*$, the \textit{prefix-closure} of $\setGeneric'$ is: $\overline{\setGeneric'} = \{\sigma_1 \in \setGeneric^* | \exists_{\sigma_2 \in \setGeneric^*} (\sigma_1 \cdot \sigma_2 \in \setGeneric')\}$.
	We extend the notion of a prefix-closure on bags of sequences.
	Let $\setGeneric' \subseteq \setGeneric^*$ and $\bag_{\setGeneric'}: \setGeneric' \rightarrow \mathbb{N}$ we define $\overline{B}_{\setGeneric'} : \overline{\setGeneric'} \rightarrow \mathbb{N}$, such that:
	$
	\overline{B}_{\setGeneric'}(\sigma) = B_{\setGeneric}(\sigma) + \sum_{\sigma \cdot \langle e \rangle \in \overline{\setGeneric'}} \overline{B}_{\setGeneric'}(\sigma \cdot \langle e \rangle)
	$.
	For example, $\bag_2 = [\langle a,b \rangle^5, \langle a,c \rangle^3]$ yields $\overline{\bag}_2 = [\epsilon^{8}, \langle a \rangle^{8},\langle a,b \rangle^5, \langle a,c \rangle^3]$.

	Given set $\setGeneric$ and a range of values $R \subseteq \mathbb{R}$. 
	Vectors are denoted as $\vec{z} \in R^{|\setGeneric|}$, where $\vec{z}(e) \in R$ and $e \in \setGeneric$. 
	We assume vectors to be \textit{column vectors}. 
	For vector multiplication we assume that vectors agree on their indices.
	Throughout the paper we assume a \textit{total ordering on sets of the same domain}.
	Given $\setGeneric = \{e_1, e_2, ..., e_n\}$ and $\vec{z}_1, \vec{z}_2 \in R^{|\setGeneric|}$ we have $\vec{z}_1^{\intercal} \vec{z}_2 = \sum_{i=1}^{n} \vec{z}_1(e_i) \vec{z}_2(e_i)$. 
	A \textit{Parikh vector} $\vec{p}$ represents the number of occurrences of an element within a sequence, i.e. $\vec{p} : \setGeneric^* \rightarrow \mathbb{N}^{|\setGeneric|}$ with $\vec{p}(\sigma) = (\parikh{\sigma}{e_1}, \parikh{\sigma}{e_2}, ..., \parikh{\sigma}{e_n})$ where $\parikh{\sigma}{e_i}= |\{i' \in \{1,2, ..., |\sigma|\} \mid \sigma(i') = e_i\}|$.
	
	\subsection{Event Logs and Workflow Nets}
	\label{subsec:sec:preliminaries_subsec:event_logs_and_nets}
	In process discovery an event log acts as a main source of input and describes the actual execution of activities in context of a business process. 
	An example event log, adopted from~\cite{DBLP:books/sp/Aalst16}, is presented in \autoref{tab:sec:preliminaries_subsec:event_logs_and_languages_tab:example_event_log}.
	\begin{table}[tb]
		\caption{Fragment of a fictional event log (each line corresponds to an event).}
		\label{tab:sec:preliminaries_subsec:event_logs_and_languages_tab:example_event_log}
		\begin{center}
			\begin{tabular}{|c|c|c|c|}
				\hline
				\textbf{Case-id} & \textbf{Activity} & \textbf{Resource} & \textbf{Time-stamp} \\
				\hline
				... & ... & ... & ...\\
				\hline
				\textit{1} & \textit{register request ($a$)} & \textit{John} & \textit{2015-05-08:08.45} \\
				\hline
				\textit{1} & \textit{examine thoroughly ($b$)} & \textit{Lucy} & \textit{2015-05-08:09.13} \\
				\hline
				\textit{2} & \textit{register request ($a$)} & \textit{John} & \textit{2015-05-08:09.14} \\
				\hline
				\textit{2} & \textit{check ticket ($d$)} & \textit{Pete} & \textit{2015-05-08:10.11} \\
				\hline
				\textit{1} & \textit{check ticket ($d$)} & \textit{Pete} & \textit{2015-05-08:10.28} \\
				\hline
				\textit{2} & \textit{examine causally ($b$)} & \textit{Rob} & \textit{2015-05-08:10.43} \\
				\hline
				\textit{1} & \textit{decide ($e$)} & \textit{Rob} & \textit{2015-05-08:11.14} \\
				\hline
				\textit{1} & \textit{reject request ($h$)} & \textit{Rob} & \textit{2015-05-08:11.35} \\
				\hline
				... & ... & ... & ...\\
				\hline
			\end{tabular}
		\end{center}
	\end{table}
	Consider all activities related to \textit{Case-id 1}. 
	John \textit{registers a request}, after which Lucy \textit{examines it thoroughly}.
	Pete \textit{checks the ticket} after which Rob \textit{decides} to \textit{reject the request}. 
	The execution of an \textit{activity} in context of a business process is referred to as an \textit{event}.
	A sequence of events, e.g. the sequence of events related to case \textit{1}, is referred to as a \textit{trace}.

	Let $\activityUniverse$ denote the universe of all possible activities.
	An event log $\eventLog$ is a bag of sequences over $\activityUniverse$, i.e., $\eventLog \in \allBags(\activityUniverse^*)$.
	Typically, there exists $\activities_{\eventLog} \subset \activityUniverse$ of activities that are actually present in $\eventLog$.
	In some cases we refer to an event log as $\eventLog \in \allBags(\activities_{\eventLog}^*)$.
	A sequence $\sequence \in \eventLog$ represents a trace.
	We write case \textit{1} as trace $\langle$\textit{``register request",``examine thoroughly", ``check ticket", ``decide", ``reject request"}$\rangle$.
	In the remainder of the paper we use simple characters for activity names, e.g. we write case \textit{1} as $\langle a,b,d,e,h \rangle$.
	
	The goal within process discovery is to discover a process model based on an event log. 
	In this paper we consider \textit{workflow nets (WF-nets)}~\cite{DBLP:journals/jcsc/Aalst98}, based on \textit{Petri nets}~\cite{murata_1989_petri_nets}, to describe process models. 
	We first introduce Petri nets and their execution semantics, after which we define workflow nets.
	
\begin{figure}[tb]
		\centering
		\includegraphics[width=0.85\textwidth]{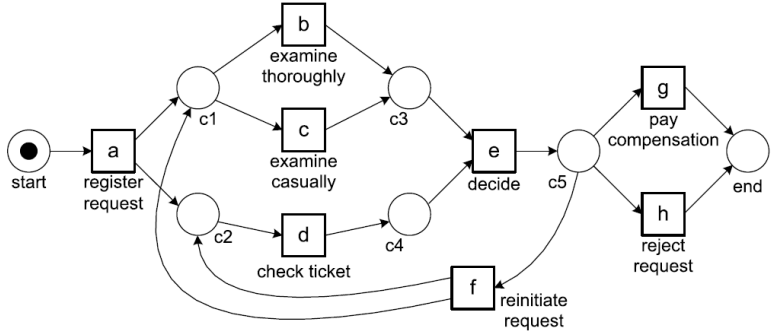}
		\caption{Example WF-net $W_1$, adopted from~\cite{DBLP:books/sp/Aalst16}.}
		\label{fig:example_wf_net}
	\end{figure}

	A Petri net is a \textit{bipartite graph} consisting of a set of vertices called \textit{places} and a set of vertices called \textit{transitions}. 
	Arcs connect places with transitions and vice versa. 
	Additionally, transitions have a (possibly unobservable) label which describes the activity that the transition represents.
	A Petri net is a quadruple $\petriNet = (\places,\transitions,\flowRelation, \labelFunc)$, where $\places$ is a set of places and $\transitions$ is a set of transitions with $\places \cap \transitions = \emptyset$. 
	$\flowRelation$ denotes the flow relation of $\petriNet$, i.e., $\flowRelation \subseteq (\places \times \transitions) \cup (\transitions \times \places)$.
	$\labelFunc$ denotes the label function, i.e. given a set of activities $\labels \subset \activityUniverse$ and an \textit{unobservable activity} $\silent \notin \labels$, it is defined as $\labelFunc \colon \transitions \to \labels \cup \{ \silent \}$.
	For a node $x \in \places \cup \transitions$, the pre-set of $x$ in $\petriNet$ is defined as $\bullet x = \{y \mid (y,x) \in \flowRelation\}$ and $x \bullet = \{y \mid (x,y) \in \flowRelation\}$ denotes the post-set of $x$.
	Graphically we represent places as \textit{circles} and transitions as \textit{boxes}.
	For every $(x,y) \in F$ we draw an \textit{arc} from $x$ to $y$.
	An example Petri net (which is also a WF-net) is depicted in \autoref{fig:example_wf_net}.
	Observe that we have $\bullet d = \{c_2\}$, $d \bullet = \{c_4\}$ and $\labelFunc(d) =$``reject request''.
	The Petri net does not contain any silent transition.
	
	The execution semantics of Petri nets are based on the concept of \textit{markings}
	A marking $\marking$ is a bag of tokens, i.e. $\marking \in \allBags(\places)$.
	Graphically, a place $\place$'s marking is visualized by drawing $\marking(\place)$ number of dots inside place $p$, e.g. place ``start'' in \autoref{fig:example_wf_net}.
	A \textit{marked Petri net} is a $2$-tuple $(\petriNet,\marking)$, where $\marking$ represents $\petriNet$'s marking.
	We let $\marking_i$ denote $\petriNet$'s \textit{initial marking}.
	Transition $\transition \in \transitions$ is \textit{enabled} in marking $\marking$ if $\forall_{\place \in \bullet \transition}(\marking(\place) > 0)$.
	Enabled transition $\transition$ in marking $\marking$, may \textit{fire}, which results in new marking $\marking'$.
	If $\transition$ fires, denoted as $(\petriNet,\marking)\xrightarrow{\transition}(\petriNet,\marking')$, then for each $\place \in \places$ we have $\marking'(\place) = \marking(\place) - 1$ if $\place \in \bullet \transition \setminus \transition \bullet$, $\marking'(\place) = \marking(\place) + 1$ if $\place \in \transition \bullet \setminus \bullet \transition$, and, $\marking'(\place) = \marking(\place)$ otherwise, e.g. in \autoref{fig:example_wf_net} we have $(\workflowNet_1,[start])\xrightarrow{a}(\workflowNet_1,[c_1, c_2])$.
	Given sequence $\sequence = \langle \transition_1, \transition_2, ..., \transition_n \rangle \in \transitions^*$, $\sequence$ is a \textit{firing sequence} of $(\petriNet,\marking)$, written as $(\petriNet,\marking)\xrightarrowdbl{\sequence}(N,M')$ if and only if for $n = |\sequence|$ there exist markings $\marking_1,\marking_2,...,\marking_{n-1}$ such that $(\petriNet,\marking)\xrightarrow{t_1}(\petriNet,\marking_1)$, $(\petriNet,\marking_1)\xrightarrow{t_2}(\petriNet,\marking_2),...,(\petriNet,\marking_{n-1})\xrightarrow{t_n}(\petriNet,\marking')$.
	We write $(\petriNet, \marking) \xrightarrowdbl{\sequence} *$ if there exists a marking $\marking'$ s.t. $(\petriNet, \marking) \xrightarrowdbl{\sequence} (\petriNet, \marking')$.
	We write $(\petriNet, \marking) \rightsquigarrow (\petriNet, \marking')$ if there exists $\sequence \in \transitions^*$ s.t. $(\petriNet, \marking) \xrightarrowdbl{\sequence}(\petriNet,\marking')$.
	We define $\petriNet$'s language as $\lang(\petriNet, \marking_i) = \{ \sequence \in \transitions^* \mid \marking_i \xrightarrowdbl{\sequence} * \}$, i.e. $\lang(\petriNet, \marking_i)$ is \textit{prefix-closed}.
	We define $\petriNet$'s \textit{labelled language} as $\lang^{\labels}(\petriNet, \marking_i) = \{ \sequence \in \labels^* \mid \exists_{\sequence' \in \transitions^*, \sequence'' \in (\labels \cup \{\silent\})^*}(\sequence' \in \lang_(\petriNet, \marking_i) \wedge |\sequence'| = |\sequence''| \wedge \forall_{i \in \{1,2,..., |\sequence'|\}}(\labelFunc(\sequence'(i)) = \sequence''(i)) \wedge \sequence''_{\downarrow_{\labels}} = \sequence \}$.
	Given a final marking $\marking_f$, we define $\petriNet$'s \textit{execution language} w.r.t. $\marking_f$ as $\lang_{\marking_f}(\petriNet, \marking_i) = \{\sequence \in \transitions^* \mid (\petriNet,\marking_i)\xrightarrowdbl{\sigma} (\petriNet,\marking_f)\}$.
	We define $\petriNet$'s \textit{labelled execution language} as $\lang^{\labels}_{\marking_f}(\petriNet, \marking_i) = \{ \sequence \in \labels^* \mid \exists_{\sequence' \in \transitions^*, \sequence'' \in (\labels \cup \{\silent\})^*}(\sequence' \in \lang_{\marking_f}(\petriNet, \marking_i) \wedge |\sequence'| = |\sequence''| \wedge \forall_{i \in \{1,2,..., |\sequence'|\}}(\labelFunc(\sequence'(i)) = \sequence''(i)) \wedge \sequence''_{\downarrow_{\labels}} = \sequence \}$

	WF-nets extend Petri nets and require the existence of a unique \textit{source-} and \textit{sink place} which describe the start, respectively end, of a case.
	Moreover, each element within the WF-net needs to be on a path from the source to the sink place.		
	\begin{definition}[Workflow net~\cite{DBLP:journals/jcsc/Aalst98}]
		\label{def:workflow_net}
		Let $N = (P,T,F, \labelFunc)$ be a Petri net.
		Let $\place_i,\place_o \in \places$ with $\place_i \neq \place_o$.
		Let $\labels \subset \activityUniverse$ be a set of activities, let $\silent \notin \labels$ and let $\labelFunc \colon \transitions \to \labels \cup \{\silent\}$.
		Tuple $\workflowNet = (\places, \transitions, \flowRelation, \place_i, \place_o, \labelFunc)$ is a workflow net (WF-net) if and only if:
		\begin{enumerate}
			\item $\bullet \place_i = \emptyset$ 
			\item $\place_o \bullet = \emptyset$ 
			\item Each element $x \in \places \cup \transitions$ is on a path from $\place_i$ to $\place_o$.
		\end{enumerate}
	\end{definition}
	
	The execution semantics defined for Petri nets can directly be applied on the elements $\places$, $\transitions$ and $\flowRelation$ of $\workflowNet = (\places, \transitions, F, \place_i, \place_o, \labelFunc)$.
	Notation-wise we substitute $\workflowNet$ for its underlying net structure $\petriNet = (\places, \transitions, \flowRelation)$, e.g. $(\workflowNet, \marking)\xrightarrow{t}(\workflowNet, \marking')$, $\lang(\workflowNet, \marking_i)$ etc.
	In context of WF-nets, we assume $\marking_i = [\place_i]$ and $\marking_f = [\place_o]$ unless mentioned otherwise.

	We compute metrics such as replay-fitness and precision, as introduced in the introduction, based on an event log.
	Several behavioural quality metrics, that do not need any form of domain knowledge, exist for WF-nets.
	Several notions of \textit{soundness} of WF-nets are defined~\cite{DBLP:journals/fac/AalstHHSVVW11}.
	For example, \textit{classical sound} WF-nets are guaranteed to be free of livelocks, deadlocks, and other anomalies that can be detected automatically.
	In this paper we consider the weaker notion of \textit{relaxed soundness}.
	Relaxed soundness requires that each transition is at some point enabled, and, after firing such transition we are able to eventually reach the final marking.
	
	\begin{definition}[Relaxed Soundness~\cite{DBLP:journals/fac/AalstHHSVVW11}]
		Let $\workflowNet = (\places, \transitions, \flowRelation, \place_i, \place_o, \labelFunc)$ be a WF-net.
		$\workflowNet$ is relaxed sound if and only if: $\forall_{\transition \in \transitions}(\exists_{\marking, \marking' \in \allBags(\places)}((\workflowNet,[\place_i]) \rightsquigarrow (\workflowNet,\marking) \wedge (\workflowNet,\marking) \xrightarrow{\transition} (\workflowNet, \marking') \wedge (\workflowNet,\marking') \rightsquigarrow (\workflowNet, [\place_o])))$.
	\end{definition}
	
	Reconsider $\workflowNet_1$ (\autoref{fig:example_wf_net}) and assume we are given an event log with one trace: $\langle a,b,d,e,h \rangle$.
	It is quite easy to see that $\workflowNet_1$ is relaxed sound.
	Moreover, replay-fitness is perfect, i.e. $\langle a,b,d,e,h \rangle$ is in the WF-net's labelled execution language.
	Precision is not perfect as the WF-net can produce a lot more traces than just $\langle a,b,d,e,h \rangle$.

	\section{Discovering Petri Net Places using Integer Linear Programming}
	\label{sec:regions}
	In this section we show how to, given an event log as an input, discover multiple places of a Petri net using language based regions.

	\subsection{Regions}
	Conceptually, a region represents a place in a Petri net that, given the prefix-closure of an event log, does not block the execution of any sequence within the prefix-closure.
	We represent a region as an assignment of binary decision variables describing the incoming and outgoing arcs of its corresponding place, as well as its marking.
	
	\begin{definition}[Region]
		\label{def:sec:regions_def:region}
		Given an event log $\eventLog$ over a set of activities $\activities_\eventLog$.
		Let $m \in \{0,1\}$ and $\vec{x},\vec{y} \in \{0,1\}^{|\activities_\eventLog|}$.
		A triple $r = (m,\vec{x},\vec{y})$ is a region if and only if:
		\begin{equation}
		\label{eq:sec:regions_def:region_eq:region}
		\forall_{\sequence = \sequence' \cdot \langle a \rangle \in \overline{\eventLog}} (m + \vec{p}(\sigma')^{\intercal}\vec{x} - \vec{p}(\sigma)^{\intercal}\vec{y} \geq 0 )
		\end{equation}
	\end{definition}
	
	Variable $m$ indicates whether or not the region's corresponding place contains a token, $\vec{x}$ denotes incoming arcs and $\vec{y}$ denotes outgoing arcs.
	Consider event log $\eventLog_1 = [\langle a,b,d,e,g \rangle^{10}, \langle a,c,d,e,f,d,b,e,g \rangle^{12}, \langle a,d,c,e,h \rangle^{9}, \langle a,b,d,e,f,c,d,e,g \rangle^{11}, \langle a,d,c,e,f,b,d,e,h \rangle^{13}]$.
	In \autoref{tab:sec:regions_tab:linear_inequalities_example} we depict a part of the corresponding set of linear inequalities based on \autoref{def:sec:regions_def:region}.
	\begin{table}[tb]
		\caption{Linear inequalities corresponding to event log $L_1$ based on \autoref{eq:sec:regions_def:region_eq:region}.}
		\label{tab:sec:regions_tab:linear_inequalities_example}
		\begin{center}
			\resizebox{1\textwidth}{!}{
				\begin{tabular}{ll}
					$m - \vec{y}(a) \geq 0$ & $\langle a \rangle$ \\
					$m + \vec{x}(a) - \vec{y}(a) - \vec{y}(b) \geq 0$ & $\langle a,b \rangle$ \\
					$m + \vec{x}(a) - \vec{y}(a) - \vec{y}(c) \geq 0$ & $\langle a,c \rangle$ \\
					$m + \vec{x}(a) + \vec{x}(b) - \vec{y}(a) - \vec{y}(b) - \vec{y}(d) \geq 0$ & $\langle a,b,d \rangle$ \\ 
					$\vdots$ & $\vdots$ \\
					$m + \vec{x}(a) + \vec{x}(b) + \vec{x}(c) + 2 \vec{x}(d) + 2 \vec{x}(e) + \vec{x}(f) - \vec{y}(a) - \vec{y}(b) - \vec{y}(c) - 2 \vec{y}(d) - 2 \vec{y}(e) - \vec{y}(f) - \vec{y}(h) \geq 0$ & $\langle a,d,c,e,f,b,d,e,h \rangle$ \\
				\end{tabular}
			}
		\end{center}
	\end{table}
	For every non-empty sequence in $\overline{L_1}$, i.e. $\langle a \rangle$, $\langle a,b \rangle$, ..., $\langle a,d,c,e,f,b,d,e,h \rangle$ there is an associated linear inequality in terms of the variables $m$, $\vec{x}$ and $\vec{y}$.
	For example, $\langle a \rangle$ leads to $m - \vec{y}(a)$, $\langle a, b \rangle$ leads to $m + \vec{x}(a) - \vec{y}(a) - \vec{y}(b)$ etc.
	Note that the inequalities abstract from the ordering of activities in traces, e.g. $\langle a,c,d,e \rangle$ and $\langle a,d,c,e \rangle$ both map to $m + \vec{x}(a) + \vec{x}(c) + \vec{x}(d) - \vec{y}(a) - \vec{y}(c) - \vec{y}(d) - \vec{y}(e)$.
	
	A region $\region$ is translated to a Petri net place $\place$ as follows.
	Given a Petri net that has a unique transition $\transition_{\activity}$ for each $\activity \in \activities_{\eventLog}$ such that $\labelFunc(\transition_{\activity}) = \activity$.
	If, for $\activity \in \activities_{\eventLog}$ $\vec{x}(\activity) = 1$, we add $\transition_{\activity}$ to $\bullet \place$.
	Symmetrically, if for $\activity \in \activities_{\eventLog}$ $\vec{y}(\activity) = 1$, we add $\transition_{\activity}$ to $\place \bullet$.
	Finally, if $m = 1$, place $\place$ is initially marked.
	Since translating a region to a place is deterministic, we are also able to translate a place to a region, e.g. place $c_2$ in \autoref{fig:example_wf_net} corresponds to a region with $\vec{x}(a) = 1$, $\vec{x}(f) = 1$, $\vec{y}(d) = 1$ and all other variables set to zero.

	From a formal perspective, consider an event log $\eventLog$ over a set of activities $\activities_{\eventLog}$, a Petri net $\petriNet = (\places, \transitions, \flowRelation, \labelFunc)$ and marking $M_i$ s.t. $\overline{\eventLog} \subseteq \lang^{\labels}(\petriNet, \marking_i)$.
	After adding a place, based on any possible region w.r.t. $\eventLog$, to $\petriNet$ (and updating $\marking_i$ if $m = 1$) still $\overline{\eventLog} \subseteq \lang^{\labels}(\petriNet, \marking_i)$ holds.
	However, adding the region potentially decreases the size of $\lang^{\labels}(\petriNet, \marking_i) \setminus \eventLog$.

	Triples $r_{\vec{0}} = (0, \vec{0}, \vec{0})$ and $r_{\vec{1}} = (1, \vec{1}, \vec{1})$ are always regions and hence are \textit{trivial regions}.
	We let $\regions(\eventLog)$ denote the set of \textit{non-trivial regions} based on event log $\eventLog$.

	\subsection{A Basic ILP Formulation}
	
	Set $\regions(\eventLog)$ represents a huge set of regions.
	However, when using $\eventLog_1$ as an input for process discovery, our goal is to find (a very similar WF-net to) the WF-net in \autoref{fig:example_wf_net}.
	Several regions exist in $R(\eventLog_1)$ that are not a place in \autoref{fig:example_wf_net}.
	For example, a variable assignment with $\vec{x}(a) = 1$ and $\vec{y}(e) = 1$ (all other variables zero), i.e. representing a place connecting transitions $a$ and $e$, is a region.
	We therefore need means to search through the \textit{solution space} to find regions that are of interest.
	
	Firstly, we are only interested in \textit{minimal regions}, i.e. regions that are not expressible as a non-negative linear combination of two other regions, because non-minimal regions correspond to implicit places~\cite{DBLP:journals/fuin/derWerfDHS09}.
	Hence, when applying region-based techniques in terms of process discovery, we only search for minimal regions.
	Finding all minimal regions does however not suffice, e.g. the aforementioned region with $\vec{x}(a) = 1$ and $\vec{y}(e) = 1$ is minimal yet implicit.	
	To this end we define an  \textbf{I}nteger \textbf{L}inear \textbf{P}rogramming (ILP)~\cite{DBLP:books/daglib/0090562} formulation using the region definition as a \textit{constraint body}~\cite{DBLP:journals/fuin/derWerfDHS09}.
	ILP is a mathematical optimization problem defined over a set of integer variables. 
	The objective and constraints of an ILP-problem are an expression in terms of the variables of linear form.
	Before introducing the basic ILP-formulation for the purpose of process discovery, we reformulate regions in terms of matrices.
	
	\begin{definition}[Region (Matrix Form)]
		\label{def:sec:ilp_def:region_matrix_form}
		Given an event log $\eventLog$ over a set of activities $\activities_{\eventLog}$, let $m \in \{0,1\}$ and let $\vec{x},\vec{y} \in \{0,1\}^{|\activities_{\eventLog}|}$. 
		Let $\mathbf{M}$ and $\mathbf{M}'$ be two $|\overline{L} \setminus \{\epsilon\} | \times |\activities_{\eventLog}|$ matrices with $\mathbf{M}(\sigma,a) = \vec{p}(\sigma)(a)$ and $\mathbf{M}'(\sigma,a) = \vec{p}(\sigma')(a)$ (where $\sigma = \sigma' \cdot \langle a' \rangle \in \overline{L}$). 
		Tuple $r = (m, \vec{x}, \vec{y})$ is a region if and only if:
		\begin{equation}
		\label{eg:sec:ilp_eq:region_matrix_form}
		m \vec{1} + \mathbf{M}'\vec{x} - \mathbf{M} \vec{y} \geq \vec{0}
		\end{equation}
	\end{definition}
	
	We additionally define matrix $\mathbf{M}_{L}$ which is an $|\eventLog| \times |\activities_{\eventLog}|$ matrix with $\mathbf{M}_{\eventLog}(\sequence, \activity) = \vec{p}(\sequence)(a)$ for $\sequence \in \eventLog$, i.e., $\mathbf{M}_L$ is the equivalent of $\mathbf{M}$ for all traces in the event log.
	We define a general process discovery ILP-formulation that guarantees to find a non-trivial region with the additional property that the corresponding place is always empty after replaying each trace within the event log.
	
	\begin{definition}[Process Discovery ILP-formulation]
		\label{def:sec:ilp_def:ilp_formulation}
		Given an event log $\eventLog$ over a set of activities $\activities_{\eventLog}$ and corresponding matrices $\mathbf{M}$, $\mathbf{M}'$ and $\mathbf{M}_L$.
		Let $c_m \in \mathbb{R}$ and $\vec{c_x}, \vec{c_y} \in \mathbb{R}^{|\activities_\eventLog|}$.
		The process discovery ILP-formulation, $ILP_{L}$, is defined as:
		\begin{center}
			\begin{tabular}{r l r l l}
				\textbf{minimize} & & $z = c_m m + \vec{c_x}^{\intercal} \vec{x} +\vec{c_y}^{\intercal} \vec{y}$ & & objective function\\
				\textbf{such that} & & $m \vec{1} + \mathbf{M}' \vec{x} - \mathbf{M} \vec{y} \geq \vec{0}$ & & theory of regions\\
				\textbf{\textit{and}} & & $m \vec{1} + \mathbf{M}_L(\vec{x} - \vec{y}) = \vec{0}$ & & corresp. place is empty after each trace \\
				& & $\vec{1}^{\intercal} \vec{x} + \vec{1}^{\intercal}\vec{y} \geq 1$ & & at least one arc connected\\
				& & $\vec{0} \leq \vec{x} \leq \vec{1}$ & & i.e. $\vec{x} \in \{0,1\}^{|A|}$\\
				& & $\vec{0} \leq \vec{y} \leq \vec{1}$ & & i.e. $\vec{y} \in \{0,1\}^{|A|}$\\
				& & $0 \leq m \leq 1$ & & i.e. $m \in \{0,1\}$\\
			\end{tabular}
		\end{center}
	\end{definition}
	
	\autoref{def:sec:ilp_def:ilp_formulation} acts as a basic formulation for process discovery using ILP. 
	To actually use the formulation we need to instantiate $c_m$, $\vec{c_x}$ and $\vec{c_y}$, i.e. the \textit{objective coefficients}, with meaningful values. 
	By varying the actual values for the objective coefficients we are able to let the ILP favour different solutions. 
	In~\cite{DBLP:journals/fuin/derWerfDHS09} an objective function is proposed that minimizes the number of incoming arcs and maximizes the number of outgoing arcs to a place.
	In~\cite{DBLP:conf/apn/ZelstDA15} the aforementioned objective function is extended such that it minimizes the time a token resides in the corresponding place.
	Both objective functions are expressible as a more general function which favours minimal regions~\cite{DBLP:conf/apn/ZelstDA15}.
	In general we are able to use any objective function, as long as it favours minimal regions.
	Hence, in this paper we assume that one uses such objective function.

	Using the basic formulation with some objective function instantiation only yields one, optimal, result.
	Hence we need a more structured approach for finding multiple Petri net places, using the ILP formulation presented as a basis.
	
	\subsection{Exploiting Causalities}
	We need to find multiple regions that \textit{together} form places of a WF-net, in line with the behaviour present within the event log.
	One of the most suitable techniques to find multiple regions in a controlled, structured manner, is by exploiting \textit{causal relations} present within an event log.
	A causal relation between activities $a$ and $b$ implies that activity $a$ causes $b$, i.e. $b$ is likely to follow (somewhere) after activity $a$. 
	
	Several approaches exist to compute causalities relations~\cite{DBLP:journals/topnoc/DongenMW09}.
	The $\alpha$-Miner~\cite{DBLP:journals/tkde/AalstWM04} defines causal relation $a \to_{\eventLog} b$ from activity $a$ to activity $b$ if, within some event log $\eventLog$, we find traces of the form $\langle ..., a,b, ... \rangle$ though we do not find traces of the form $\langle ..., b,a, ... \rangle$.
	Within the Heuristics Miner~\cite{DBLP:journals/icae/WeijtersA03,DBLP:conf/cidm/WeijtersR11} this relation was further developed to take frequencies into account as well.
	Given these multiple definitions, we assume the existence of a \textit{causal relation oracle} which, given an event log, produces a set of pairs $(a,b)$ indicating that activity $a$ has a causal relation with (to) activity $b$.
	\begin{definition}[Causal relation oracle]
		A causal relation oracle $\causalOracle$ maps a bag of traces to a set of activity pairs, i.e. $\causalOracle : \allBags(\activityUniverse^*) \to \powerSet(\activityUniverse \times \activityUniverse)$.
	\end{definition}

	A causal oracle maps an event log onto its activities, i.e. $\causalOracle(\eventLog) \in \powerSet(\activities_\eventLog \times \activities_\eventLog)$.
	It defines a directed graph with $\activities_\eventLog$ as vertices and each pair $(a,b) \in \causalOracle(\eventLog)$ as an arc between $a$ and $b$.
	Later we exploit this graph-based view, for now we refer to $\causalOracle(\eventLog)$ as a collection of pairs.
	
	When adopting a causal-based ILP process discovery strategy, we try to find net places that represent a causality found in the event log. 
	Given an event log $\eventLog$, for each pair $(a,b) \in \causalOracle(\eventLog)$ we enrich the constraint body with three constraints: 1.) $m = 0$, 2.) $\vec{x}(a) = 1$ and 3.) $\vec{y}(b) = 1$.
	The three constraints ensure that if we find a solution to the ILP it corresponds to a place which is not marked and connects transition $a$ to transition $b$.
	Given pair $(a,b) \in \causalOracle(\eventLog)$ we denote the corresponding extended causality based ILP-formulation as $ILP_{(L,a \to b)}$.
	
	After solving $ILP_{(L,a \to b)}$ for each $(a,b) \in \causalOracle(L)$, we end up with a set of regions that we are able to transform into places in a resulting Petri net.
	Since we enforce $m=0$ for each causality, none of these places is initially marked.
	Moreover, due to constraints based on $m \vec{1} + \mathbf{M}_L(\vec{x} - \vec{y}) = \vec{0}$, the resulting place is empty after replaying each trace in the input event log within the net.
	Since we additionally enforce $\vec{x}(a) = 1\ \text{and}\ \vec{y}(b) = 1$, if we find a solution to the ILP, the corresponding place has both input and output arcs and is not eligible for being a source/sink place.
	Hence, the approach as-is does not allow us to find WF-nets.
	In the next section we show that a simple pre-processing step performed on the event log, together with specific instances of $\causalOracle(L)$, allows us to discover WF-nets which are relaxed sound.

	\section{Discovering Relaxed Sound Workflow Nets}
	\label{sec:disc_wf_nets}
	Reconsider example event log $\eventLog_1$, i.e. $\eventLog_1 = [\langle a,b,d,e,g \rangle^{10}, \langle a,c,d,e,f,d,b,e,g \rangle^{12}, \langle a,d,c,e,h \rangle^{9}, \langle a,b,d,e,f,c,d,e,g \rangle^{11}, \langle a,d,c,e,f,b,d,e,h \rangle^{13}]$.
	Let $\activities_f \subseteq \activities_\eventLog$ denote the set of \textit{final activities}, i.e. activities $a_f$ s.t. there exists a trace of the form $\langle ..., a_f \rangle$ in the event log.
	For example, for $\eventLog_1$, $\activities_f = \{g,h\}$.
	After solving each $ILP_{\eventLog, a \to b}$ instance based on $\causalOracle(\eventLog)$ and adding corresponding places, we know that when we exactly replay any trace from $\eventLog_1$, after firing $g$ or $h$, the net is empty.
	Since $g$ and $h$ never co-occur in a trace, it is trivial to add a sink place $\place_o$, s.t. after replay each trace in $\eventLog_1$, $\place_o$ is the only place marked, i.e. $\bullet \place_o = \{f,g\}$ and $\place_o \bullet = \emptyset$ (place ``end'' in \autoref{fig:example_wf_net}).
	In general, such decision is not trivial.
	However, a trivial case for adding a sink $\place_o$ is the case when there is only one end activity that uniquely occurs once, at the end of each trace, i.e.  $\activities_f = \{\activity_f\}$ and there exists no trace of the form $\langle ...,a_f, ..., a_f \rangle$.
	In such case we have $\bullet \place_o = \{\activity_f\}$, $\place_o \bullet = \emptyset$.
	
	A similar rationale holds for adding a source place.
	We define a set $\activities_s$ that denotes the set of \textit{start activities}, i.e. activities $a_s$ s.t. there exists a trace of the form $\langle a_s, ... \rangle$ in the event log.
	For each activity $a_s$ in $\activities_s$ we know that for some traces in the event log, these are the first ones to be executed.
	Thus, we know that the source place $\place_i$ must connect, in some way, to the elements of $\activities_s$.
	Like in the case of final transitions, creating a source place is trivial when $\activities_s = \{\activity_s\}$ and there exists no trace of the form $\langle a_s, ..., a_s, ... \rangle$, i.e. the start activity uniquely occurs once in each trace.
	In such case we create place $\place_i$ with $\bullet \place_i = \emptyset$, $\place_i \bullet = \{\activity_s\}$.
	
	In order to be able to find a source and a sink place, it suffices to guarantee that sets $\activities_s$ and $\activities_f$ are of size one and their elements always occur uniquely at the start, respectively, end of a trace.
	We formalize this idea through the notion of \textit{unique start/end event logs}, after which we show that transforming an arbitrary event log to such unique start/end event log is trivial.
	
	\begin{definition}[Unique start/end event log]
		\label{unique_start_end_event_log}
		Let $\eventLog$ be an event log over a set of activities $\activities_\eventLog$.
		$\eventLog$ is a \textbf{U}nique \textbf{S}tart/\textbf{E}nd event \textbf{L}og (USE-Log) if there exist $a_s,a_f \in \activities_{\eventLog}$ s.t. $a_s \neq a_f$, $\forall_{\sequence \in \eventLog}(\sequence(1) = a_s \wedge \forall_{i \in \{2,3,...,|\sequence|\}}(\sequence(i) \neq a_s))$ and $\forall_{\sequence \in \eventLog}(\sequence(|\sequence|) = a_f \wedge \forall_{i \in \{1,2,...,|\sequence|-1\}}(\sequence(i) \neq a_f))$.
	\end{definition}
	
	Since the set of activities $\activities_\eventLog$ is finite, it is trivial to transform any event log to a USE-log.
	Assume we have an event log $\eventLog$ over $\activities_\eventLog$ that is not a USE-log.
	We generate two ``fresh'' activities $a_s,a_f \in \activityUniverse$ s.t. $a_s,a_f \notin \activities_{\eventLog}$ and create a new event log $L'$ over $\activities_{\eventLog} \cup \{a_s, a_f\}$, by adding $\langle a_s \rangle \cdot \sequence \cdot \langle a_f \rangle$ to $\eventLog$ for each $\sequence \in \eventLog$.
	We let $\useProj : \allBags(\activityUniverse^*) \to \allBags(\activityUniverse^*)$ denote such USE-transformation.
	We omit $\activity_s$ and $\activity_f$ from the domain of $\useProj$ and assume that given some USE-transformation the two symbols are known.
	
	Clearly, after applying a USE-transformation, finding a unique source and sink place is trivial.
	It also provides an additional advantage considering the ability to find WF-nets.
	In fact, an ILP instance $ILP_{\eventLog, a \to b}$ always has a solution if $\eventLog$ is a USE-log.
	We provide a proof of this property in \autoref{lemma:solution_exists}, after which we present an algorithm that, given specific instantiations of $\causalOracle$, discovers WF-nets.
		
	\begin{lemma}[A USE-Log based causality has a solution]
		\label{lemma:solution_exists}
		Let $\eventLog$ be an event log over a set of activities $\activities_{\eventLog}$. 
		Let $\useProj : \allBags(\activityUniverse^*) \to \allBags(\activityUniverse^*)$ denote a USE-transformation function and let $\activity_s$, $\activity_f$ denote the start and end activities.
		For every $(a,b) \in \causalOracle(\useProj(\eventLog))$ with $a \neq \activity_f$ and $b \neq \activity_s$, $ILP_{(\useProj(\eventLog), a \to b)}$ has a solution.
		\begin{proof}[Constructive]
			
			We consider the case $a \neq \activity_s$ and $b \neq \activity_f$.
			We show that variable assignment $\vec{x}(\activity_s) = \vec{x}(a) = \vec{x}(b) = \vec{y}(a)= \vec{y}(b) = \vec{y}(\activity_f) = 1$, all other variables $0$ (\autoref{fig:trivial_a_b}), adheres to all constraints of $ILP_{(\useProj(\eventLog), a \to b)}$.
					
			Consider constraints of the form $\forall_{\sequence = \sequence' \cdot \langle a \rangle \in \overline{\useProj(\eventLog)}} (m + \vec{p}(\sigma')^{\intercal}\vec{x} - \vec{p}(\sigma)^{\intercal}\vec{y} \geq 0)$ ($m \vec{1} + \mathbf{M}' \vec{x} - \mathbf{M} \vec{y} \geq \vec{0}$) and let $\sequence = \sequence' \cdot \langle x \rangle \in \overline{\useProj(\eventLog)}$.
			
			\textit{Case I:} $x \neq a, x \neq b, x \neq \activity_f$. Since $x \neq \activity_f$ we know $\vec{p}(\sequence)(\activity_f) = 0$.
			Moreover, since $x \neq a, x \neq b$, we know that $\vec{p}(\sequence')(a) = \vec{p}(\sequence)(a)$ and $\vec{p}(\sequence')(b) = \vec{p}(\sequence)(b)$, and hence $\vec{p}(\sequence')(a)\vec{x}(a) - \vec{p}(\sequence)(a)\vec{y}(a) = 0$ and $\vec{p}(\sequence')(b)\vec{x}(b) - \vec{p}(\sequence)(b)\vec{y}(b) = 0$.
			Since $\vec{x}(\activity_s) = 1$ and $\activity_s$ occurs uniquely at the start of each trace, if $\sequence' = \epsilon$ such constraint equals $0$, and, $1$ otherwise.
			
			\textit{Case  II:} $x = a$. We know $\vec{p}(\sequence)(\activity_f) = 0$ and $\vec{p}(\sequence')(b) = \vec{p}(\sequence)(b)$. 
			Now $\vec{p}(\sequence')(a) = \vec{p}(\sequence)(a) - 1$ and thus $\vec{p}(\sequence')(a)\vec{x}(a) - \vec{p}(\sequence)(a)\vec{y}(a) = -1$.
			Since $\activity_s \in \sequence'$ we have  $\vec{p}(\sequence')(\activity_s)\vec{x}(\activity_s) = 1$, and thus the constraint equals $0$.
			
			\textit{Case III:} $x = b$
			Similar to \textit{Case II}.
			
			\textit{Case IV:} $x = a_f$.
			We again have $\vec{p}(\sequence')(a) = \vec{p}(\sequence)(a)$ and $\vec{p}(\sequence')(b) = \vec{p}(\sequence)(b)$.
			Since $\vec{p}(\sequence)(\activity_f)\vec{y}(\activity_f) = \vec{p}(\sequence')(\activity_s)\vec{x}(\activity_s) = 1$, each constraint equals $0$.
			
			The constraints of the form:  $\vec{1}^{\intercal} \vec{x} + \vec{1}^{\intercal}\vec{y} \geq 1$, $\vec{0} \leq \vec{x} \leq \vec{1}$, $\vec{0} \leq \vec{y} \leq \vec{1}$ and $0 \leq m \leq 1$ are trivially satisfied.
			From \textit{Case IV} combined with $\vec{x}(\activity_f) = 0$ it follows that all constraints of the form $\forall_{\sequence \in \useProj(\eventLog)}(m + \vec{p}(\sequence)^{\intercal}(\vec{x} - \vec{y}) = 0)$ ($m \vec{1} + \mathbf{M}_L(\vec{x} - \vec{y}) = \vec{0}$) hold.
			Finally the assignment adheres to $m = 0$, $\vec{x}(a) = 1$ and $\vec{y}(b) = 1$.
			
			In case we have $a = a_s$ and $b \neq a_f$ the region $\vec{x}(a_s) = \vec{x}(a) = \vec{y}(a) = \vec{y}(a_f) = 1$, all other variables $0$ (\autoref{fig:trivial_a_s_af_a}), is a solution.
			The proof is similar to the proof of the previous case.
			
			In case we have $a \neq a_s$ and $b = a_f$ the region $\vec{x}(a_s) = \vec{x}(b) = \vec{y}(b) = \vec{y}(a_f) = 1$, all other variables $0$ (\autoref{fig:trivial_a_s_af_a}), is a solution.
			Again the proof is similar to the proof in the first case.
			
			Finally in case we have $a = a_s$ and $b = a_f$ the region $\vec{x}(a_s) = \vec{y}(a_f) = 1$, all other variables $0$ (\autoref{fig:trivial_a_s_af}), is a solution.
			Again the proof is similar to the proof in the first case.			
		\end{proof}
	\end{lemma}

	\begin{figure}[tb]
		\centering
		\begin{subfigure}[b]{0.325\textwidth}
			\centering			
			\begin{tikzpicture}
			[   
			node distance=1cm,
			on grid,>=stealth',
			bend angle=17.5,
			auto,
			every place/.style= {minimum size=4mm},
			every transition/.style = {minimum size = 6mm}
			]
			
			\node [place] (p1) [label = below left:$p$]{};
			
			\node [transition] (as) [label = center:$a_s$, left = of p1] {}
			edge [post] node[auto] {} (p1);
			
			\node [transition] (af) [label = center:$a_f$, right = of p1] {}
			edge [pre] node[auto] {} (p1);
			
			\node [transition] (a) [label = center:$a$, above = of p1] {}
			edge [post, bend left = 15] node[auto] {} (p1)
			edge [pre, bend right = 15] node[auto] {} (p1);
			
			\node [transition] (b) [label = center:$b$, below = of p1] {}
			edge [pre, bend left = 15] node[auto] {} (p1)
			edge [post, bend right = 15] node[auto] {} (p1);		
			\end{tikzpicture}
			\caption{Solution in case $a \neq a_s$ and $b \neq a_f$.}
			\label{fig:trivial_a_b}
		\end{subfigure}
		\begin{subfigure}[b]{0.325\textwidth}
			\centering			
			\begin{tikzpicture}
			[   
			node distance=1cm,
			on grid,>=stealth',
			bend angle=17.5,
			auto,
			every place/.style= {minimum size=4mm},
			every transition/.style = {minimum size = 6mm}
			]
			
			\node [place] (p1) [label = below left:$p$]{};
			
			\node [transition] (as) [label = center:$a_s$, left = of p1] {}
			edge [post] node[auto] {} (p1);
			
			\node [transition] (af) [label = center:$a_f$, right = of p1] {}
			edge [pre] node[auto] {} (p1);
			
			\node [transition] (a) [label = center:$a / b$, above = of p1] {}
			edge [post, bend left = 15] node[auto] {} (p1)
			edge [pre, bend right = 15] node[auto] {} (p1);	
			\end{tikzpicture}
			\caption{Solution in case $a = a_s$ and $b \neq a_f$ or $a \neq a_s$ and $b = a_f$.}
			\label{fig:trivial_a_s_af_a}
		\end{subfigure}
		\begin{subfigure}[b]{0.325\textwidth}
			\centering			
			\begin{tikzpicture}
			[   
			node distance=1cm,
			on grid,>=stealth',
			bend angle=17.5,
			auto,
			every place/.style= {minimum size=4mm},
			every transition/.style = {minimum size = 6mm}
			]
			
			\node [place] (p1) [label = below left:$p$]{};
			
			\node [transition] (as) [label = center:$a_s$, left = of p1] {}
			edge [post] node[auto] {} (p1);
			
			\node [transition] (af) [label = center:$a_f$, right = of p1] {}
			edge [pre] node[auto] {} (p1);
			\end{tikzpicture}
			\caption{Solution in case $a = a_s$ and $b = a_f$.}
			\label{fig:trivial_a_s_af}
		\end{subfigure}
		\caption{Visualizations of trivial solutions to $ILP_{(\useProj(\eventLog), a \to b)}$ in terms of Petri net places.}
		\label{fig:trivial}
	\end{figure}

	In~\autoref{alg:ilp_disc} we present an ILP-Based process discovery approach that uses a USE-log internally in order to find multiple Petri net places.
	For every $(a,b) \in \causalOracle(\useProj(\eventLog))$ with $a \neq \activity_f$ and $b \neq \activity_s$ it solves $ILP_{(\useProj(\eventLog), a \to b)}$.
	Moreover, it finds a unique source and sink place.
	
	\begin{algorithm}[tb]
		\small
		\caption{$\texttt{ILP-Based Process Discovery}$\label{alg:ilp_disc}}
		\SetKwInOut{Input}{input}\SetKwInOut{Output}{output}
		\Input{$\eventLog \in \allBags(\activities_{\eventLog}^*)$, $\causalOracle \colon \allBags(\activityUniverse^*) \to \powerSet(\activityUniverse \times \activityUniverse)$} 
		\Output{$\workflowNet = (\places, \transitions, \flowRelation, \place_i, \place_o, \labelFunc)$} 
		\Begin{
			\nl $\places, \transitions, \flowRelation \gets \emptyset$\;
			\nl let $a_s, a_f \notin \activities_\eventLog$\;
			\nl $\transitions \gets \{ \transition_{\activity} \mid \activity \in \activities_\eventLog \cup \{\activity_s, \activity_f\}  \}$\;
			\nl \ForEach{$(a,b) \in \causalOracle(\useProj(\eventLog))$} {
				\nl $(m,\vec{x},\vec{y}) \gets $ solution to $ILP_{(\useProj(\eventLog), a \to b)}$\;
				\nl let $\place_{(a,b)} \notin \places$\;
				\nl $\places \gets \places \cup \place_{(a,b)}$\;
				\nl \ForEach{$a' \in \activities_\eventLog \cup \{\activity_s, \activity_f\}$}{
					\nl \If{$\vec{x}(a') = 1$}{
						\nl $\flowRelation \gets \flowRelation \cup \{(\transition_{a'}, \place_{(a,b)}\}$\;
					}
					\nl \If{$\vec{y}(a') = 1$} {
						\nl $\flowRelation \gets \flowRelation \cup \{(\place_{(a,b)}, \transition_{a'})\}$\;
					}
				}
			}
			\nl let $\place_i, \place_o \notin \places$\;
			\nl $\places \gets \places \cup \{ \place_i, \place_o \}$\;
			\nl $\flowRelation \gets \flowRelation \cup \{(\place_i, \transition_{\activity_s})\}$\;			
			\nl $\flowRelation \gets \flowRelation \cup \{(\transition_{\activity_f}, \place_o)\}$\;
			\nl let $\labelFunc \colon \transitions \to \activities \cup \{\silent\}$\;
			\nl \ForEach{$\activity \in \activities_\eventLog$}{
				\nl $\labelFunc(\transition_{\activity}) \gets \activity$ \;
			}
			\nl $\labelFunc(\transition_{\activity_s}), \labelFunc(\transition_{\activity_f}) \gets \silent$\;
			\nl $\mathbf{return}$ $(\places, \transitions, \flowRelation, \place_i, \place_o, \labelFunc)$\;
		}			
	\end{algorithm}	
	The algorithm constructs an initially empty Petri net $N = (P,T,F)$.
	Subsequently for each $a \in \activities_{\eventLog} \cup \{a_s, a_f\}$ a transition $\transition_a$ is added to $\transitions$.
	For each causal pair in the USE-variant of input event log $\eventLog$, a place $\place_{(a,b)}$ is discovered by solving $ILP_{(\useProj(\eventLog), a \to b)}$ after which $\places$ and $\flowRelation$ are updated accordingly.
	The algorithm adds an initial place $\place_i$ and connects it to $\transition_{\activity_s}$ and similarly creates sink place $\place_o$ which is connected to $\transition_{\activity_f}$.
	For transition $\transition_\activity$ related to $\activity \in \activities_{\eventLog}$, we have $\labelFunc(\transition_\activity) = \activity$, whereas $\labelFunc(\transition_{\activity_s}) = \labelFunc(\transition_{\activity_f})= \silent$.

	The algorithm is guaranteed to always find a solution to $ILP_{(\useProj(\eventLog), a \to b)}$, hence for each causal relation a place is found.
	Additionally, a unique source and sink place are constructed.
	However, the algorithm does not guarantee that we find a connected component, i.e. requirement 3 of \autoref{def:workflow_net}.
	In fact, the nature of $\causalOracle$ determines whether or not we discover a WF-net.
	In~\autoref{th:wf_nets} we characterize this nature and prove, by exploiting \autoref{lemma:solution_exists}, that we are able to discover WF-nets.
		
	\begin{theorem}[There exist sufficient conditions for finding WF-nets]
		\label{th:wf_nets}
		Let $\eventLog$ be an event log over a set of activities $\activities_\eventLog$.
		Let $\useProj : \allBags(\activityUniverse^*) \to \allBags(\activityUniverse^*)$ denote a USE-transformation function.
		Let $\activity_s, \activity_f$ denote the unique start- and end activity of $\useProj(\eventLog)$.
		Let $\causalOracle : \allBags(\activityUniverse^*) \to \powerSet(\activityUniverse \times \activityUniverse)$ be a causal oracle and consider $\causalOracle(\useProj(\eventLog))$ as a \textit{directed graph}.
		If each $\activity \in \activities_L$ is on a path from $\activity_s$ to $\activity_f$ in $\causalOracle(\useProj(\eventLog))$, and there is no path from $\activity_s$ to itself, nor a path from $\activity_f$ to itself, then $\texttt{ILP-Based Process Discovery}(\eventLog, \causalOracle)$ returns a WF-net.
		\begin{proof}[On the structure of $\causalOracle(\useProj(\eventLog))$]
			By the requirements on $\causalOracle(\useProj(\eventLog))$ and \autoref{lemma:solution_exists}, we know that for each $(a,b) \in \causalOracle(\useProj(\eventLog))$ a corresponding place will be found that has a transition labelled with $a$ as an input and a transition labelled $b$ as an output.
			Hence every path in $\causalOracle(\useProj(\eventLog))$ corresponds to a path in the resulting net and as a consequence, every transition is on a path from $\activity_s$ to $\activity_f$.
			As every place that is added has input transition ($\vec{x}(a)=1$) and an output transition ($\vec{y}(b) = 1$), every place is also on a path from $\activity_s$ to $\activity_f$.
			By construction this then also holds from $\place_i$ to $\place_o$.
		\end{proof}	
	\end{theorem}

	\autoref{th:wf_nets} proves that if we use a causal structure that, when interpreting it as a graph, has the property that each $\activity \in \activities_\eventLog$ is on a path from $\activity_s$ to $\activity_f$, the result of $\texttt{ILP-Based Process Discovery}(\eventLog, \causalOracle)$ is a WF-net.
	Although this seems a rather strict property of the causal structure, there exists a specific causal graph definition that guarantees this property~\cite{DBLP:conf/cidm/WeijtersR11}.
	Hence we are able to use this definition as an instantiation for $\causalOracle$.

	\autoref{th:wf_nets} does not provide any behavioural guarantees, i.e. a WF-net is a purely graph-theoretical property.
	Recall that the premise of a region is that it does not block the execution of any sequence within the prefix-closure of an event log.
	Intuitively we deduce that we are therefore able to fire each transition in the WF-net at least once.
	Moreover, since we know that $\activity_f$ is the final transition of each sequence in $\useProj(\eventLog)$, and after firing the transition each place based on any $ILP_{\useProj(\eventLog), a\to  b}$ is empty, we know that we are able to mark $\place_o$.
	These two observations hint on the fact that the WF-net is \textit{relaxed sound}, which we prove in \autoref{th:relax_sound}
	
	\begin{theorem}
		\label{th:relax_sound}
		Let $\eventLog$ be an event log over a set of activities $\activities_\eventLog$.
		Let $\useProj : \allBags(\activityUniverse^*) \to \allBags(\activityUniverse^*)$ denote a USE-transformation function and let $\activity_s, \activity_f$ denote the unique start- and end activity of $\useProj(\eventLog)$.
		Let $\causalOracle : \allBags(\activityUniverse^*) \to \powerSet(\activityUniverse \times \activityUniverse)$ be a causal oracle.
		Let $\workflowNet = (\places, \transitions, \flowRelation, \place_i, \place_o, \labelFunc) = \texttt{ILP-Based Process Discovery}(\eventLog, \causalOracle)$.
		If $\workflowNet$ is a WF-net, then $\workflowNet$ is relaxed sound.
		\begin{proof}[By construction of traces in the event log]
			Recall that $\workflowNet$ is relaxed sound if and only if: $\forall_{\transition \in \transitions}(\exists_{\marking, \marking' \in \allBags(\places)}((\workflowNet,[\place_i]) \rightsquigarrow (\workflowNet,\marking) \wedge (\workflowNet,\marking) \xrightarrow{\transition} (\workflowNet, \marking') \wedge (\workflowNet,\marking') \rightsquigarrow (\workflowNet, [\place_o])))$.			
		
			Observe that $\transition_{\activity_s}$ is trivially enabled in $\marking_i = [\place_i]$ since $\bullet \transition_{\activity_s} = \{\place_i\}$.
			Consider arbitrary $\transition \in \transitions \setminus \{\transition_{\activity_s},\transition_{\activity_f}\}$.
			We know $\exists_{\sequence \in \useProj(\eventLog)}(\sequence = \langle \activity_s \rangle \cdot \sequence' \cdot \langle \labelFunc(t) \rangle \cdot \sequence'' \cdot \langle \activity_f \rangle)$.
			
			Let $\langle \transition'_1, \transition'_2, ..., \transition'_n \rangle$ s.t. $\langle \labelFunc(\transition'_1), \labelFunc(\transition'_2), ..., \labelFunc(\transition'_n) \rangle = \sequence'$.
			The fact that each place $\place \in \places \setminus \{\place_i, \place_o\}$ corresponds to a region yields that we may deduce $[\place_i] \xrightarrow{\transition_{\activity_s}} \marking'_1$,$\marking'_1 \xrightarrow{\transition'_1} \marking'_2$,...,$\marking'_n \xrightarrow{\transition'_n} \marking'$ s.t. $\marking' \supseteq \bullet \transition$ (if there exists $\place \in \bullet \transition$ s.t. $\marking'(\place) = 0$, then $\place$ does not correspond to a region).
			Hence for any $\transition \in \transitions \setminus \{\transition_{\activity_s},\transition_{\activity_f}\}$ there exists a marking reachable from $[\place_i]$ that enables $\transition$.
			
			Now let $\langle \transition''_1, \transition''_2, ..., \transition''_n \rangle$ s.t. $\langle \labelFunc(\transition''_1), \labelFunc(\transition''_2), ..., \labelFunc(\transition''_n) \rangle = \sequence''$.
			Note that also, again by the fact that each place $\place \in \places \setminus \{\place_i, \place_o\}$ corresponds to a region, we may deduce $\marking' \xrightarrow{\transition''_1} \marking''_1$, $\marking''_1\xrightarrow{\transition''_2} \marking''_2$,...,$\marking''_{n-1} \xrightarrow{\transition''_n} \marking''_n$.
			Clearly we have $\marking''_n \xrightarrow{\transition_{\activity_f}} \marking_f$ with $\marking_f(\place_o) = 1$ since $\transition_{\activity_f} \bullet = \{ \place_o \}$, and this is the first time we fire $\transition_{\activity_f}$, i.e., $\activity_f \notin \langle \activity_s \rangle \cdot \sequence' \cdot \langle \labelFunc(t) \rangle \cdot \sequence''$.
			Clearly $\marking_f(\place_i) = 0$ and because of constraints of the form $m \vec{1} + \mathbf{M}_L(\vec{x} - \vec{y}) = \vec{0}$ we have $\forall_{\place \in \places \setminus \{\place_i, \place_o\}}(\marking_f(\place) = 0)$.
			Hence $\marking_f = [\place_o]$ and thus after firing $\transition$ there exists a firing sequence that leads to marking $[\place_o]$ which proves $\workflowNet$ is relaxed sound.
		\end{proof}
	\end{theorem}

	We have shown that with a few pre- and post-processing steps and a specific class of causal structures we are able to guarantee to find WF-nets that are relaxed sound.
	These results are interesting since several process mining techniques require WF-nets as an input.
	The ILP problems solved still require their solutions to allow for all possible behaviour in the event log.
	As a result, the algorithm incorporates all infrequent exceptional behaviour and still results in over-fitting complex WF-nets.
	Hence, in the upcoming section we show how to efficiently prune the ILP constraint body to identify and eliminate infrequent exceptional behaviour.
	
	\section{Dealing with Infrequent Behaviour}
	\label{sec:eliminating_replay_fitness}
	In this section we present an efficient pruning technique that identifies and eliminates constraints related to infrequent exceptional behaviour.
	We first present the impact of infrequent exceptional behaviour after which we present the pruning technique.
	
	\subsection{The Impact of Infrequent Exceptional Behaviour}
	\label{sec:curse_of_perfect_replay_fitness}
	In \autoref{sec:motivation} we already indicated the impact of infrequent behaviour on the results of ILP-based process discovery.
	In this section we highlight the main cause of ILP-based discovery's inability to handle infrequent behaviour and we devise a filtering mechanism that exploits the nature of the underlying body of constraints.
	
\begin{figure}[tb]
		\centering
		\begin{subfigure}[b]{\textwidth}
			\centering
			\includegraphics[width=0.75\textwidth]{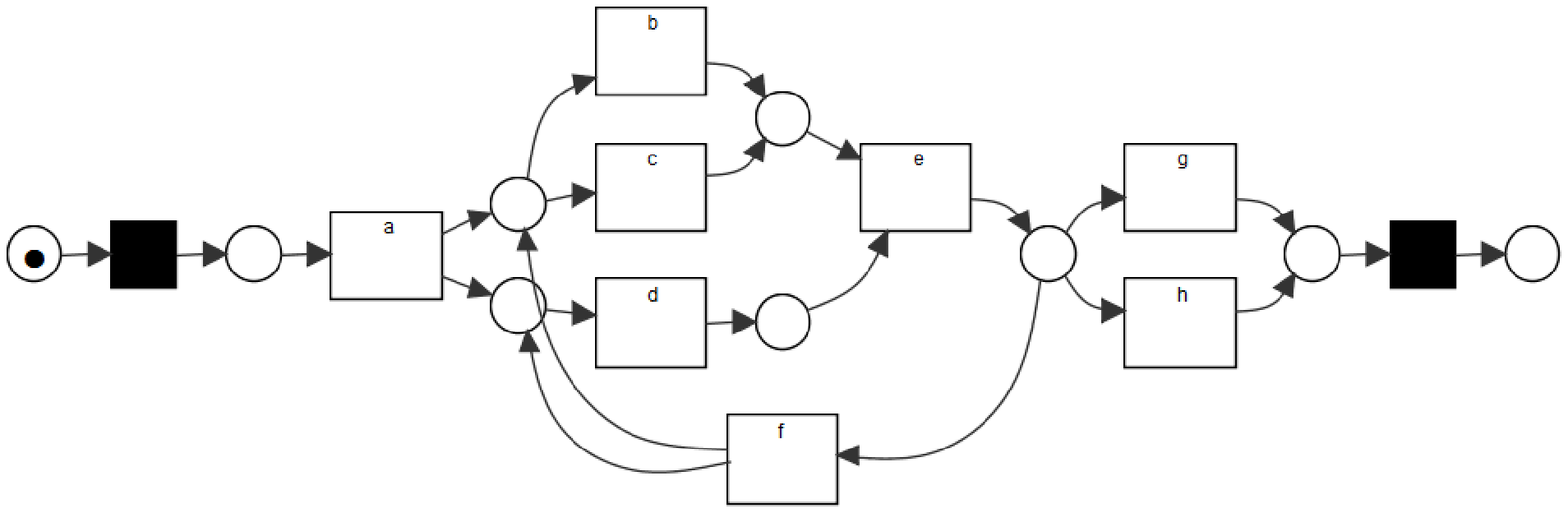}
			\caption{Result based on event log $\eventLog_1$}
			\label{fig:ilp_example_no_noise}
		\end{subfigure}
		\begin{subfigure}[b]{\textwidth}
			\centering
			\includegraphics[width=0.75\textwidth]{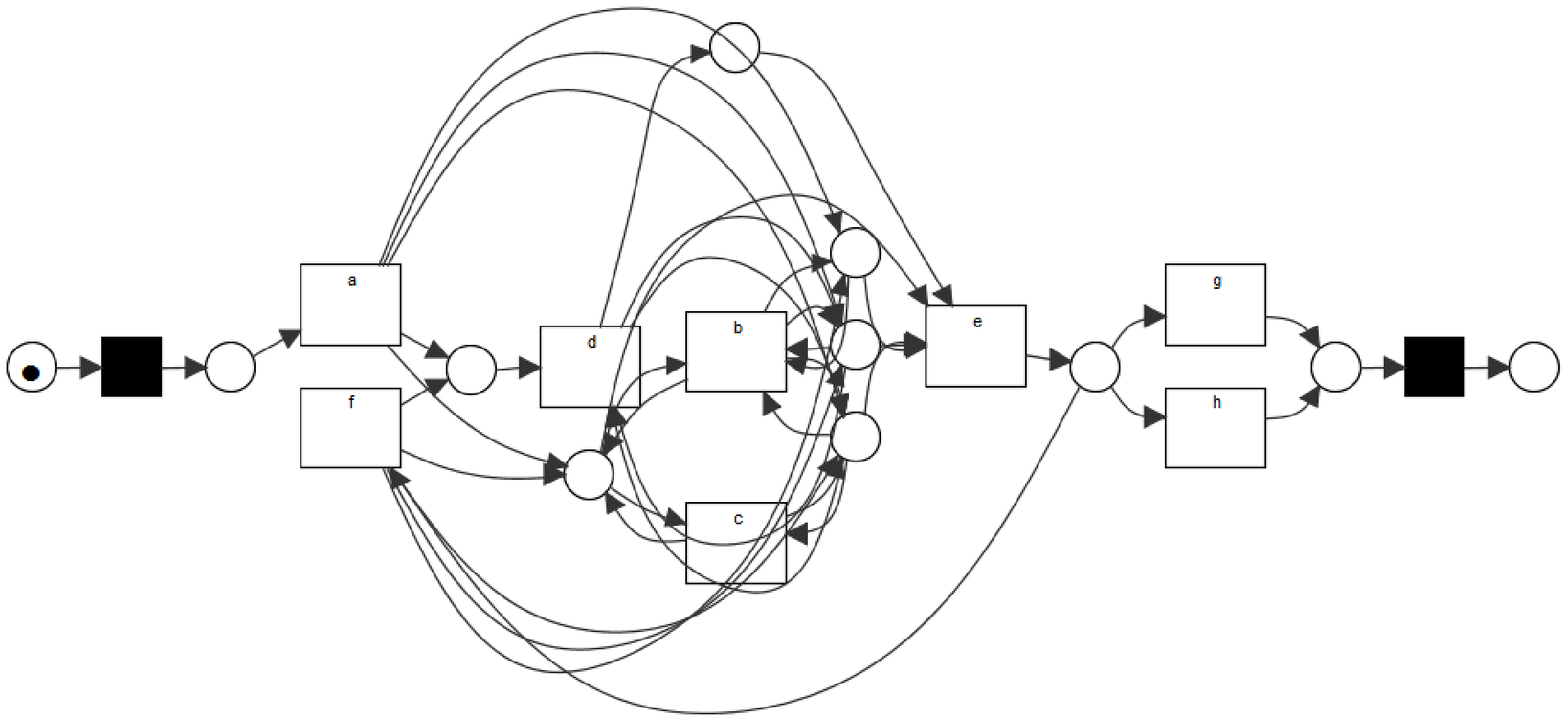}
			\caption{Result based on event log $\eventLog'_1$}
			\label{fig:ilp_example_noise}
		\end{subfigure}
		\caption{Results of applying \autoref{alg:ilp_disc} (\textit{HybridILPMiner} package in the ProM Framework~\cite{DBLP:conf/caise/VerbeekBDA10})  based on $\eventLog_1$ and $\eventLog'_1$.}
		\label{fig:ilp_example}
	\end{figure}

	Let us again reconsider example event log $L_1$, i.e., $\eventLog_1 = [\langle a,b,d,e,g \rangle^{10}, \langle a,c,d,e,f,d,b,e,g \rangle^{12}, \langle a,d,c,e,h \rangle^{9}, \langle a,b,d,e,f,c,d,e,g \rangle^{11}, \langle a,d,c,e,f,b,d,e,h \rangle^{13}]$.
	Using an implementation of \autoref{alg:ilp_disc} in ProM~\cite{DBLP:conf/caise/VerbeekBDA10}, with a suitable causal structure $\causalOracle$, we find the WF-net depicted in \autoref{fig:ilp_example_no_noise}.
	The WF-net describes the same behaviour as the model presented in \autoref{fig:example_wf_net} and has perfect replay-fitness w.r.t. $\eventLog_1$.
	However, if we create event log $\eventLog'$ by simply adding one instance of the trace $\langle a,b,c,d,e,g \rangle$, we obtain the result depicted in \autoref{fig:ilp_example_noise}.
	Due to one exceptional trace, the model allows us, after executing $a$ or $f$ to execute an arbitrary number of $b$- and $c$-labelled transitions.
	This is undesirable since precision of the resulting process model drops significantly.
	Thus, the addition of one exceptional trace results in a less comprehensible WF-net and reduces the precision of the resulting WF-net.
	
		When analysing the two models we observe that they share some equal places, e.g. both models have a place $\place_{(\{a,f\},\{d\})}$ with $\bullet \place_{(\{a,f\},\{d\})} = \{a,f\}$ and $\place_{(\{a,f\},\{d\})} \bullet = \{d\}$.
	However, the two places $\place_{(\{a,f\},\{b,c\})}$ with $\bullet \place_{(\{a,f\},\{b,c\})} = \{a,f\}$ and $\place_{(\{a,f\},\{b,c\})} \bullet = \{b,c\}$ and $\place_{(\{b,c\},\{e\})}$ with $\bullet \place_{(\{b,c\},\{e\})} = \{b,c\}$ and $\place_{(\{b,c\},\{e\})} \bullet = \{e\}$ in \autoref{fig:ilp_example_no_noise}, are not present in \autoref{fig:ilp_example_noise}.
	These are ``replaced'' by the less desirable places containing self-loops in \autoref{fig:ilp_example_noise}.
	This is caused by the fact that $L'_1$ contains all traces present in $L_1$, combined with the additional constraints depicted in \autoref{tab:sec:curse_of_perfect_replay_tab:linear_inequalities_added}.
	
	For place  $\place_{(\{a,f\},\{b,c\})}$ in \autoref{fig:ilp_example_no_noise} we define a corresponding tuple $r = (m, \vec{x}, \vec{y})$ with $\vec{x}(a) = 1$, $\vec{x}(f) = 1$, $\vec{y}(b)= 1$ and $\vec{y}(c) = 1$ (all other variables 0).
	The additional constraints in \autoref{tab:sec:curse_of_perfect_replay_tab:linear_inequalities_added} all evaluate to $-1$ for $\region$, e.g. constraint $m + \vec{x}(a_s) + \vec{x}(a) + \vec{x}(b) - \vec{y}(a_s) - \vec{y}(a) - \vec{y}(b) - \vec{y}(c)$ evaluates to $0 + 0 + 1 + 0 - 0 - 0 - 1 - 1 = -1$.
	In case of place $\place_{(\{b,c\},\{e\})}$ we observe that the corresponding tuple $r = (m, \vec{x}, \vec{y})$ with $\vec{x}(b) = 1$, $\vec{x}(c) = 1$ and $\vec{y}(e) = 1$, yields a value of $1$ for all constraints generated by trace $\langle a,b,c,d,e,g \rangle$.
	For all constraints having a ``$\geq 0$ right hand side'' this is valid, however, for constraint $m + \vec{x}(a_s) + \vec{x}(a) + \vec{x}(b) + \vec{x}(c) + \vec{x}(d) + \vec{x}(e) + \vec{x}(g) + \vec{x}(a_f) - \vec{y}(a_s) - \vec{y}(a) - \vec{y}(b) - \vec{y}(c) - \vec{y}(d) - \vec{y}(e) - \vec{y}(g) - \vec{y}(a_f) = 0$ this is not valid.
	
	\begin{table}[tb]
		\begin{center}
			\caption{Some of the newly added constraints based on trace $\langle a,b,c,d,e,g \rangle$ in event log $L'_1$, starting from prefix $\langle a,b,c \rangle$ which is not present in $\overline{\eventLog_1}$.}
			\label{tab:sec:curse_of_perfect_replay_tab:linear_inequalities_added}
			\resizebox{\textwidth}{!}{
				
				\begin{tabular}{l}
					$m + \vec{x}(a_s) + \vec{x}(a) + \vec{x}(b) - \vec{y}(a_s) - \vec{y}(a) - \vec{y}(b) - \vec{y}(c) \geq 0$\\
					$m + \vec{x}(a_s) + \vec{x}(a) + \vec{x}(b) + \vec{x}(c) - \vec{y}(a_s) - \vec{y}(a) - \vec{y}(b) - \vec{y}(c) - \vec{y}(d) \geq 0$\\
					$\vdots$\\
					$m + \vec{x}(a_s) + \vec{x}(a) + \vec{x}(b) + \vec{x}(c) + \vec{x}(d) + \vec{x}(e) + \vec{x}(g) - \vec{y}(a_s) - \vec{y}(a) - \vec{y}(b) - \vec{y}(c) - \vec{y}(d) - \vec{y}(e) - \vec{y}(g) - \vec{y}(a_f) \geq 0$\\
					$m + \vec{x}(a_s) + \vec{x}(a) + \vec{x}(b) + \vec{x}(c) + \vec{x}(d) + \vec{x}(e) + \vec{x}(g) + \vec{x}(a_f) - \vec{y}(a_s) - \vec{y}(a) - \vec{y}(b) - \vec{y}(c) - \vec{y}(d) - \vec{y}(e) - \vec{y}(g) - \vec{y}(a_f) = 0$ \\
				\end{tabular} 
			}
		\end{center}
	\end{table}
	
	The example shows that the addition of $\langle a,b,c,d,e,g \rangle$ yields constraints that invalidate places $\place_{(\{a,f\},\{b,c\})}$ and $\place_{(\{b,c\},\{e\})}$.
	As a result the WF-net based on event log $\eventLog'_1$ contains places with self-loops on both $b$ and $c$ which greatly reduces its precision and simplicity.
	Due to the relative infrequency of trace $\langle a,b,c,d,e,g \rangle$ it is arguably acceptable to trade-off the perfect replay-fitness guarantee of ILP-Based process discovery and return the WF-net of \autoref{fig:ilp_example_no_noise}, given $\eventLog'_1$.
	Hence, we need filtering techniques and/or trace clustering techniques in order to remove exceptional behaviour.
	However, apart from simple pre-processing, we aim at adapting the ILP-based process discovery approach itself to be able to cope with infrequent behaviour.
	
	By manipulating the constraint body such that it no longer allows for all behaviour present in the input event log, we are able to deal with infrequent behaviour within event logs.
	Given the problems that arise because of the presence of exceptional traces, a natural next step is to leave out the constraints related to the problematic traces.
	An advantage of filtering the constraint body is the fact that the constraints are based on the prefix-closure of the event log.
	Thus, even if all traces are unique yet they do share prefixes, we are able to filter.
	Additionally, leaving out constraints decreases the size of the ILP's constraint body, which has a potential positive effect on the time needed to solve an ILP. 
	We devise a graph-based filtering technique, i.e., \textit{sequence encoding filtering}, that allows us to prune constraints based on trace frequency information.
	
	\subsection{Sequence Encoding Graphs}
	As a first step towards sequence encoding filtering we define the relationship between sequences and constraints.
	We do this in terms of \textit{sequence encodings}.
	A sequence encoding is a vector-based representation of a sequence in terms of region theory, i.e., representing the sequence's corresponding constraint.
	\begin{definition}[Sequence encoding]
		Given a set of activities $A = \{a_1, a_2, ..., a_n\}$. 
		$\vec{\phi} : A^* \rightarrow \mathbb{N}^{2|A| + 1}$ denotes the sequence encoding function mapping every $\sigma \in A ^*$ to a $2 \cdot |A| + 1$-sized vector. 
		We define $\vec{\phi}$ as:
		\[
		\begin{tabular}{cc}
		$\vec{\phi}(\sigma' \cdot \langle a \rangle) =
		\begin{pmatrix}
		1\\ 
		\vec{p}(\sigma')\\
		-\vec{p}(\sigma' \cdot \langle a \rangle)
		\end{pmatrix}$ &
		$\vec{\phi}(\epsilon) =
		\begin{pmatrix}
		1\\ 
		0\\
		\vdots \\
		0
		\end{pmatrix}$
		\end{tabular}
		\]
	\end{definition}
	
	As an example of a sequence encoding vector consider sequence $\langle a_s, a,b \rangle$ originating from $\overline{\useProj(\eventLog'_1)}$, for which we have $\vec{\phi}(\langle a_s, a,b \rangle)^\intercal = (1,1,1,0,0,0,0,0,0,0,0,-1,-1,-1,0,0,0,0,0,0,0)$. 
	Sequence encoding vectors directly correspond to region theory based constraints, e.g. if we are given $m \in \{0,1\}$ and $\vec{x}, \vec{y} \in \{0,1\}^{|A|}$ and create a vector $\vec{r}$ where $\vec{r}(1) = m$, $\vec{r}(2) = \vec{x}(a_s)$, $\vec{r}(3) = \vec{x}(a)$, ..., $\vec{r}(10) = \vec{x}(h)$, $\vec{r}(11) = \vec{x}(a_f)$, $\vec{r}(12) = \vec{y}(a_f)$, ..., $\vec{r}(21) = \vec{y}(a_f)$, then $\vec{\phi}(\langle a_s, a,b \rangle)^{\intercal}\vec{r} = m + \vec{x}(a_s) + \vec{x}(a) - \vec{y}(a_s) - \vec{y}(a) - \vec{y}(b)$.
	As compact notation for $\sigma = \sigma' \cdot \langle a \rangle$ we write $\vec{\phi}(\sigma)$ as a pair of the bag representation of the Parikh vector of $\sigma'$ and $a$, i.e. $\vec{\phi}(\langle a_s, a, b \rangle)$ is written as $([a_s,a], b)$ whereas $\vec{\phi}(\langle a_s, a,b,c \rangle)$ is written as $([a_s,a,b],c)$. 
	For $\vec{\phi}(\epsilon)$ we write $([], \bot)$.
	
	Consider the prefix-closure of $\useProj(L'_1)$ which generates the linear inequalities presented in \autoref{tab:sec:eliminating_replay_fitness_tab:bag_prefix_closure_noise}. 
	\begin{table}[!t]
		\caption{Schematic overview of sequence encodings based on $\overline{\useProj(L'_1)}$.}
		\label{tab:sec:eliminating_replay_fitness_tab:bag_prefix_closure_noise}
		\begin{center}
			\resizebox{\textwidth}{!}{
				\begin{tabular}{|l|l|l|l|}
					\hline
					$\sigma \in \overline{\useProj(\eventLog'_1)}$ & $\vec{\phi}(\sigma)^{\intercal}$, i.e. $(m, \vec{x}(a_s), \vec{x}(a), ..., \vec{y}(h), \vec{y}(a_f))$ & $\vec{\phi}(\sigma)$ (shorthand) & $\overline{\useProj(\eventLog'_1)}(\sigma) $ \\
					\hline
					$\epsilon$ & $(1,0,0,0,0,0,0,0,0,0,0,0,0,0,0,0,0,0,0,0,0)$ & $([],\bot)$ & $56$\\
					\hline
					$\langle a_s \rangle$ & $(1,0,0,0,0,0,0,0,0,0,0,-1,0,0,0,0,0,0,0,0,0)$ & $([],a_s)$ & $56$\\
					\hline
					$\langle a_s, a \rangle$ & $(1,1,0,0,0,0,0,0,0,0,0,-1,-1,0,0,0,0,0,0,0,0)$ & $([a_s],a)$ & $56$ \\
					\hline
					$\langle a_s,a,b \rangle$ & $(1,1,1,0,0,0,0,0,0,0,0,-1,-1,-1,0,0,0,0,0,0,0)$ & $([a_s,a],b)$ & $22$ \\
					$\langle a_s,a,c \rangle$ & $(1,1,1,0,0,0,0,0,0,0,0,-1,-1,0,-1,0,0,0,0,0,0)$ & $([a_s,a],c)$ & $12$ \\
					$\langle a_s,a,d \rangle$ & $(1,1,1,0,0,0,0,0,0,0,0,-1,-1,0,0,-1,0,0,0,0,0)$ & $([a_s,a],d)$ &  $22$ \\
					\hline
					$\langle a_s,a,b,c \rangle$ & $(1,1,1,1,0,0,0,0,0,0,0,-1,-1,-1,-1,0,0,0,0,0,0)$ & $([a_s,a,b],c)$ & $1$ \\
					$\langle a_s,a,b,d \rangle$ & $(1,1,1,1,0,0,0,0,0,0,0,-1,-1,-1,0,-1,0,0,0,0,0)$ & $([a_s,a,b],d)$ & $21$ \\
					$\langle a_s,a,c,d \rangle$ & $(1,1,1,0,1,0,0,0,0,0,0,-1,-1,0,-1,-1,0,0,0,0,0)$ & $([a_s,a,c],b)$ & $12$ \\
					$\langle a_s,a,d,c \rangle$ & $(1,1,1,0,0,1,0,0,0,0,0,-1,-1,0,-1,-1,0,0,0,0,0)$ & $([a_s,a,d],c)$ &  $22$ \\ 
					\hline
					$\langle a_s,a,b,c,d \rangle$ & $(1,1,1,1,1,0,0,0,0,0,0,-1,-1,-1,-1,-1,0,0,0,0,0)$ & $([a_s,a,b,c],d)$ & $1$ \\
					$\langle a_s,a,b,d,e \rangle$ & $(1,1,1,1,0,1,0,0,0,0,0,-1,-1,-1,0,-1,-1,0,0,0,0)$ & $([a_s,a,b,d],e)$ & $21$ \\
					$\langle a_s,a,c,d,e \rangle$ & $(1,1,1,0,1,1,0,0,0,0,0,-1,-1,0,-1,-1,-1,0,0,0,0)$ & $([a_s,a,c,d],e)$ & $12$ \\
					$\langle a_s,a,d,c,e \rangle$ & $(1,1,1,0,1,1,0,0,0,0,0,-1,-1,0,-1,-1,-1,0,0,0,0)$ & $([a_s,a,c,d],e)$ &  $22$ \\ 
					\hline
					$\langle a_s,a,b,c,d,e \rangle$ & $(1,1,1,1,1,1,0,0,0,0,0,-1,-1,-1,-1,-1,-1,0,0,0,0)$ & $([a_s,a,b,c,d],e)$ & $1$ \\
					$\langle a_s,a,b,d,e,f \rangle$ & $(1,1,1,1,0,1,1,0,0,0,0,-1,-1,-1,0,-1,-1,-1,0,0,0)$ & $([a_s,a,b,d,e],f)$ & $11$ \\
					$\langle a_s,a,b,d,e,g \rangle$ & $(1,1,1,1,0,1,1,0,0,0,0,-1,-1,-1,0,-1,-1,0,-1,0,0)$ & $([a_s,a,b,d,e],g)$ & $10$ \\
					$\langle a_s,a,c,d,e,f \rangle$ & $(1,1,1,0,1,1,1,0,0,0,0,-1,-1,0,-1,-1,-1,-1,0,0,0)$ & $([a_s,a,c,d,e],f)$ & $12$ \\
					$\langle a_s,a,d,c,e,f \rangle$ & $(1,1,1,0,1,1,1,0,0,0,0,-1,-1,0,-1,-1,-1,-1,0,0,0)$ & $([a_s,a,c,d,e],f)$ &  $13$ \\ 
					$\langle a_s,a,d,c,e,h \rangle$ & $(1,1,1,0,1,1,1,0,0,0,0,-1,-1,0,-1,-1,-1,0,0,-1,0)$ & $([a_s,a,c,d,e],h)$ &  $9$ \\ 
					\hline
					$\langle a_s,a,b,c,d,e,g \rangle$ & $(1,1,1,1,1,1,1,0,0,0,0,-1,-1,-1,-1,-1,-1,0,-1,0,0)$ & $([a_s,a,b,c,d,e],g)$ & $1$ \\
					$\langle a_s,a,b,d,e,f,c \rangle$ & $(1,1,1,1,0,1,1,1,0,0,0,-1,-1,-1,-1,-1,-1,-1,0,0,0)$ & $([a_s,a,b,d,e,f],c)$ & $11$ \\
					$\langle a_s,a,b,d,e,g,a_f \rangle$ & $(1,1,1,1,0,1,1,0,1,0,0,-1,-1,-1,0,-1,-1,0,-1,0,-1)$ & $([a_s,a,b,d,e,g],a_f)$ & $10$\\
					$\langle a_s,a,c,d,e,f,d \rangle$ & $(1,1,1,0,1,1,1,1,0,0,0,-1,-1,0,-1,-2,-1,-1,0,0,0)$ & $([a_s,a,c,d,e,f],d)$ & $12$ \\
					$\langle a_s,a,d,c,e,f,b \rangle$ & $(1,1,1,0,1,1,1,1,0,0,0,-1,-1,-1,-1,-1,-1,-1,0,0,0)$ & $([a_s,a,c,d,e,f],b)$ &  $13$ \\ 
					$\langle a_s,a,d,c,e,h,a_f \rangle$ & $(1,1,1,0,1,1,1,0,0,1,0,-1,-1,0,-1,-1,-1,0,0,-1,-1)$ & $([a_s,a,c,d,e,h],a_f)$ &  $9$ \\ 
					\hline
					$\langle a_s,a,b,c,d,e,g,a_f \rangle$ & $(1,1,1,1,1,1,1,0,1,0,0,-1,-1,-1,-1,-1,-1,0,-1,0,-1)$ & $([a_s,a,b,c,d,e,g],a_f)$ & $1$ \\
					$\langle a_s,a,b,d,e,f,c,d \rangle$ & $(1,1,1,1,1,1,1,1,0,0,0,-1,-1,-1,-1,-2,-1,-1,0,0,0)$ & $([a_s,a,b,c,d,e,f],d)$ & $11$ \\
					$\langle a_s,a,c,d,e,f,d,b \rangle$ & $(1,1,1,0,1,2,1,1,0,0,0,-1,-1,-1,-1,-2,-1,-1,0,0,0)$ & $([a_s,a,c,d^2,e,f],b)$ & $12$ \\
					$\langle a_s,a,d,c,e,f,b,d \rangle$ & $(1,1,1,1,1,1,1,1,0,0,0,-1,-1,-1,-1,-2,-1,-1,0,0,0)$ & $([a_s,a,b,c,d,e,f],d)$ &  $13$ \\
					\hline
					$\langle a_s,a,b,d,e,f,c,d,e \rangle$ & $(1,1,1,1,1,2,1,1,0,0,0,-1,-1,-1,-1,-2,-2,-1,0,0,0)$ & $([a_s,a,b,c,d^2,e,f],e)$ & $11$ \\
					$\langle a_s,a,c,d,e,f,d,b,e \rangle$ & $(1,1,1,1,1,2,1,1,0,0,0,-1,-1,-1,-1,-2,-2,-1,0,0,0)$ & $([a_s,a,b,c,d^2,e,f],e)$ & $12$ \\
					$\langle a_s,a,d,c,e,f,b,d,e \rangle$ & $(1,1,1,1,1,2,1,1,0,0,0,-1,-1,-1,-1,-2,-2,-1,0,0,0)$ & $([a_s,a,b,c,d^2,e,f],e)$ &  $13$ \\
					\hline
					$\langle a_s,a,b,d,e,f,c,d,e,g \rangle$ & $(1,1,1,1,1,2,2,1,0,0,0,-1,-1,-1,-1,-2,-2,-1,-1,0,0)$ & $([a_s,a,b,c,d^2,e^2,f],g)$ & $11$ \\
					$\langle a_s,a,c,d,e,f,d,b,e,g \rangle$ & $(1,1,1,1,1,2,2,1,0,0,0,-1,-1,-1,-1,-2,-2,-1,-1,0,0)$ & $([a_s,a,b,c,d^2,e^2,f],g)$ & $12$ \\
					$\langle a_s,a,d,c,e,f,b,d,e,h \rangle$ & $(1,1,1,1,1,2,2,1,0,0,0,-1,-1,-1,-1,-2,-2,-1,0,-1,0)$ & $([a_s,a,b,c,d^2,e^2,f],h)$ &  $13$ \\
					\hline
					$\langle a_s,a,b,d,e,f,c,d,e,g,a_f \rangle$ & $(1,1,1,1,1,2,2,1,1,0,0,-1,-1,-1,-1,-2,-2,-1,-1,0,-1)$ & $([a_s,a,b,c,d^2,e^2,f,g],a_f)$ & $11$ \\
					$\langle a_s,a,c,d,e,f,d,b,e,g,a_f \rangle$ & $(1,1,1,1,1,2,2,1,1,0,0,-1,-1,-1,-1,-2,-2,-1,-1,0,-1)$ & $([a_s,a,b,c,d^2,e^2,f,g],a_f)$ & $12$ \\
					$\langle a_s,a,d,c,e,f,b,d,e,h,a_f \rangle$ & $(1,1,1,1,1,2,2,1,0,1,0,-1,-1,-1,-1,-2,-2,-1,0,-1,-1)$ & $([a_s,a,b,c,d^2,e^2,f,h],a_f)$ &  $13$ \\
					\hline
				\end{tabular}
			}
		\end{center}
	\end{table}
	The table shows each sequence present in $\overline{\useProj(\eventLog'_1)}$ accompanied by its $\vec{\phi}$-value and the number of occurrences of the sequence in $\overline{\useProj(\eventLog'_1)}$, e.g.  $\overline{\useProj(\eventLog'_1)}(\langle a_s,a \rangle) = 56$.
	Observe that there is a relation between the occurrence of a sequence and its corresponding postfixes, i.e. after the $56$ times that sequence $\langle a_s,a \rangle$ occurred, $\langle a_s,a,b \rangle$ occurred $22$ times, $\langle a_s,a,c \rangle$ occurred $12$ times and $\langle a_s,a,d \rangle$ occurred $22$ times (note: $56 = 22 + 12 + 22$). 
	Due to coupling of sequences to constraints, i.e. by means of sequence encoding, we can now apply the aforementioned reasoning to constraints as well. 
	The frequencies in $\overline{\useProj(\eventLog'_1)}$ allow us to decide whether the presence of a certain constraint is in line with predominant behaviour in the event log. 
	For example, in \autoref{tab:sec:eliminating_replay_fitness_tab:bag_prefix_closure_noise}, $\vec{\phi}(\langle a_s,a,b,c \rangle)$ seems to relate to \textit{infrequent behaviour} as it appears only once. 
	
	To apply filtering, we construct a weighted directed graph in which each sequence encoding acts as a vertex. 
	We connect two vertices by means of an arc if the source constraint corresponds to a sequence that is a prefix of a sequence corresponding to the target constraint, i.e., we connect $\vec{\phi}(\langle a_s, a \rangle)$ to $\vec{\phi}(\langle a_s, a,b \rangle)$ as $\langle a_s, a \rangle$ is a prefix of $\langle a_s,a,b \rangle$.
	Arc weight is based on trace frequency in the input event log.
	\begin{figure}[b!]
		\begin{center}
			\resizebox{\textwidth}{!}{
				\begin{tikzpicture}[->,>=stealth',shorten >=1pt,auto,node distance=2.25cm,main node/.style={circle,fill=white,draw,minimum size=1pt, inner sep=0pt}]
				
				\node[main node] (eps) [label=above:{$([], \bot)$}]{};
				\node[main node] (n1) [label=above right:{$([],a_s)\ \ \ \ $}, below right = of eps]{};
				\node[main node] (n2) [label=above right:{$([a_s],a)$}, below right = of n1]{};
				\node[main node] (n3) [label=above:{$([a_s,a],b)$}, left = of n2]{};
				\node[main node] (n4) [label= above left:{$([a_s,a],c)$}, below = of n2]{};
				\node[main node] (n5) [label=above:{$([a_s,a],d)$}, right = of n2]{};
				\node[main node] (n6) [label=above :{$([a_s,a,b],c)$}, left = of n3]{};
				\node[main node] (n7) [label=above left:{$([a_s,a,b],d)$}, below = of n3]{};
				\node[main node] (n8) [label= below:{$([a_s,a,c],d)$}, right = of n4]{};
				\node[main node] (n9) [label=above right:{$([a_s,a,d],c)$}, right = of n5]{};
				\node[main node] (n10) [label=above left:{$([a_s,a,b,c],d)$}, left = of n6]{};
				\node[main node] (n11) [label=above left:{$([a_s,a,b,d],e)$}, below left = of n7]{};
				\node[main node] (n12) [label= above right:{$([a_s,a,c,d],e)$}, right = of n8]{};
				\node[main node] (n13) [label=above left:{$([a_s,a,b,c,d],e)$}, below left = of n10]{};
				\node[main node] (n14) [label=above left:{$([a_s,a,b,d,e],f)$}, below left = of n11]{};
				\node[main node] (n15) [label=above right:{$([a_s,a,b,d,e],g)$}, below = of n11]{};
				\node[main node] (n16) [label= above left:{$([a_s,a,c,d,e],f)$}, below left = of n12]{};
				\node[main node] (n17) [label= above right:{$([a_s,a,c,d,e],h)$}, below right = of n12]{};
				\node[main node] (n18) [label=above left:{$([a_s,a,b,c,d,e],g)$}, below left = of n13]{};
				\node[main node] (n19) [label=above left:{$([a_s,a,b,d,e,f],c)$}, below left = of n14]{};
				\node[main node] (n20) [label=right:{$([a_s,a,b,d,e,g],a_f)$}, below left = of n15]{};
				\node[main node] (n21) [label= above left:{$([a_s,a,c,d,e,f],b)$}, below left = of n16]{};
				\node[main node] (n22) [label= above right:{$([a_s,a,c,d,e,f],d)$}, below right = of n16]{};
				\node[main node] (n23) [label= above right:{$([a_s,a,c,d,e,h],a_f)$}, below right = of n17]{};
				\node[main node] (n24) [label=above left:{$([a_s,a,b,c,d,e,g],a_f)$}, below left = of n18]{};
				\node[main node] (n25) [label=above left:{$([a_s,a,b,d,e,f,c],d)$}, below = of n19]{};
				\node[main node] (n26) [label= above left:{$([a_s,a,b,c,d,e,f],d)$}, below left = of n21]{};
				\node[main node] (n27) [label= above right:{$([a_s,a,c,d^2,e,f],b)$}, below = of n22]{};
				\node[main node] (n28) [label= right:{$([a_s,a,b,c,d^2,e,f],e)$}, below = of n26]{};
				\node[main node] (n29) [label= below left:{$([a_s,a,b,c,d^2,e^2,f],g)$}, below left = of n28]{};
				\node[main node] (n30) [label= below:{$([a_s,a,b,c,d^2,e^2,f],h)$}, below right = of n28]{};
				\node[main node] (n31) [label= left:{$([a_s,a,b,c,d^2,e^2,f,g],a_f)$}, left = of n29]{};
				\node[main node] (n32) [label= right:{$([a_s,a,b,c,d^2,e^2,f,h],a_f)$}, right = of n30]{};

				\path[->]
				(eps) edge node [right, label=above:{$\ \ 56$}] {} (n1)
				(n1) edge node [right, label=above:{$\ \ 56$}] {} (n2)
				(n2) edge node [right, label=above:{$22$}] {} (n3)
				(n2) edge node [right, label=above:{$\ \ 12$}] {} (n4)
				(n2) edge node [right, label=above:{$\ \ \ 22$}] {} (n5)
				(n3) edge node [right, label=above:{$\ \ 21$}] {} (n7)
				(n4) edge node [right, label=above:{$12$}] {} (n8)
				(n5) edge node [right, label=above:{$22$}] {} (n9)
				(n7) edge node [right, label=above:{$\ \ 21$}] {} (n11)
				(n8) edge node [right, label=above:{$\ \ 12$}] {} (n12)
				(n9) edge node [right, label=above:{$\ \ 22$}] {} (n12)
				(n11) edge node [right, label=above:{$11$}] {} (n14)
				(n11) edge node [right, label=above:{$\ \ 10$}] {} (n15)
				(n12) edge node [right, label=above:{$25$}] {} (n16)
				(n12) edge node [right, label=above:{$9$}] {} (n17)
				(n14) edge node [right, label=above:{$11$}] {} (n19)
				(n15) edge node [right, label=above:{$10$}] {} (n20)
				(n16) edge node [right, label=above:{$\ \ 13$}] {} (n21)
				(n16) edge node [right, label=above:{$12$}] {} (n22)
				(n17) edge node [right, label=above:{$9$}] {} (n23)
				(n19) edge node [right, label=above:{$\ \ 11$}] {} (n25)
				(n21) edge node [right, label=above:{$13$}] {} (n26)
				(n22) edge node [right, label=above:{$\ \ 12$}] {} (n27)
				(n25) edge node [right, label=above:{$11$}] {} (n28)
				(n26) edge node [right, label=above:{$\ \ 13$}] {} (n28)
				(n27) edge node [right, label=above:{$12$}] {} (n28)
				(n28) edge node [right, label=above:{$23$}] {} (n29)
				(n28) edge node [right, label=above:{$13$}] {} (n30)
				(n29) edge node [right, label=above:{$23$}] {} (n31)
				(n30) edge node [right, label=above:{$13$}] {} (n32);

				\path[->, dotted]
				(n3) edge node [right, label=above:{$1$}]{} (n6)
				(n6) edge node [right, label=above:{$1$}]{} (n10)
				(n10) edge node [right, label=above:{$1$}]{} (n13)
				(n13) edge node [right, label=above:{$1$}]{} (n18)
				(n18) edge node [right, label=above:{$1$}]{} (n24);
				\end{tikzpicture}
			}
		\end{center}
		\caption{An example sequence encoding graph $G'_1$, based on example event log $L'_1$.}
		\label{fig:sec:eliminating_replay_fitness_fig:seq_enc_graph_1}
	\end{figure}
	
	\begin{definition}[Sequence encoding graph]
		Given event log $\eventLog$ over set of activities $\activities_{\eventLog}$. A sequence encoding graph is a directed graph $G = (V,E, \psi)$ where $V = \{\vec{\phi}(\sigma) \mid \sigma \in \overline{L} \}$, $E \subseteq V \times V$ s.t. $(\vec{\phi}(\sigma'),\vec{\phi}(\sigma)) \in E \Leftrightarrow \exists_{a \in A}(\sigma' \cdot \langle a \rangle = \sigma)$ and $\psi : E \rightarrow \mathbb{N}$ where:
		\[
		\psi(v_1,v_2) = \sum_{\tiny \begin{matrix} \sigma \in \overline{L}\\ \vec{\phi}(\sigma) = v_2 \end{matrix}}\overline{L}(\sigma) - \sum_{\tiny \begin{matrix}\sigma' \in \overline{L}\\
			\sigma' \cdot \langle a \rangle \in \overline{L} \\ \vec{\phi}(\sigma' \cdot \langle a \rangle) = v_2 \\ \vec{\phi(\sigma')} \neq v_1 \end{matrix}} \overline{L}(\sigma')
		\]
	\end{definition}
	Consider the sequence encoding graph in \autoref{fig:sec:eliminating_replay_fitness_fig:seq_enc_graph_1}, based on $\useProj(\eventLog'_1)$, as an example.
	By definition, $([], \bot)$ is the root node of the graph and connects to all one-sized sequences.
	Within the graph we observe the relation among different constraints, combined with their absolute frequencies based on $\eventLog'_1$
	
	\subsection{Filtering}
	\label{subsec:sec:eliminating_replay_fitness_subsec:filtering}
	Given a sequence encoding graph we are able to filter out constraints.
	In \autoref{alg:sec:eliminating_replay_fitness_alg:sef_bfs} we devise a simple breadth-first traversal algorithm, i.e. \texttt{Sequence Encoding Filtering - Breadth First Search} (\texttt{SEF-BFS}), that traverses the sequence encoding graph and concurrently constructs a set of ILP constraints. 
	The algorithm needs a function as an input that is able to determine, given a vertex in the sequence encoding graph, what portion of adjacent vertices remains in the graph and which are removed.

	\begin{definition}[Sequence encoding filter]
		Given event log $\eventLog$ over set of activities $\activities_{\eventLog}$ and a corresponding sequence encoding graph $G = (V,E, \psi)$. A sequence encoding filter is a function $\kappa : V \rightarrow \powerSet(V)$.
	\end{definition}
	Note that $\kappa$ is an abstract function and might be parametrized As an example consider $\kappa^{\alpha}_{\max}$ which we define as:
	\[
	\kappa^{\alpha}_{\max}(v)  = \{v' \mid (v,v') \in E \wedge \psi(v,v') \geq  (1 - \alpha) \cdot \max_{v'' \in V} \psi(v,v'') \},\ \alpha \in [0,1]
	\]
	Other instantiations of $\kappa$ are possible as well and hence $\kappa$ is a parameter of the general approach.
	It is however desirable that $\kappa(v) \subseteq \{v' \mid (v,v') \in E\}$, i.e. it only considers vertices reached by $v$ by means of an arc.
	Given an instantiation of $\kappa$, it is straightforward to construct a filtering algorithm based on breadth-first graph traversal, i.e. \texttt{SEF-BFS}.
	
	\begin{algorithm}[tb]
		\small
		\caption{SEF-BFS\label{alg:sec:eliminating_replay_fitness_alg:sef_bfs}}
		\DontPrintSemicolon
		\SetKwInOut{Input}{input}\SetKwInOut{Output}{output}
		\Input{$G=(V,E,\psi)$, $\kappa : V \rightarrow \mathbb{P}(V)$}
		\Output{$C \subseteq V$}
		\Begin{
			\nl $C \leftarrow \emptyset$\;
			\nl Let $Q$ be a FIFO queue\;
			\nl $Q.enqueue(([],\bot))$\;
			\nl \While{$Q \neq \emptyset$}{
				\nl $v \leftarrow Q.dequeue()$\;
				\nl     \For{$v' \in \kappa(v)$}{
					\nl         $C \leftarrow C \cup \{v'\}$\;
					\nl         $Q.enqueue(v')$\;
				}
			}
		}
	\end{algorithm}

	The algorithm inherits its worst-case complexity of breadth first search, multiplied by the worst-case complexity of $\kappa$. 
	Thus, in case $\kappa$'s worst-case complexity is $O(1)$ then we have $O(|V| + |E|)$ for the \texttt{SEF-BFS}-algorithm.
	It is trivial to prove, by means of induction on the length of a sequence encoding's corresponding sequence, that a sequence encoding graph is acyclic.
	Hence, termination is guaranteed.
	
	As an example of executing the \texttt{SEF-BFS} algorithm, reconsider \autoref{fig:sec:eliminating_replay_fitness_fig:seq_enc_graph_1}.
	Assume we use $\kappa^{0.75}_{\max}$.
	Vertex $([],\bot)$ is initially present in $Q$ and will be analyzed.
	Since $([],a_s)$ is the only child of $([],\bot)$, it is added to $Q$.
	Vertex $([],\bot)$ is removed from the queue and is never inserted in the queue again due to the acyclic property of the graph.
	Similarly, since $([a_s],a)$ is the only child of $([], a_s)$ it is added to $Q$. 
	All children of $([a_s],a)$, i.e. $([a_s,a],b)$, $([a_s,a],c)$ and $([a_s,a],d)$, are added to the queue since the maximum corresponding arc value is $22$, and, $(1 - 0.75) * 22 = 5.5$, which is smaller than the lowest arc value $12$.
	When analysing $([a_s,a],b)$ we observe a maximum outgoing arc with value $21$ to vertex $([a_s,a,b],d)$ which is enqueued in $Q$.
	Since $(1-0.25)*21=5.25$, the algorithm does not enqueue $([a_s,a,b],c)$.
	Note that the whole path of vertices from $([a_s,a,b],c)$ to $([a_s,a,b,c,d,e,g], a_f)$ is never analysed and is stripped from the constraint body. 
	
	When applying ILP-based process discovery based on event log $\eventLog' _1$ with sequence encoding filtering and $\kappa^{0.75}_{\max}$, we obtain the WF-net depicted in \autoref{fig:ilp_example_no_noise}.
	As explained, the filter leaves out all constraints related to vertices on the path from $([a_s,a,b],c)$ to $([a_s,a,b,c,d,e,g], a_f)$.
	Hence, we find a similar model to the model found on event log $\eventLog_1$ and are able to filter out infrequent exceptional behaviour.

	\section{Evaluation}
	\label{sec:evaluation}
	\autoref{alg:ilp_disc} and \autoref{alg:sec:eliminating_replay_fitness_alg:sef_bfs} (sequence encoding filtering) are implemented in the \textit{HybridILPMiner} (\url{http://svn.win.tue.nl/repos/prom/Packages/HybridILPMiner/} package within the \textit{ProM framework}~\cite{DBLP:conf/caise/VerbeekBDA10} (\url{http://www.promtools.org}) and \texttt{RapidProM} framework~\cite{DBLP:journals/corr/AalstBZ2017}.\footnote{Experiments are performed with source code available at: \url{https://github.com/rapidprom/rapidprom-source/tree/2017_computing_ilp_1}}\footnote{Experiments are conducted on machines with 8 Intel
		Xeon CPU E5-2407 v2 \@ 2.40 GHz processors and 64 GB RAM}
	Using this implementation we validated the approach.
	In an artificial setting we evaluated the quality of models discovered and the efficiency of applying sequence encoding filtering.
	We also compare sequence encoding to the IMi~\cite{DBLP:conf/bpm/LeemansFA13} algorithm and automaton-based filtering~\cite{DBLP:journals/tkde/ConfortiRH17}.
	Finally, we assess the performance of sequence encoding filtering on real event data~\cite{https://doi.org/10.4121/uuid:270fd440-1057-4fb9-89a9-b699b47990f5,https://doi.org/10.4121/uuid:915d2bfb-7e84-49ad-a286-dc35f063a460}.

	\subsection{Model Quality}
	\label{sec:model_quality}
	
	The event logs used in the empirical evaluation of model quality are artificially generated event logs and originate from a study related to the impact of exceptional behaviour to rule-based approaches in process discovery~\cite{DBLP:journals/datamine/MarusterWAB06}. 
	Three event logs where generated out of three different process models, i.e. the \textit{ground truth event logs}.
	These event logs do not consist of any exceptional behaviour, i.e. every trace fits the originating model.
	The ground truth event logs are called \textit{a12f0n00}, \textit{a22f0n00} and \textit{a32f0n00}.
	The two digits behind the \textit{a} character indicate the number of activities present in the event log, i.e. \textit{a12f0n00} contains 12 different activities.
	From each ground truth event log, by means of trace manipulation, four other event logs are created that do contain exceptional behaviour.
	Manipulation concerns tail/head of trace removal, random part of the trace body removal and interchanging two randomly chosen events~\cite{DBLP:journals/datamine/MarusterWAB06}.
	The percentages of trace manipulation are 5\%, 10\%, 20\% and 50\%. 
	The manipulation percentage is incorporated in the last two digits of the event log's name, i.e. the 5\% manipulation version of the \textit{a22f0n00} event log is called \textit{a22f0n05}.
		
	The existence of ground truth event logs, free of exceptional behaviour, is of utmost importance for evaluation.
	We need to be able to distinguish \textit{normal} from \textit{exceptional} behaviour in an \textit{unambiguous manner}.
	Within evaluation, these event logs, combined with the quality dimension precision, allow us to judge how well a technique is able to filter out exceptional behaviour.
	Recall that precision is defined as the number of traces producible by the process model that are also present in the event log. 
	Thus if all traces producible by a process model are present in an event log, precision is maximal, i.e. the precision value is $1$. 
	If the model allows for traces that are not present in the event log, precision is lower than $1$. 
	
	If exceptional behaviour is present in an event log, the conventional ILP-based process discovery algorithm produces a WF-net that allows for all exceptional behaviour.
	As a result, the algorithm is unable to find any meaningful patterns within the event log.
	This typically leads to places with a lot of self-loops.
	The acceptance of exceptional behaviour by the WF-net, combined with the inability to find meaningful patterns yields a low level of precision, when using the ground truth log as a basis for precision computation.
	On the other hand, if we discover models using an algorithm that is more able to handle the presence of exceptional behaviour, we expect the algorithm to allow for less exceptional behaviour and find more meaningful patterns.
	Thus, w.r.t. the ground truth model, we expect higher precision values.
	
	To evaluate the sequence encoding filtering approach, we have applied the ILP-based process discovery algorithm with sequence encoding filtering using $\kappa^{\alpha}_{max}$ and $\alpha = 0,0.05,0.1,...,0.95,1$.
	Moreover, we performed similar experiments for the IMi~\cite{DBLP:conf/bpm/LeemansFA13}\footnote{\url{http://svn.win.tue.nl/repos/prom/Packages/InductiveMiner/}} and the automaton based event log filter of~\cite{DBLP:journals/tkde/ConfortiRH17}\footnote{\url{http://svn.win.tue.nl/repos/prom/Packages/NoiseFiltering/}}.
	After applying the automaton based filter we applied ILP-based process discovery as a process discovery algorithm.
	We measured precision~\cite{DBLP:series/lnbip/Munoz-Gama16} and replay-fitness~\cite{DBLP:journals/widm/AalstAD12} based on the \textit{ground truth event logs}.
	
\begin{figure}[b]
		\begin{subfigure}[b]{0.325\textwidth}
			\centering
			\includegraphics[width=\textwidth,clip,trim=0cm 2cm 0cm 3cm]{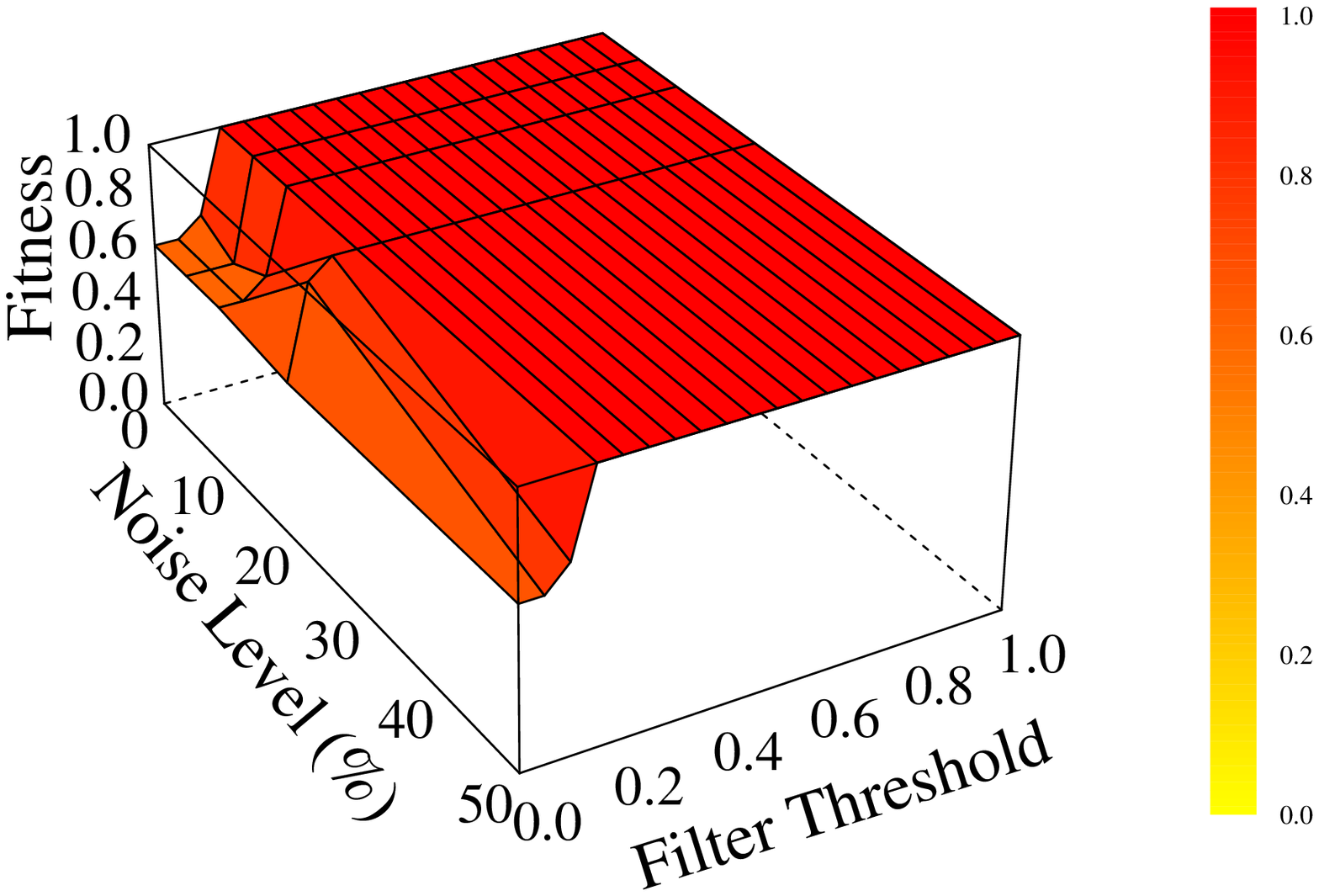}
			\caption{Sequence Encoding}
			\label{fig:a12_ilp_fitness}
		\end{subfigure}
		\begin{subfigure}[b]{0.325\textwidth}
			\centering
			\includegraphics[width=\textwidth,clip,trim=0cm 2cm 0cm 3cm]{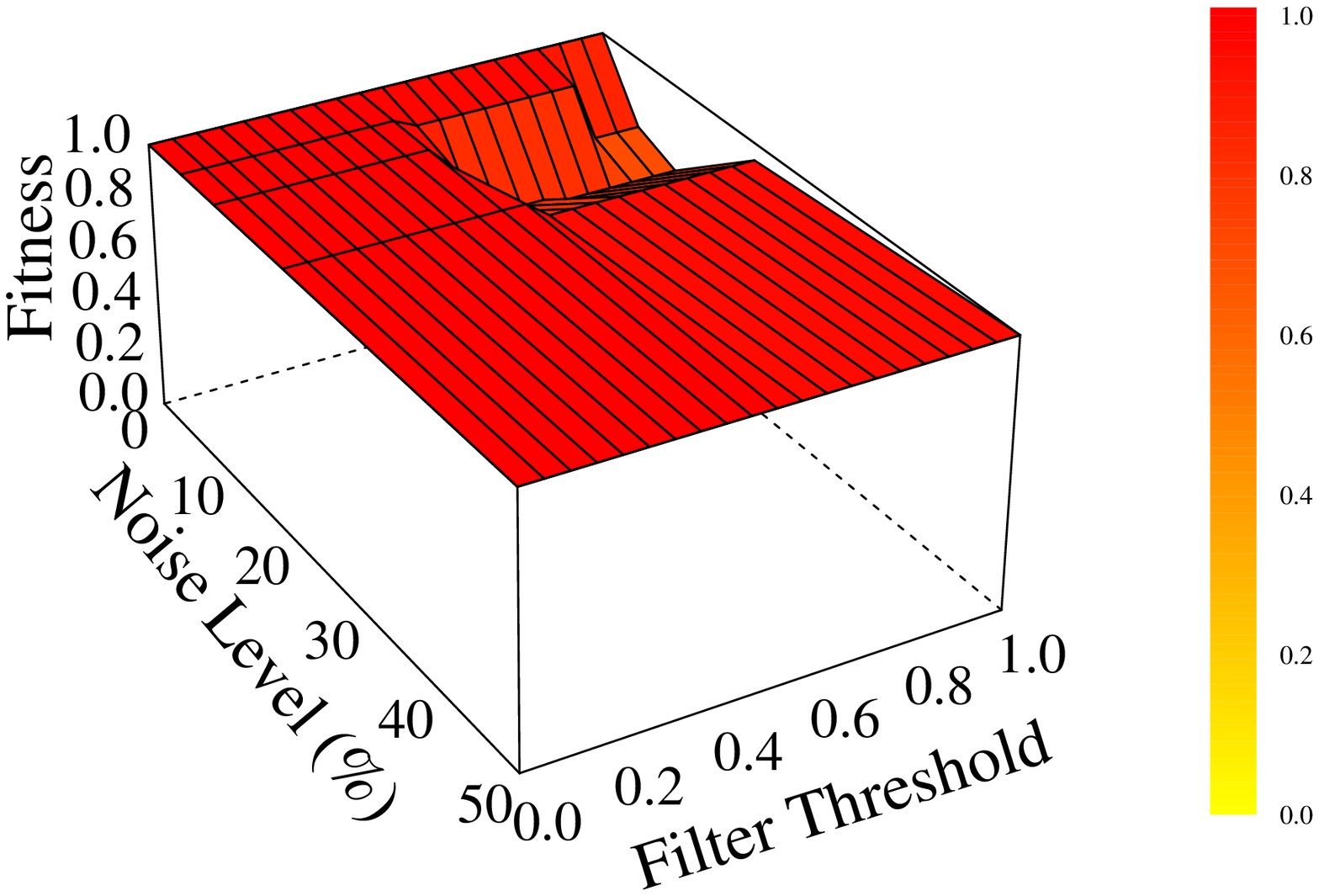}
			\caption{IMi~\cite{DBLP:conf/bpm/LeemansFA13}}
			\label{fig:a12_im_fitness}
		\end{subfigure}
		\begin{subfigure}[b]{0.325\textwidth}
			\centering
			\includegraphics[width=\textwidth,clip,trim=0cm 2cm 0cm 3cm]{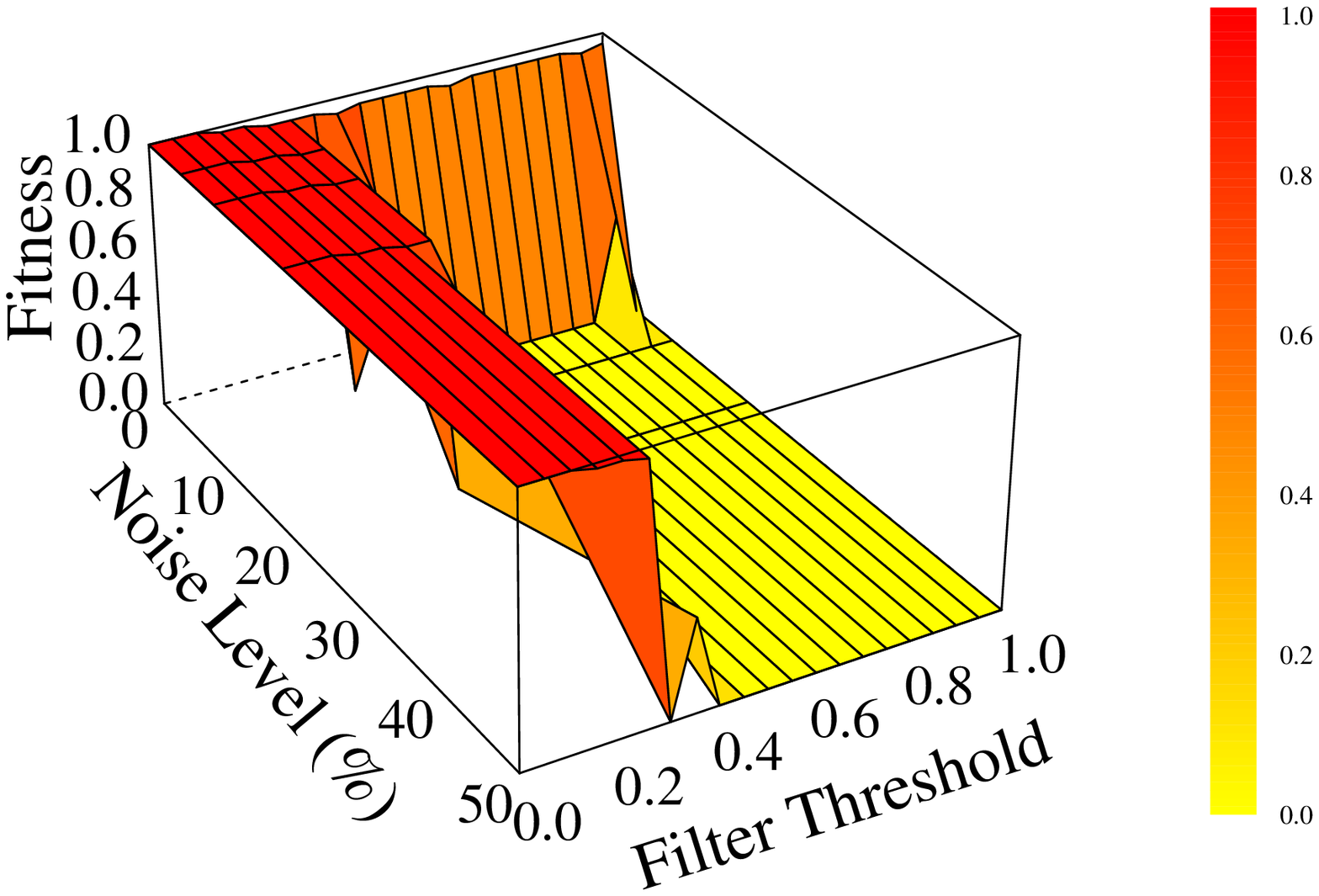}
			\caption{ILP with Automaton Filter~\cite{DBLP:journals/tkde/ConfortiRH17}}
			\label{fig:a12_filter_automata_fitness}
		\end{subfigure}
		\caption{Replay-fitness measurements based on \textit{a12f0nXX}.}
		\label{fig:a12_fitness}
	\end{figure}

	In \autoref{fig:a12_fitness} we present the replay-fitness results of the experiments with the \textit{a12f0nXX} event logs.
	In the charts we plot replay-fitness against the noise level and filter threshold.
	We additionally use a colour scheme to highlight the differences in value.
	Sequence Encoding Filtering has low replay-fitness values for all event logs when using a filter threshold of $0$.
	The replay-fitness value of the models found quickly rises to $1$ and remains $1$ for all filter threshold above $0.2$.
	In case of IMi, for a filter value of $1.0$ (comparable to $0.0$ for sequence encoding) we observe some values of $1$ for replay-fitness.
	Non-perfect replay-fitness seems to be more local, concentrated around noise levels $5\%$ and $10\%$ with corresponding threshold levels in-between $0.4$ and $0.8$. 
	Finally, automaton-based filtering rapidly loses perfect replay-fitness when the filter threshold exceeds $0.2$.
	Only for a noise-level of $0$ it seems to retain high replay-values.
	Upon inspection it turns out the filter returns empty event logs for the corresponding threshold and noise levels.
	
\begin{figure}[tb]
		\begin{subfigure}[b]{0.325\textwidth}
			\centering
			\includegraphics[width=\textwidth,clip,trim=0cm 2cm 0cm 3cm]{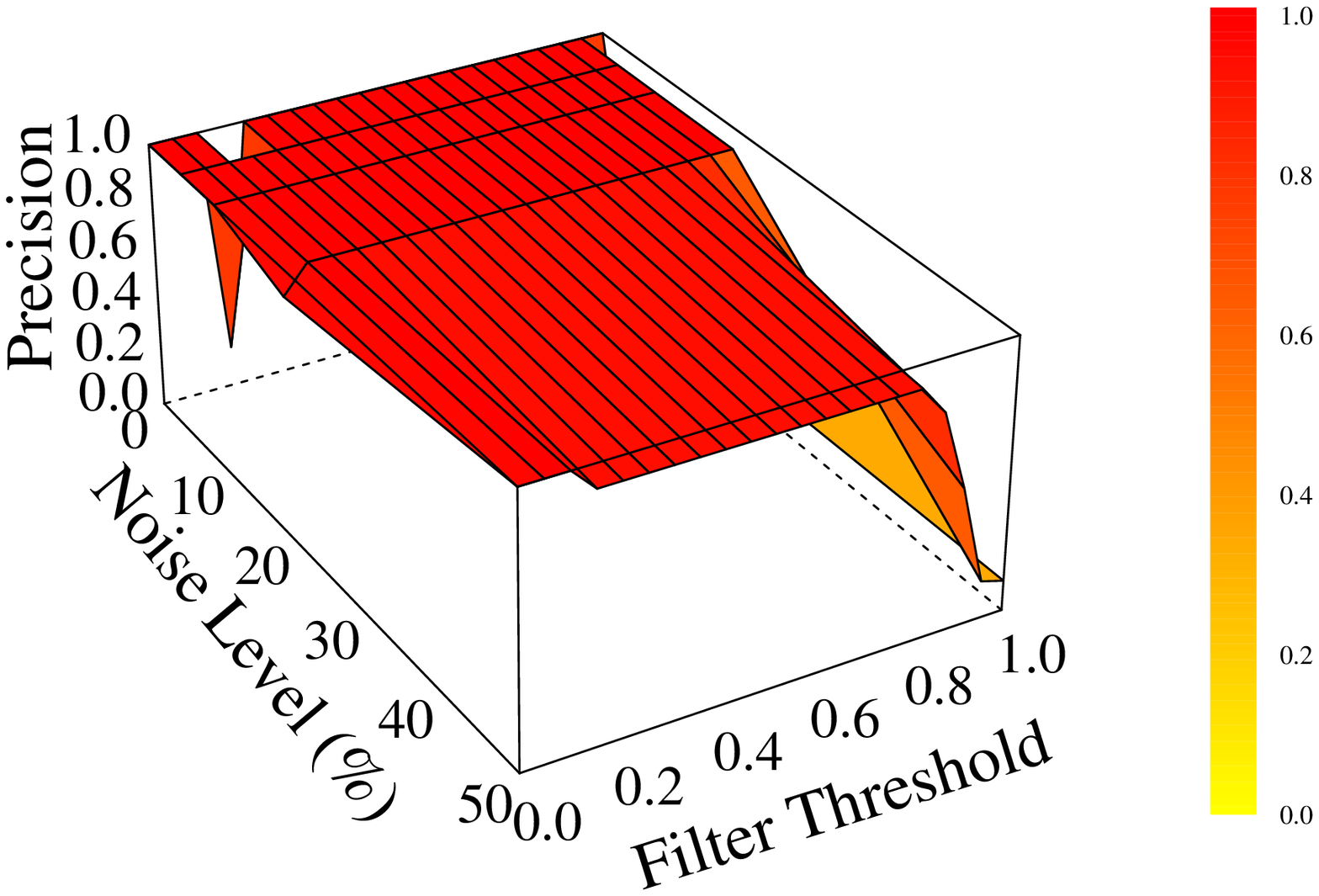}
			\caption{Sequence Encoding}
			\label{fig:a12_ilp_precision}
		\end{subfigure}
		\begin{subfigure}[b]{0.325\textwidth}
			\centering
			\includegraphics[width=\textwidth,clip,trim=0cm 2cm 0cm 3cm]{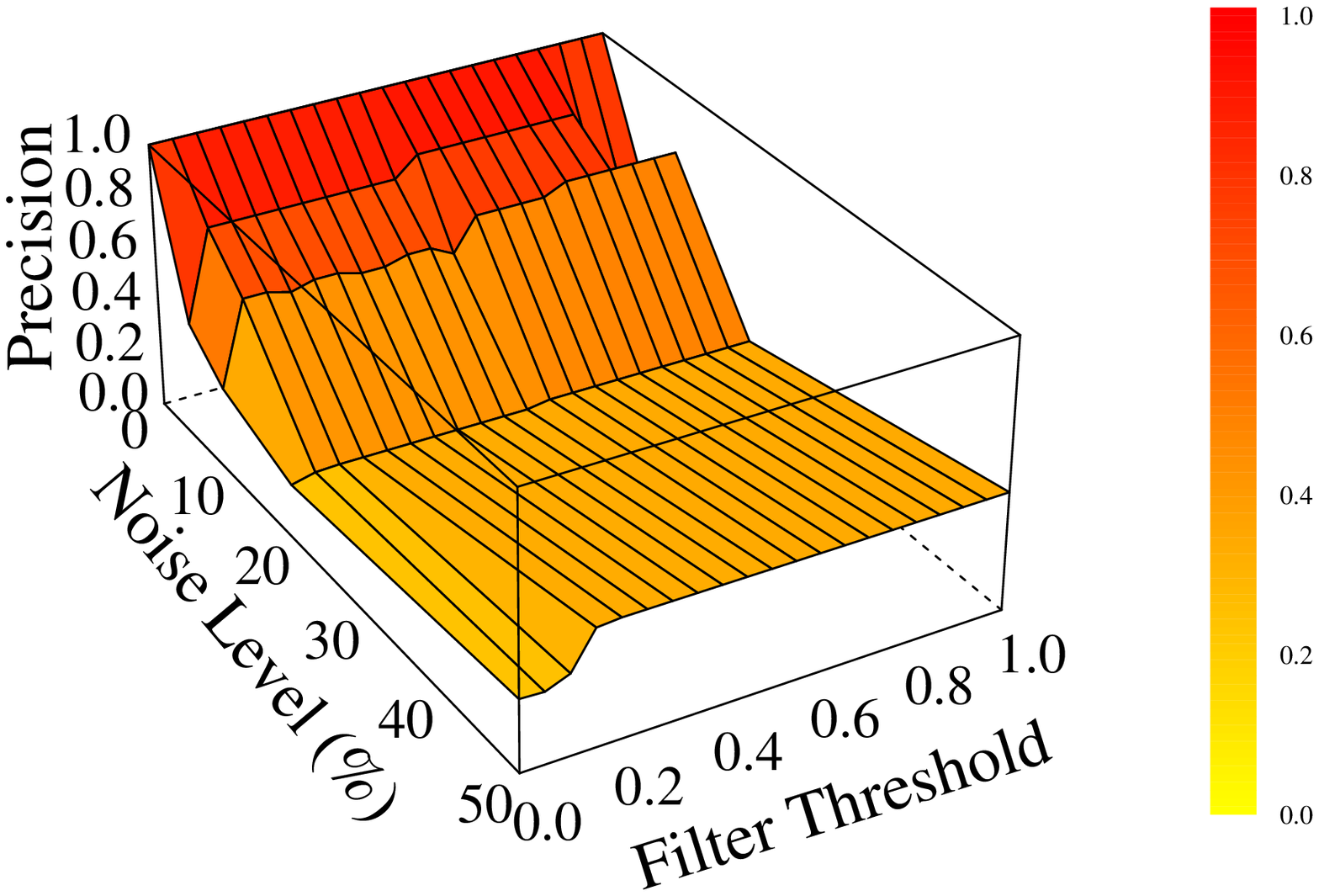}
			\caption{IMi~\cite{DBLP:conf/bpm/LeemansFA13}}
			\label{fig:a12_im_precision}
		\end{subfigure}
		\begin{subfigure}[b]{0.325\textwidth}
			\centering
			\includegraphics[width=\textwidth,clip,trim=0cm 2cm 0cm 3cm]{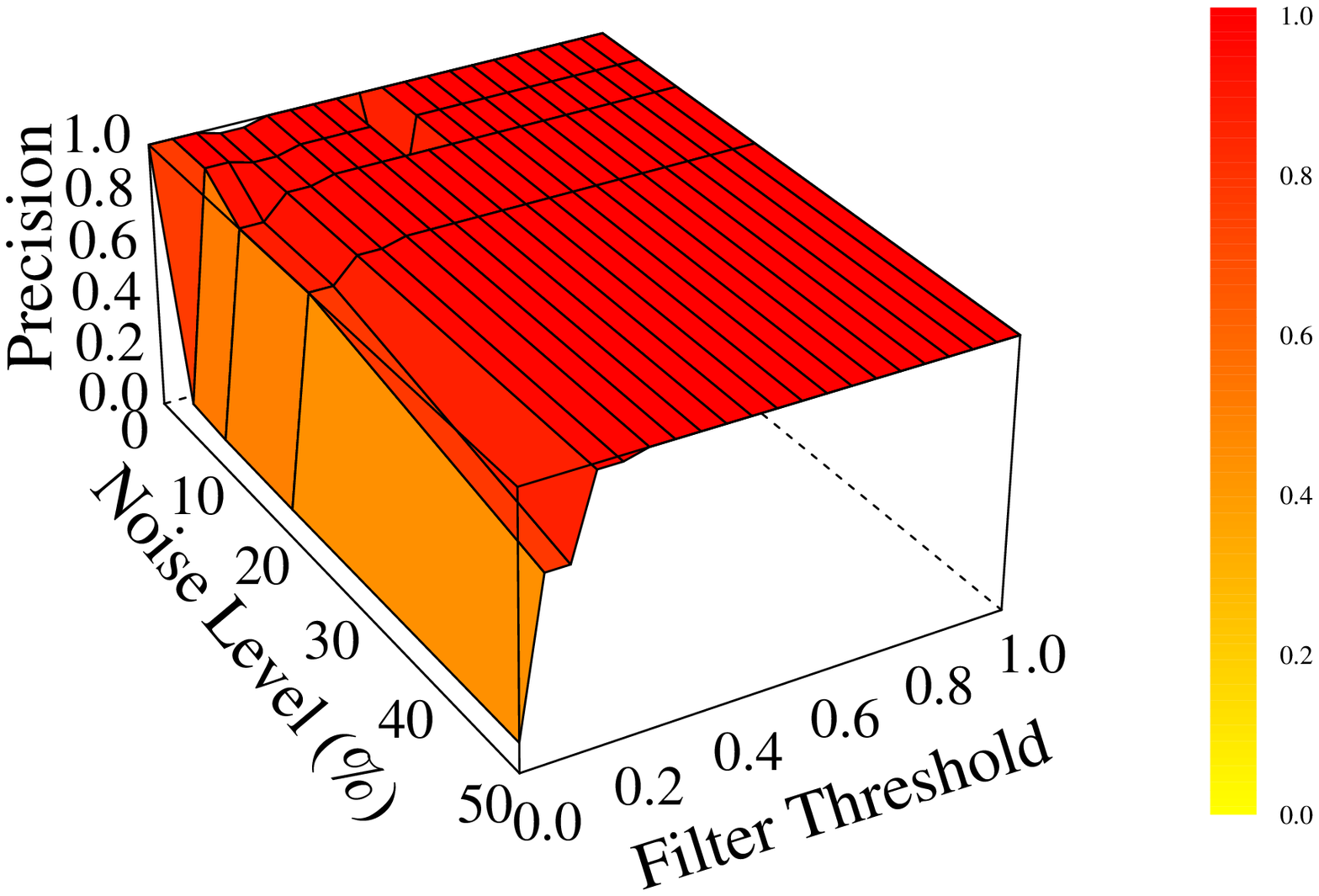}
			\caption{ILP with Automaton Filter~\cite{DBLP:journals/tkde/ConfortiRH17}}
			\label{fig:a12_filter_automata_precision}
		\end{subfigure}
		\caption{Precision measurements based on \textit{a12f0nXX}.}
		\label{fig:a12_precision}
	\end{figure}

	In \autoref{fig:a12_precision} we present the precision results of the experiments with the \textit{a12f0nXX} event logs.
	For sequence encoding the chart shows expected behaviour, i.e. with high noise levels and high filter thresholds precision is low.
	There is however an unexpected drop in precision for noise-level $0$ with a filter threshold around $0.2$.
	The IMi filter behaves a bit more unexpected since the drop in precision seems mainly depending on the noise level rather than the filter setting.
	We expect the precision to be higher in case a filter threshold of $1.0$ is chosen.
	There is only a slight increase for the $50\%$ noise log when comparing a filter threshold of $0$ to a filter threshold of $1$.
	Finally, precision of the automaton filter behaves as expected, i.e., precision rapidly increases together with an increase in the filter threshold.

\begin{figure}[tb]
		\begin{subfigure}[b]{0.325\textwidth}
			\centering
			\includegraphics[width=\textwidth,clip,trim=0cm 2cm 0cm 3cm]{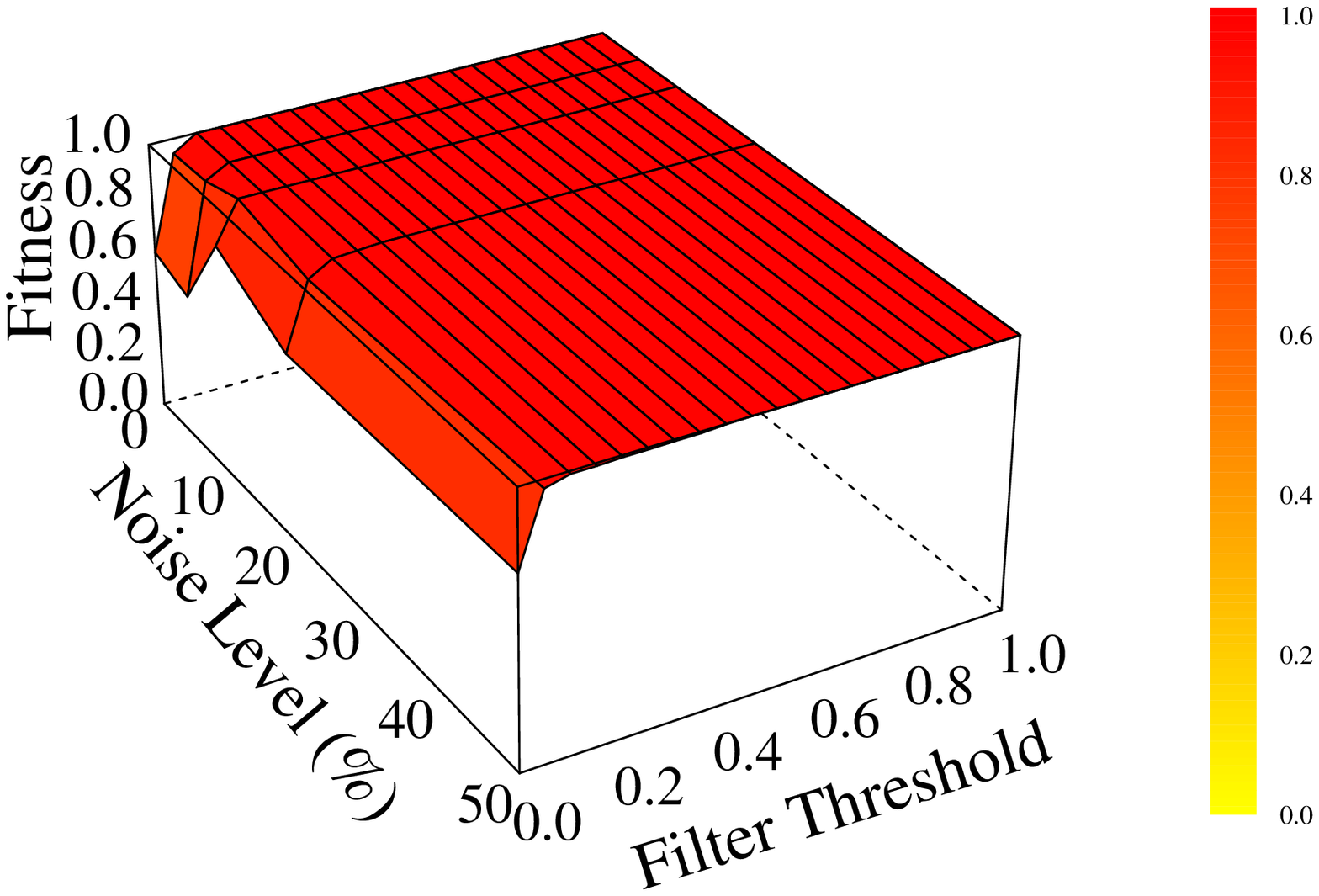}
			\caption{Sequence Encoding}
			\label{fig:a22_ilp_fitness}
		\end{subfigure}
		\begin{subfigure}[b]{0.325\textwidth}
			\centering
			\includegraphics[width=\textwidth,clip,trim=0cm 2cm 0cm 3cm]{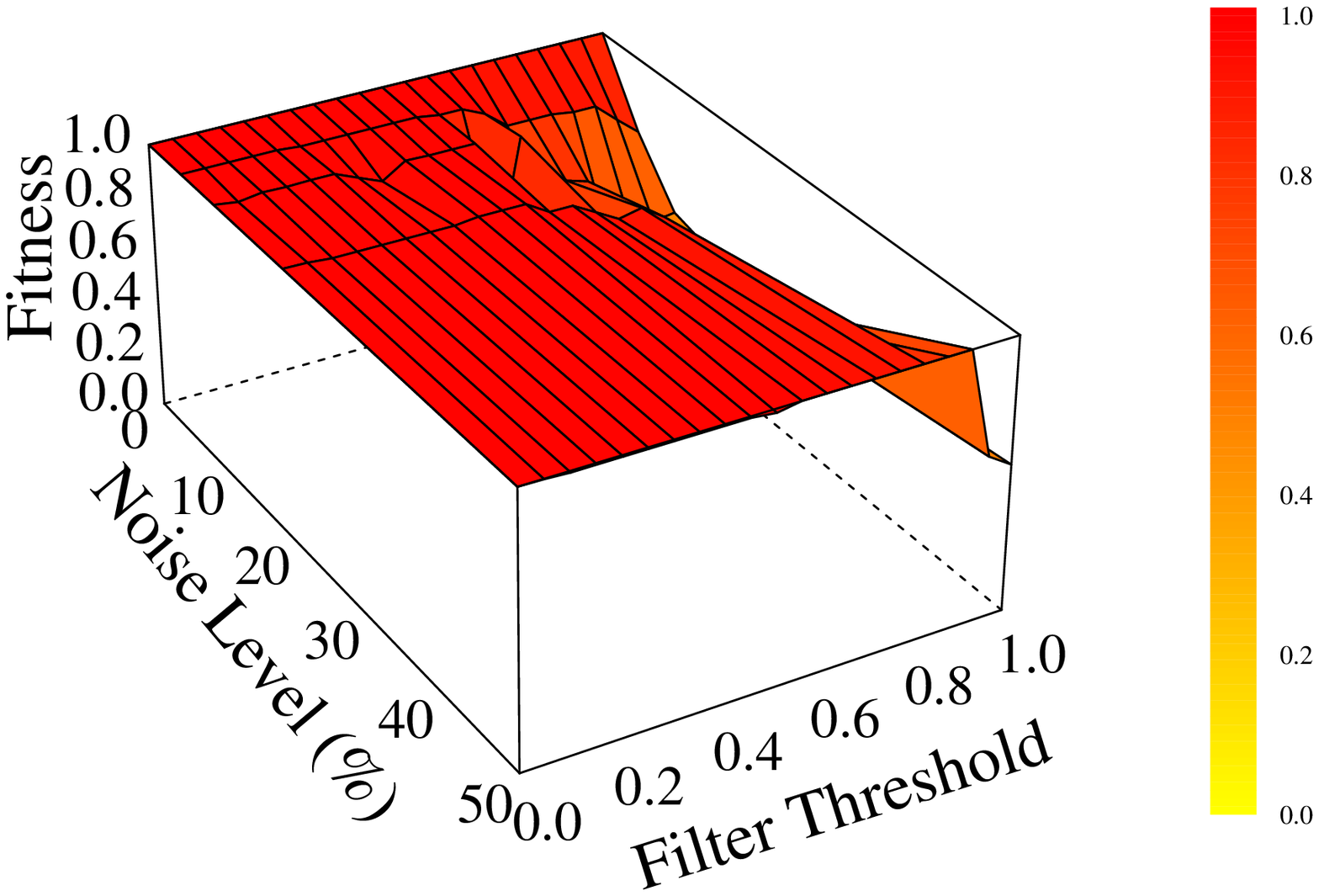}
			\caption{IMi~\cite{DBLP:conf/bpm/LeemansFA13}}
			\label{fig:a22_im_fitness}
		\end{subfigure}
		\begin{subfigure}[b]{0.325\textwidth}
			\centering
			\includegraphics[width=\textwidth,clip,trim=0cm 2cm 0cm 3cm]{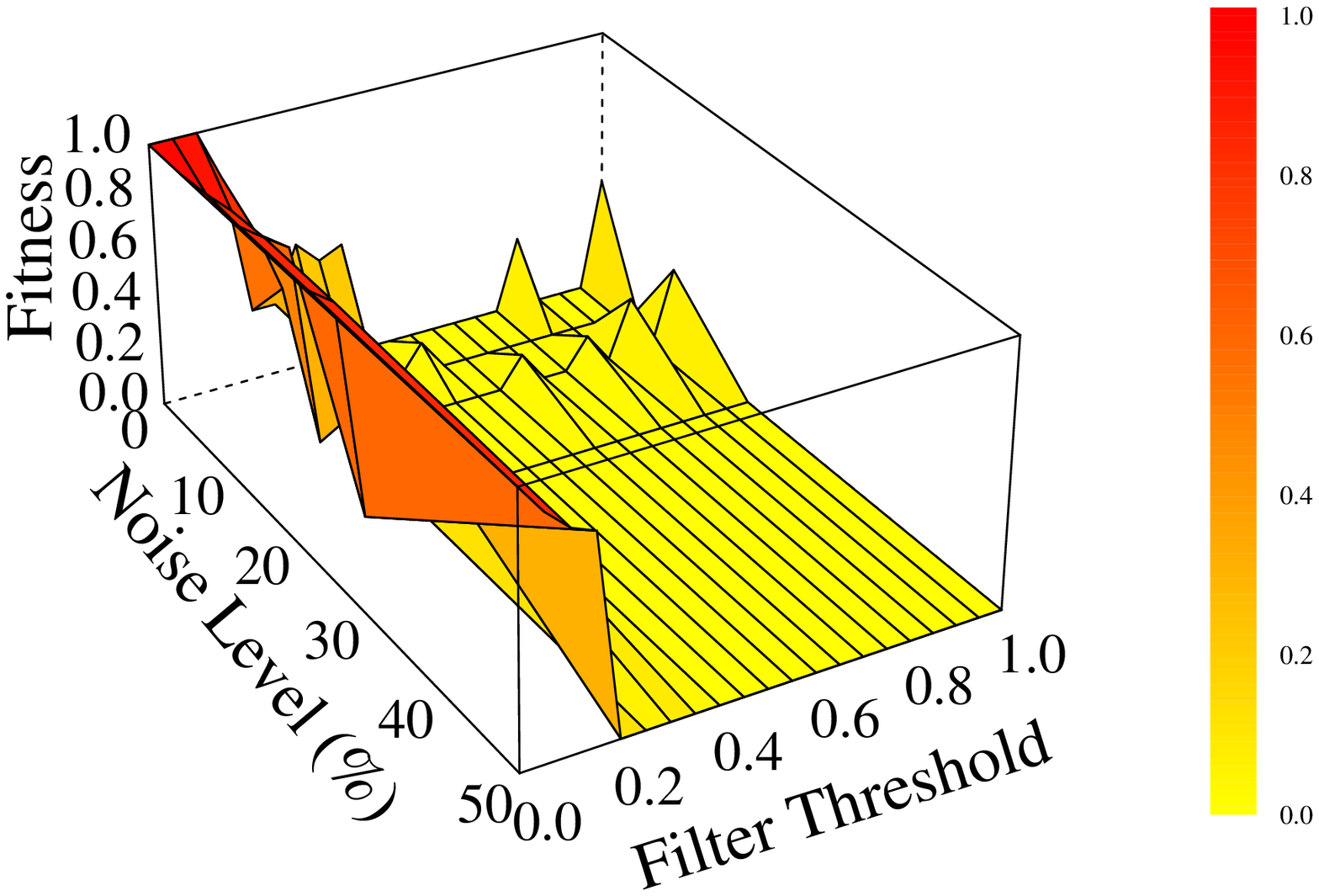}
			\caption{ILP with Automaton Filter~\cite{DBLP:journals/tkde/ConfortiRH17}}
			\label{fig:a22_filter_automata_fitness}
		\end{subfigure}
		\caption{Replay-fitness measurements based on \textit{a22f0nXX}.}
		\label{fig:a22_fitness}
	\end{figure}

	The replay-fitness results of the experiments with the \textit{a22f0nXX} event logs are presented in \autoref{fig:a22_fitness}.
	The charts very similar behaviour to the results reported for the \textit{a12f0nXX} event logs.
	The sequence encoding filter in \autoref{fig:a22_ilp_fitness} has a replay-fitness value of around $0.6$ when applying it as rigorous as possible, i.e. using $\alpha = 0$.
	This implies that the filter even removes behaviour that is present in the ground-truth event log.
	For increasing filter thresholds the replay-fitness value reaches a value of $1$ rapidly, i.e., the model is able to reproduce all traces in the event log.
	For IMi (\autoref{fig:a22_im_fitness}) we observe similar behaviour (note that the filter threshold works inverted w.r.t. sequence encoding filtering, i.e. a value of $1$ implies most rigorous filtering).
	However, replay-fitness drops a little earlier compared to sequence encoding filtering.
	Finally, automaton based filtering, depicted in \autoref{fig:a22_im_fitness}, rapidly drops to $0$.
	Again this is due to the fact that the filter tends to return empty logs for high threshold values.
	Hence, the filter seems to be very sensitive around a threshold value in-between $0$ and $0.2$.

\begin{figure}[tb]
		\begin{subfigure}[b]{0.325\textwidth}
			\centering
			\includegraphics[width=\textwidth,clip,trim=0cm 2cm 0cm 3cm]{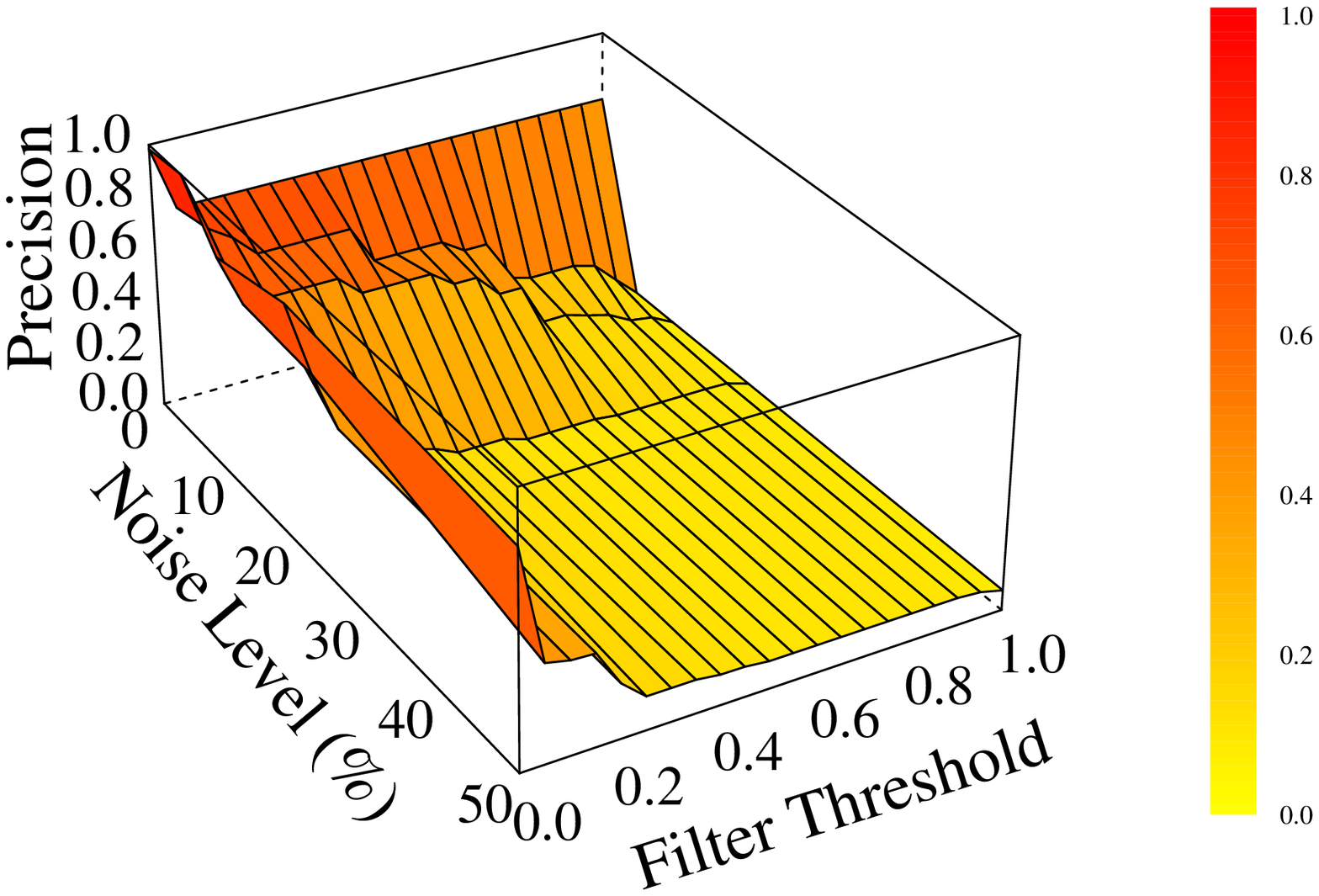}
			\caption{Sequence Encoding}
			\label{fig:a22_ilp_precision}
		\end{subfigure}
		\begin{subfigure}[b]{0.325\textwidth}
			\centering
			\includegraphics[width=\textwidth,clip,trim=0cm 2cm 0cm 3cm]{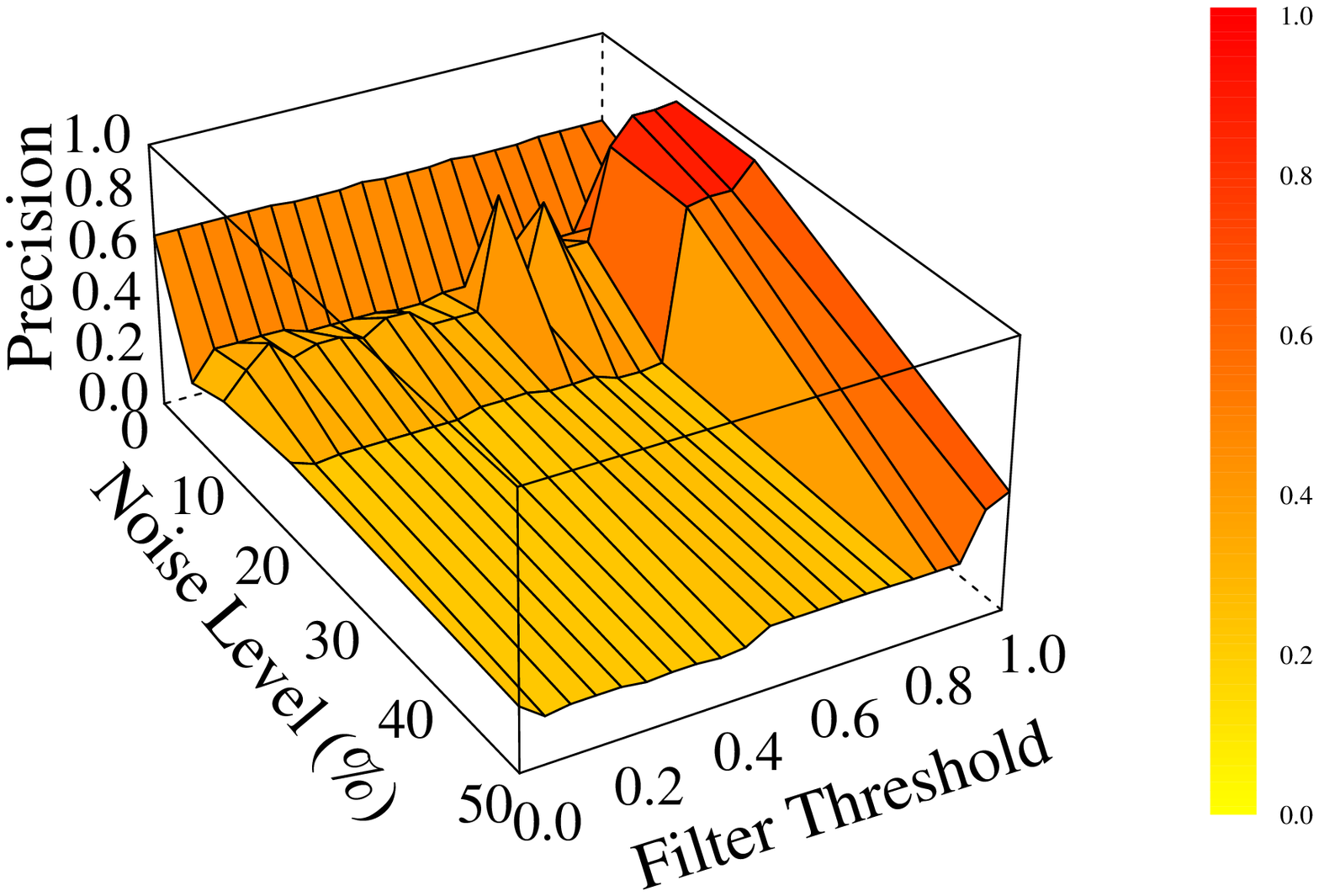}
			\caption{IMi~\cite{DBLP:conf/bpm/LeemansFA13}}
			\label{fig:a22_im_precision}
		\end{subfigure}
		\begin{subfigure}[b]{0.325\textwidth}
			\centering
			\includegraphics[width=\textwidth,clip,trim=0cm 2cm 0cm 3cm]{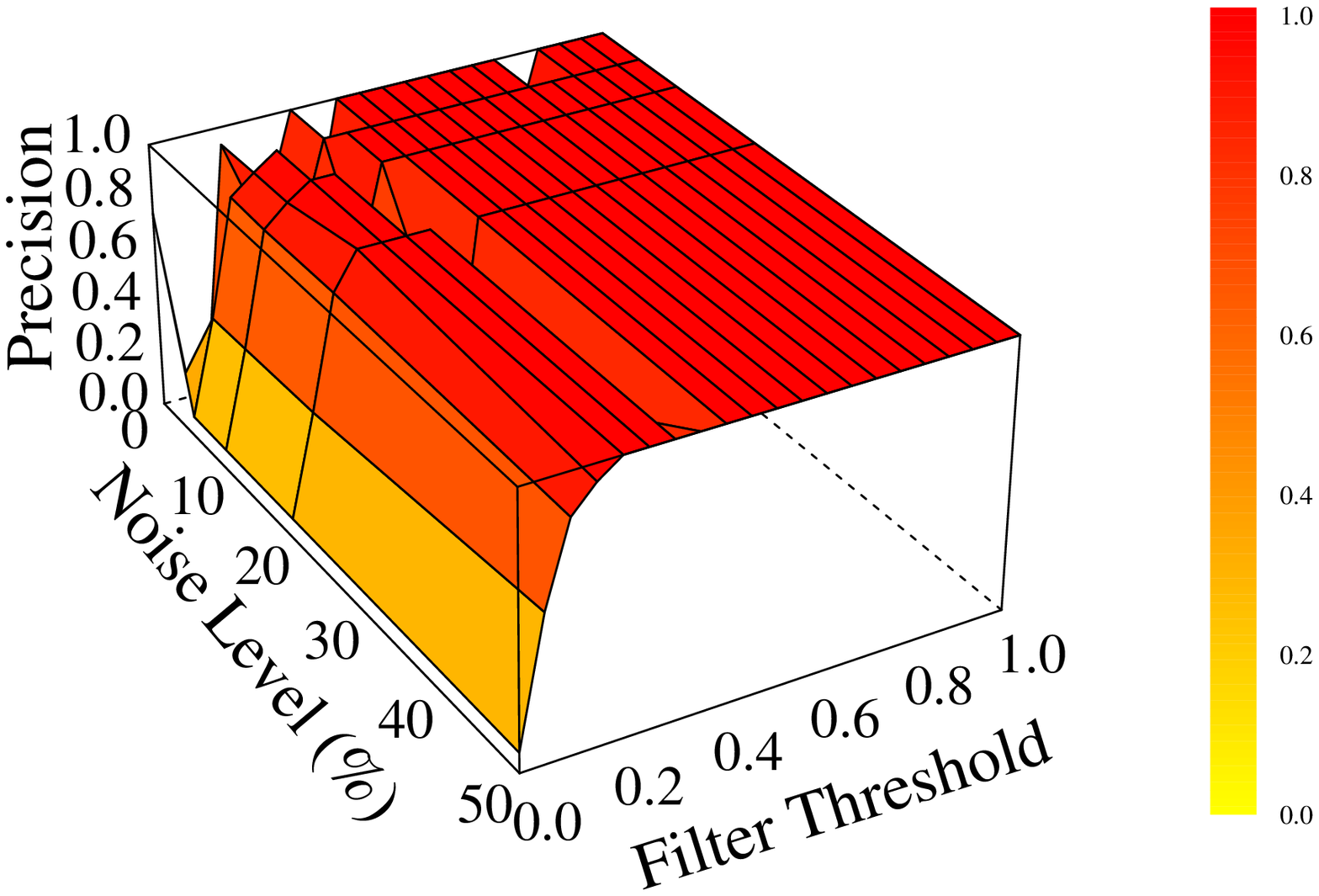}
			\caption{ILP with Automaton Filter~\cite{DBLP:journals/tkde/ConfortiRH17}}
			\label{fig:a22_filter_automata_precision}
		\end{subfigure}
		\caption{Precision measurements based on \textit{a22f0nXX}.}
		\label{fig:a22_precision}
	\end{figure}

	The precision results of the experiments with the \textit{a22f0nXX} event logs are presented in \autoref{fig:a22_precision}.
	For both the sequence encoding (\autoref{fig:a22_ilp_precision}) and IMi (\autoref{fig:a22_im_precision}) we observe a precision value of around $0.6$ based on the event logs without any noise.
	This is due to the fact that the originating model contains a loop which leads to imprecision.
	We observe that both sequence encoding filtering as well as IMi follow the same pattern in terms of precision.
	However, the drop in precision of sequence encoding filtering is more smooth than the drop in precision of IMi, i.e. there exist some spikes within the graph.
	Hence, the applying filtering within IMi seems to be less deterministic.
	Finally, the precision results for the automaton based filter are as expected.
	With a low threshold value we have very low precision, except when we have a $0\%$ noise level.
	Towards a threshold level of $0.2$ precision increases after which it maximizes out to a value of $1$.
	This is in line with the replay-fitness measurements.
	
\begin{figure}[tb]
		\begin{subfigure}[b]{0.5\textwidth}
			\centering
			\includegraphics[width=\textwidth,clip,trim=0cm 2cm 0cm 3cm]{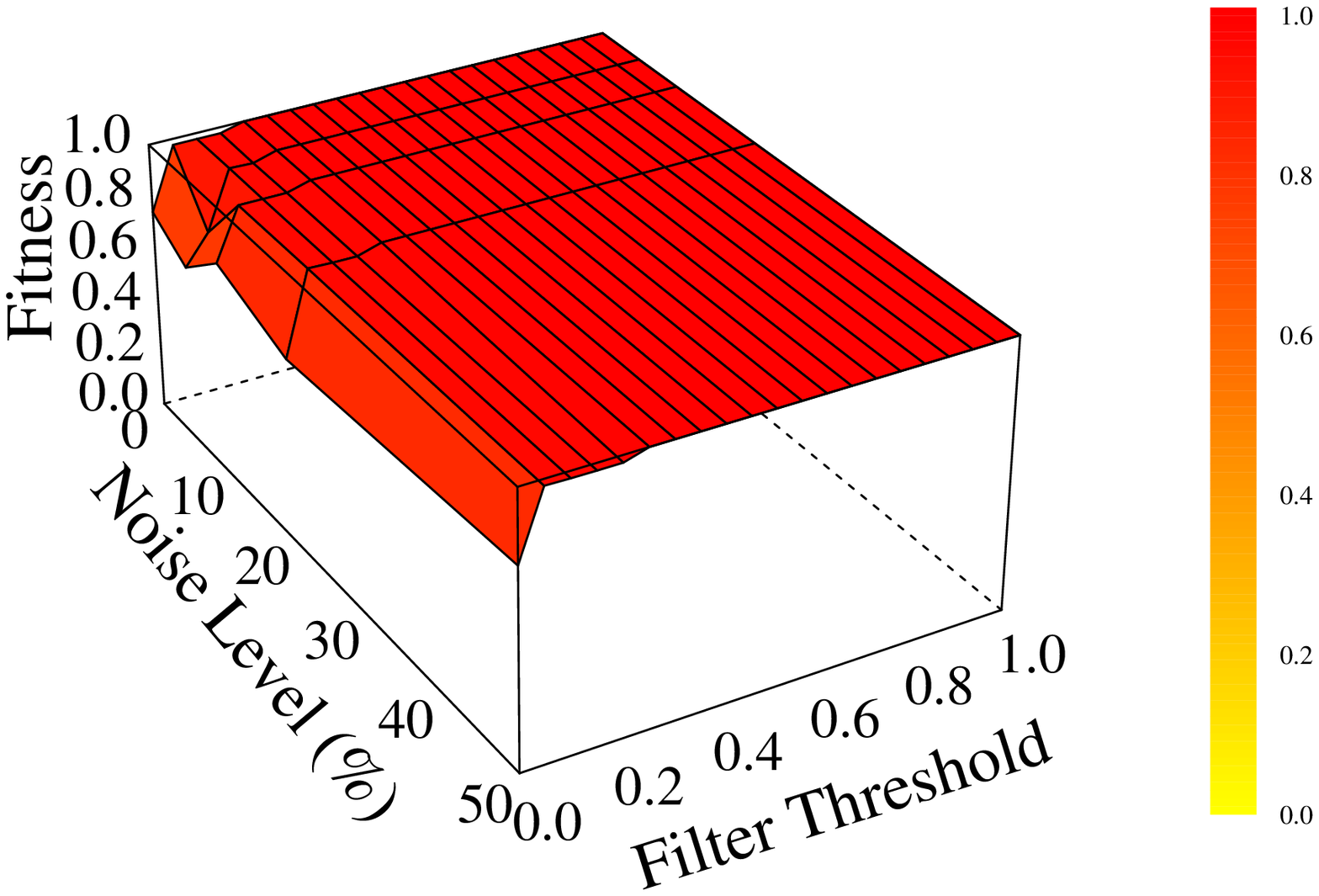}
			\caption{Sequence Encoding}
			\label{fig:a32_ilp_fitness}
		\end{subfigure}
		\begin{subfigure}[b]{0.5\textwidth}
			\centering
			\includegraphics[width=\textwidth,clip,trim=0cm 2cm 0cm 3cm]{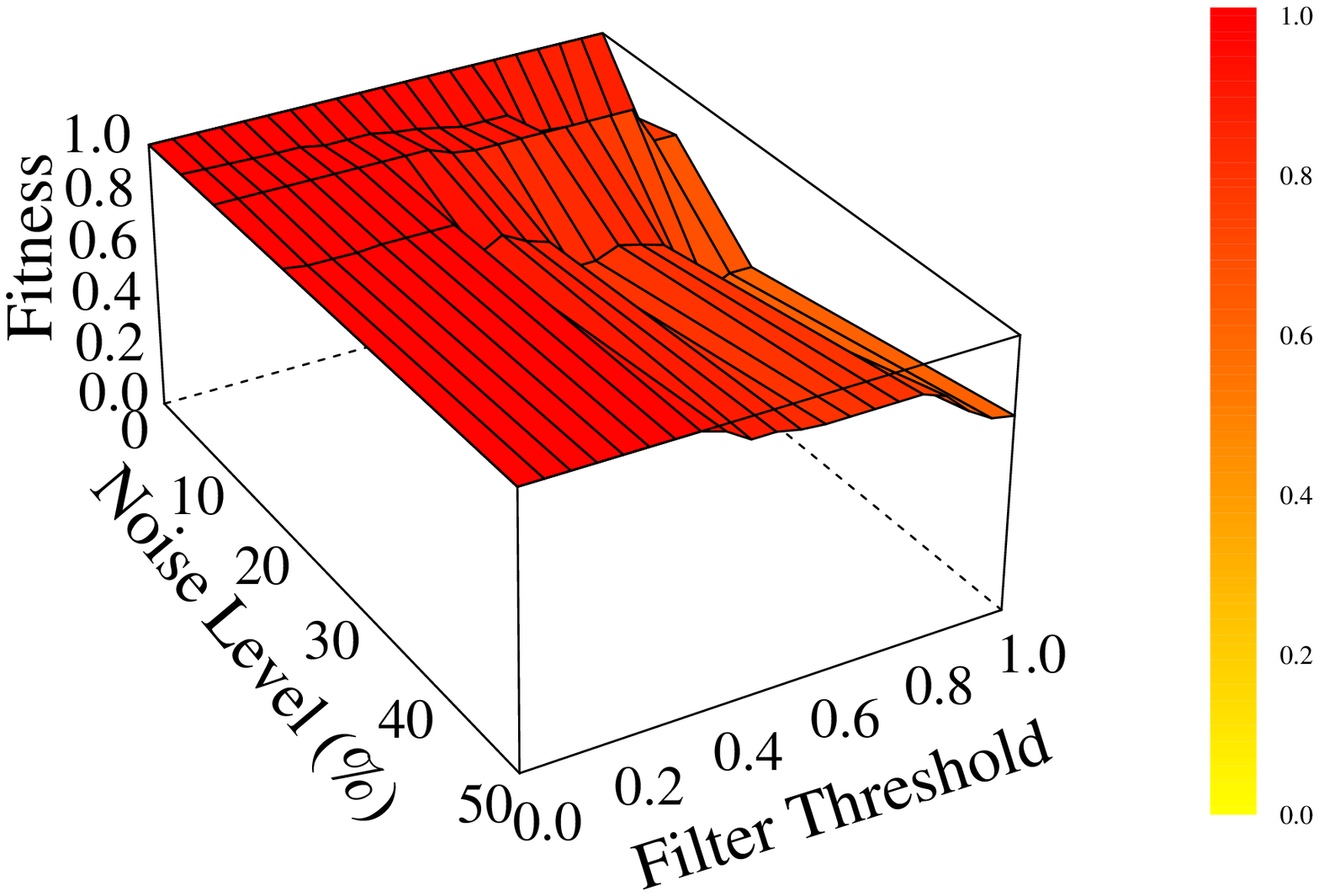}
			\caption{IMi~\cite{DBLP:conf/bpm/LeemansFA13}}
			\label{fig:a32_im_fitness}
		\end{subfigure}
		\caption{Replay-fitness measurements based on \textit{a32f0nXX}.}
		\label{fig:a32_fitness}
	\end{figure}

	In \autoref{fig:a32_fitness} we present the replay-fitness results of the experiments with the \textit{a32f0nXX} event logs.
	Due to excessive computation time the automaton based filter~\cite{DBLP:journals/tkde/ConfortiRH17} is left out of the analysis.
	We observe that sequence encoding filtering behaves similar to the experiments performed with the \textit{a12f0nXX} and \textit{a22f0nXX} event logs.
	The replay-fitness again quickly increase to $1$ for increasing filter threshold values.
	We observe that IMi seems to filter out more behaviour related to the underlying system model when the filter threshold increases.
	
\begin{figure}[tb]
		\begin{subfigure}[b]{0.5\textwidth}
			\centering
			\includegraphics[width=\textwidth,clip,trim=0cm 2cm 0cm 3cm]{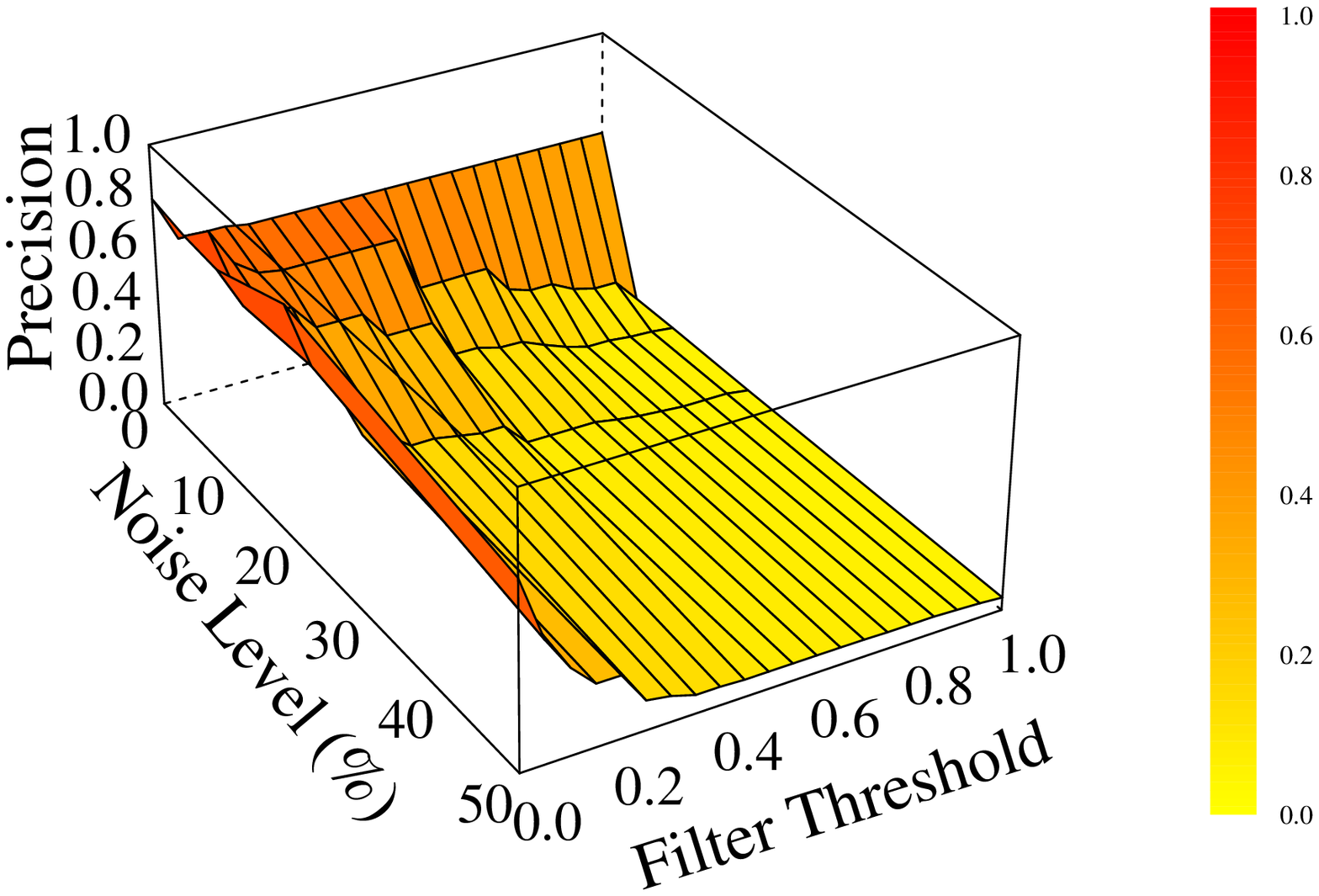}
			\caption{Sequence Encoding}
			\label{fig:a32_ilp_precision}
		\end{subfigure}
		\begin{subfigure}[b]{0.5\textwidth}
			\centering
			\includegraphics[width=\textwidth,clip,trim=0cm 2cm 0cm 3cm]{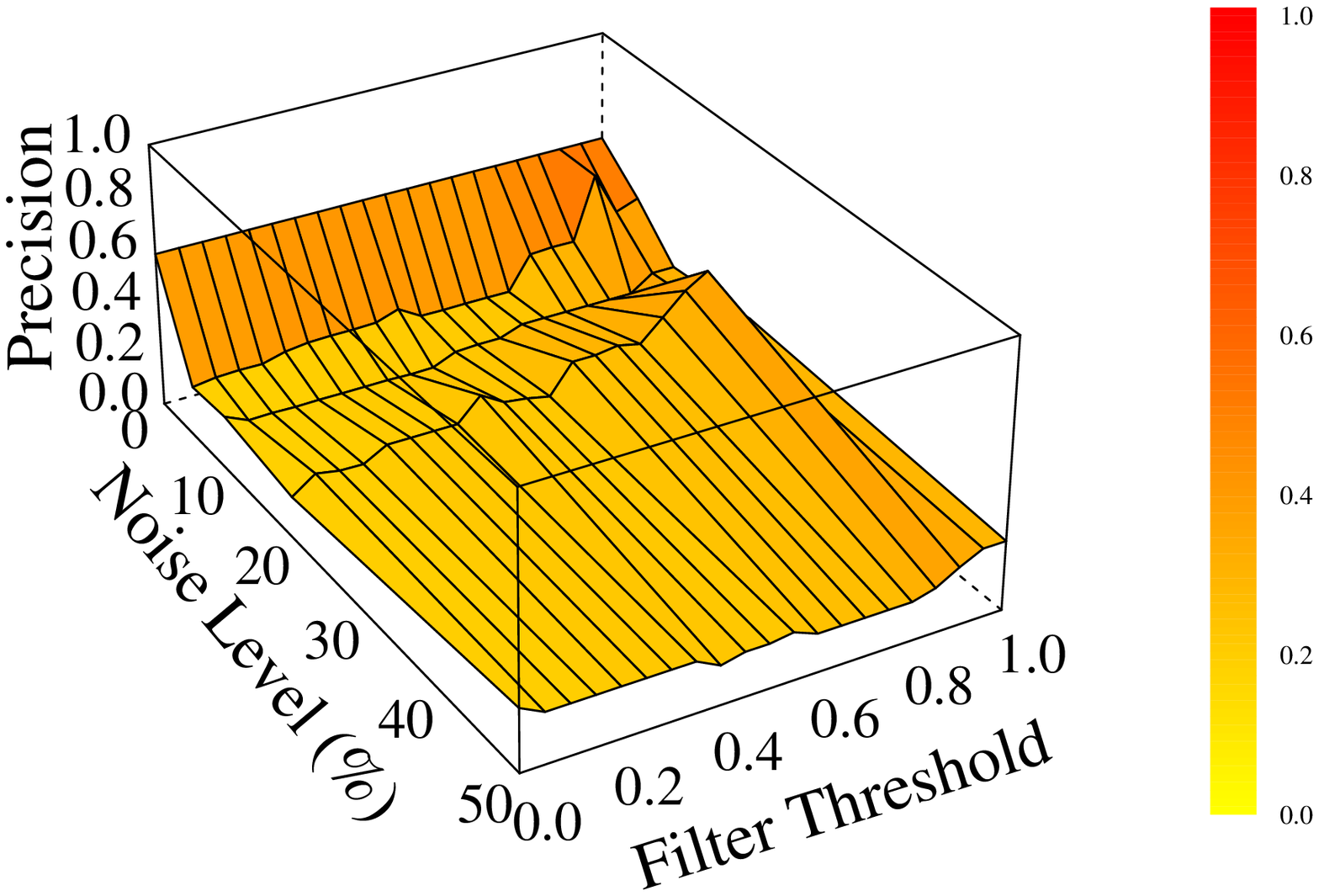}
			\caption{IMi~\cite{DBLP:conf/bpm/LeemansFA13}}
			\label{fig:a32_im_precision}
		\end{subfigure}
		\caption{Precision measurements based on \textit{a32f0nXX}.}
		\label{fig:a32_precision}
	\end{figure}

	In \autoref{fig:a32_precision} we present the precision results of the experiments with the \textit{a32f0nXX} event logs.
	Observe that, due to loop structures, the precision of a model that equals the originating model is only roughly $0.6$.
	Sequence encoding filtering shows a smooth decrease in precision when both noise and filter-thresholds are increased, which is as expected.
	With low noise levels and a low threshold sequence encoding seems to be able to filter out the infrequent behaviour, however, if there is too much noise and too little is removed we start finding WF-nets with self-loop places.
	IMi seems to result in models with a sightly higher precision compared to sequence encoding filtering.
	As is the case in the \textit{a22f0nXX} event logs, we observe spike behaviour in precision of IMi based models hinting at non-deterministic behaviour of the filter.

	Based on our experiments, we conclude that the sequence encoding filter and IMi give comparable results.
	However, the sequence encoding filter provides more expected results, i.e. IMi behaves somewhat deterministic.
	The automaton based filter does provide good results, however, sensibility of the filter threshold is much higher compared to sequence encoding filtering and IMi.

	\subsection{Computation time}
	The core of sequence encoding filtering is leaving out constraints that are likely to refer to exceptional behaviour. 
	Thus, we reduce the size of the core ILP constraint body.
	Hence, we expect a decrease in computation time when applying rigorous filtering, i.e. $\kappa^{\alpha}_{max}$ with $\alpha$ towards $0$.
	Using \texttt{RapidMiner} we repeated similar experiments to the experiments performed for model quality, and measured \textit{cpu-execution} time for the three techniques.
	However, we only use threshold values $0$, $0.25$, $0.75$ and $1$.
	
\begin{figure}[tb]
		\centering
		\includegraphics[width=0.75\textwidth]{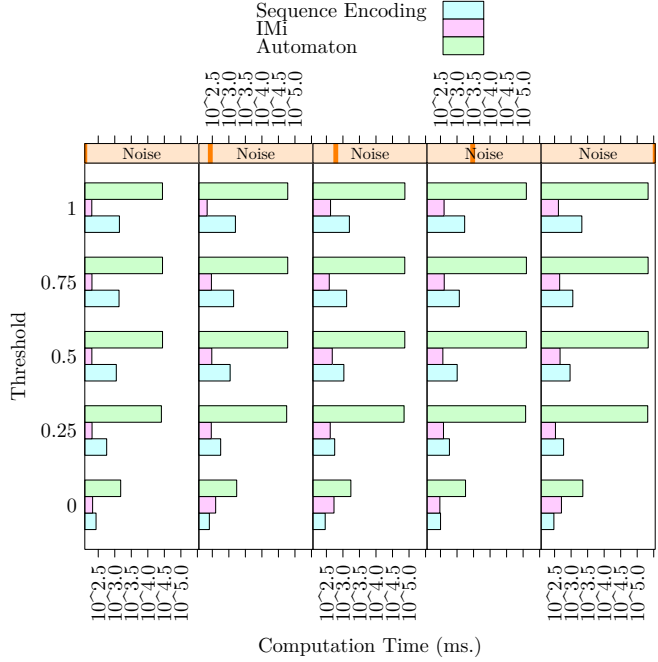}
		\caption{CPU-Execution Time (ms.) for a22f0nXX event logs (logarithmic scale).}
		\label{fig:a22_run_time}
	\end{figure}

	In \autoref{fig:a22_run_time} we present the average cpu-execution time, based on 50 experiment repetitions, needed to obtain a process model from an event log.	
	For each level of noise we depict computation time for different filter threshold settings, $0\%$ noise is depicted in the left-most figure, $50\%$ in the right-most figure.
	For IMi, we measured the inductive miner algorithm with integrated filtering.
	For sequence encoding and automaton filtering, we measure the time needed to filter, discover a causal graph and solve underlying ILP problems.	
	As we observe in \autoref{fig:a22_run_time}, IMi is fastest in all cases except for a threshold of 0 where sequence encoding tends to outperform IMi and automaton-based filtering.
	We observe that in all cases computation time increases when the amount of noise increases within the event logs.
	For sequence encoding filtering we observe that lower threshold values lead to faster computation times.
	This is as expected since a low threshold value removes more constraints from the ILP constraint body than a high threshold value.
	The automaton-based filter is slowest in all cases.
	The amount of noise seems to have little impact on the computation time of the automaton-based filter, it seems to be predominantly depending on the filter threshold.
	From \autoref{fig:a22_run_time} we conclude that IMi in general out-performs sequence encoding in terms of computation time.
	However, sequence encoding, in turn out-performs automaton-based filtering, specifically for higher threshold settings.
	
	\subsection{Application to Real-Life Event Logs}
	We additionally tested the applicability of sequence encoding filtering using real-life event logs.
	We used two event logs, one related to the administration process of handling road fines~\cite{https://doi.org/10.4121/uuid:270fd440-1057-4fb9-89a9-b699b47990f5} and one regarding the patient treatment of patients suspected to have sepsis~\cite{https://doi.org/10.4121/uuid:915d2bfb-7e84-49ad-a286-dc35f063a460}.
	
	\begin{figure}[tb]
		\begin{subfigure}[b]{0.245\textwidth}
			\centering
			\includegraphics[height=0.15\textheight]{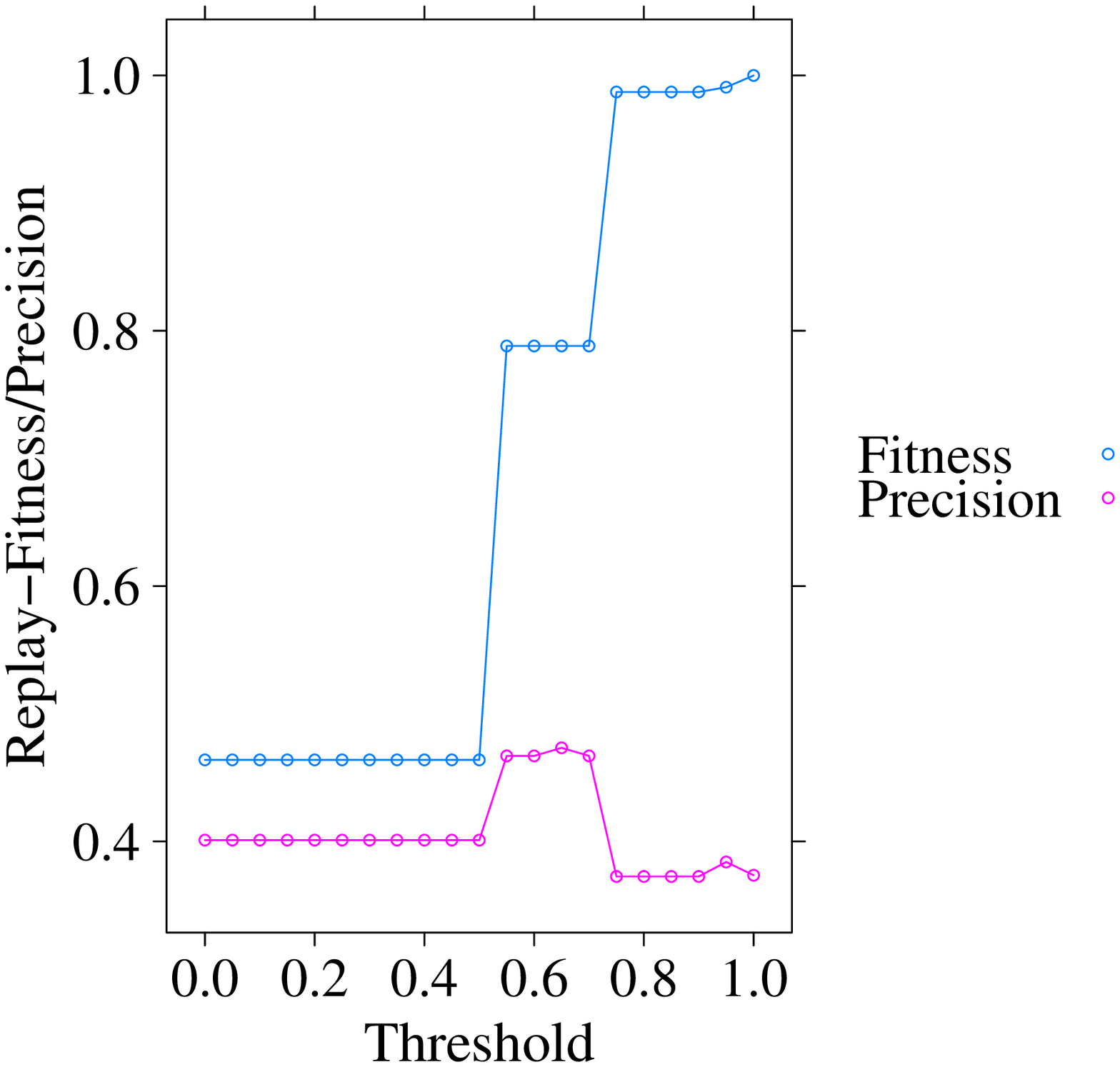}
			\caption{Fitness and Precision}
			\label{fig:road_fines_quality}
		\end{subfigure}
		\begin{subfigure}[b]{0.245\textwidth}
			\centering
			\includegraphics[height=0.15\textheight]{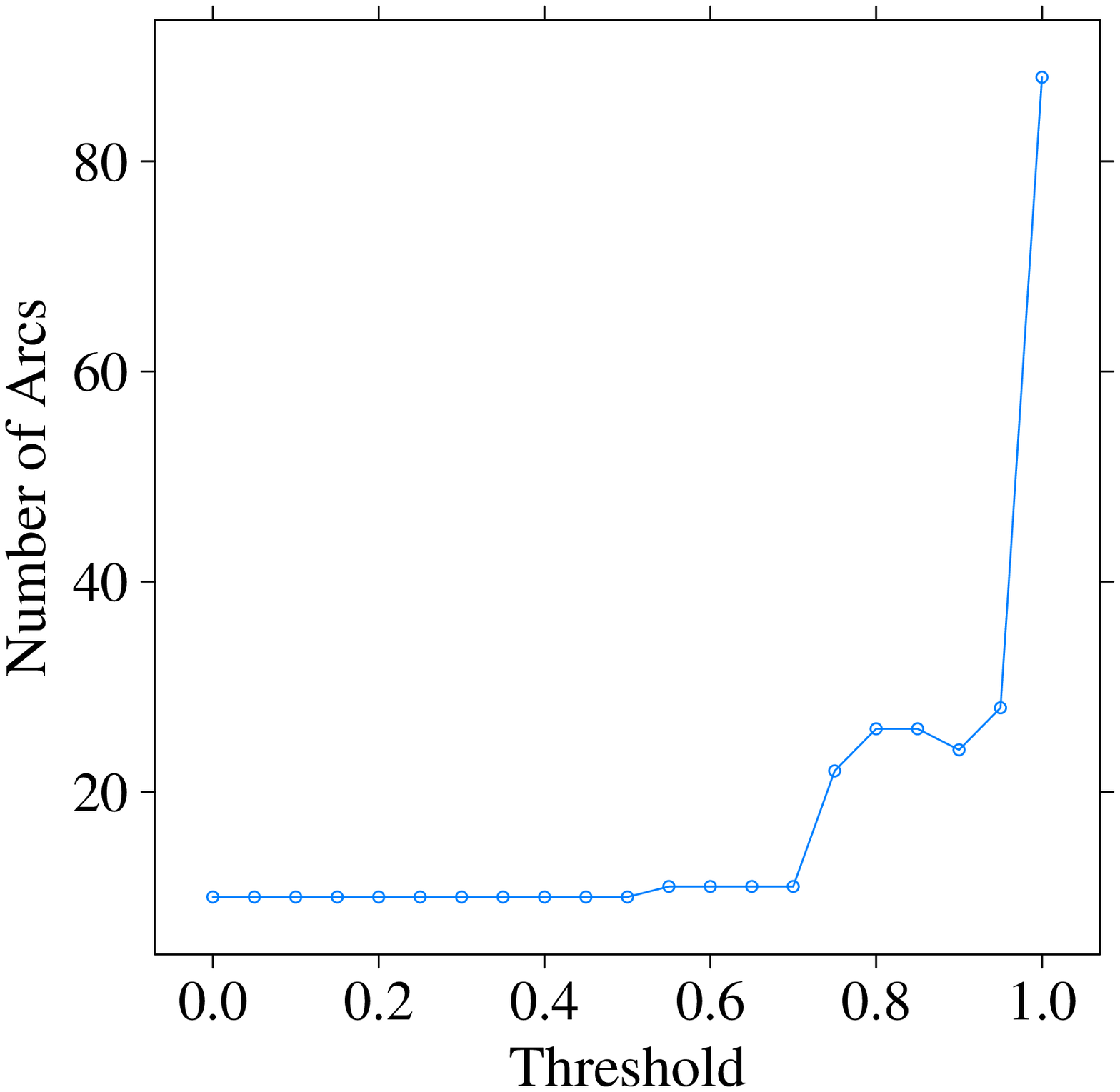}
			\caption{Number of Arcs}
			\label{fig:road_fines_complexity}
		\end{subfigure}
		\begin{subfigure}[b]{0.245\textwidth}
			\centering
			\includegraphics[height=0.15\textheight]{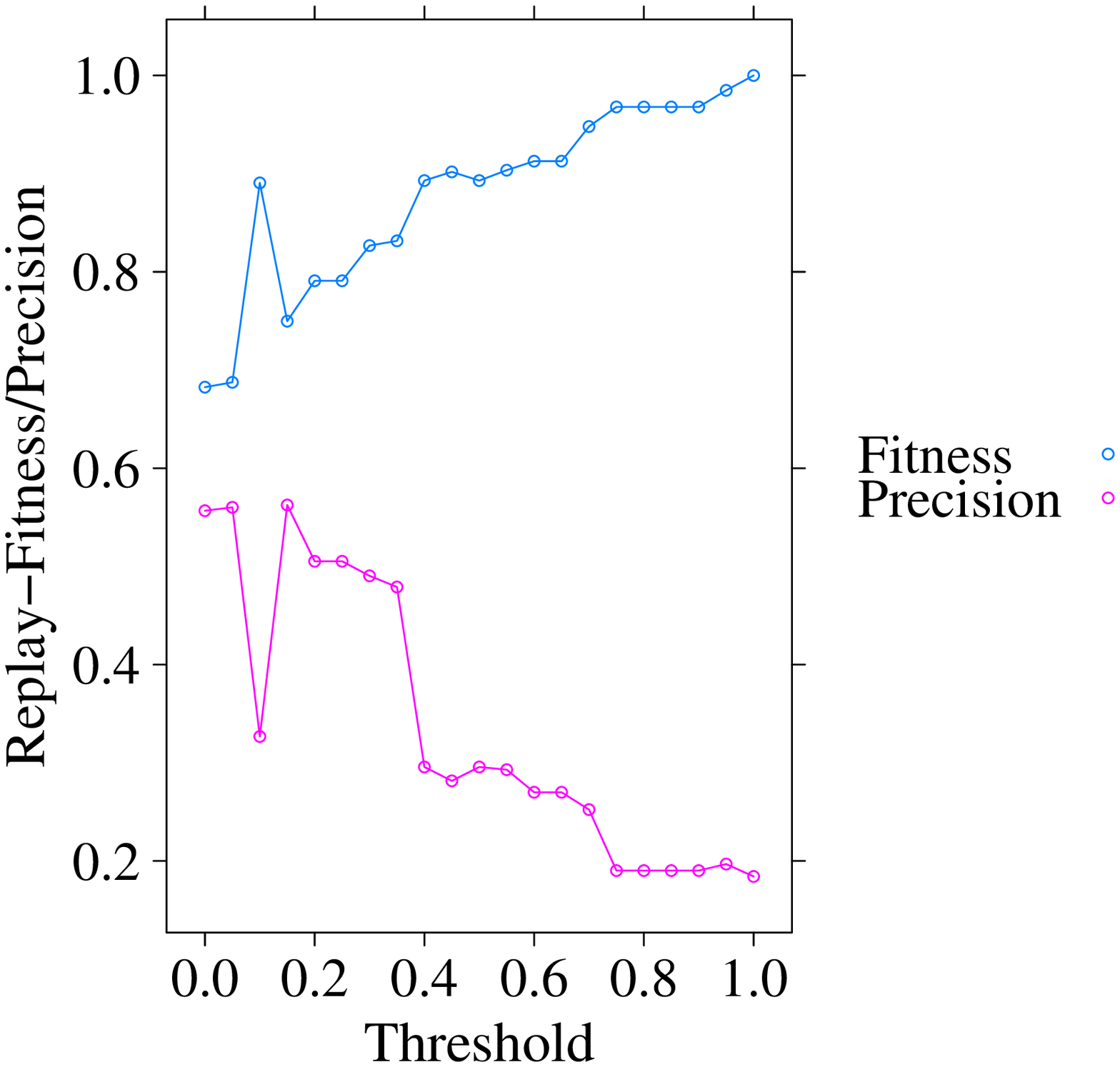}
			\caption{Fitness and Precision}
			\label{fig:sepsis_quality}
		\end{subfigure}
		\begin{subfigure}[b]{0.245\textwidth}
			\centering
			\includegraphics[height=0.15\textheight]{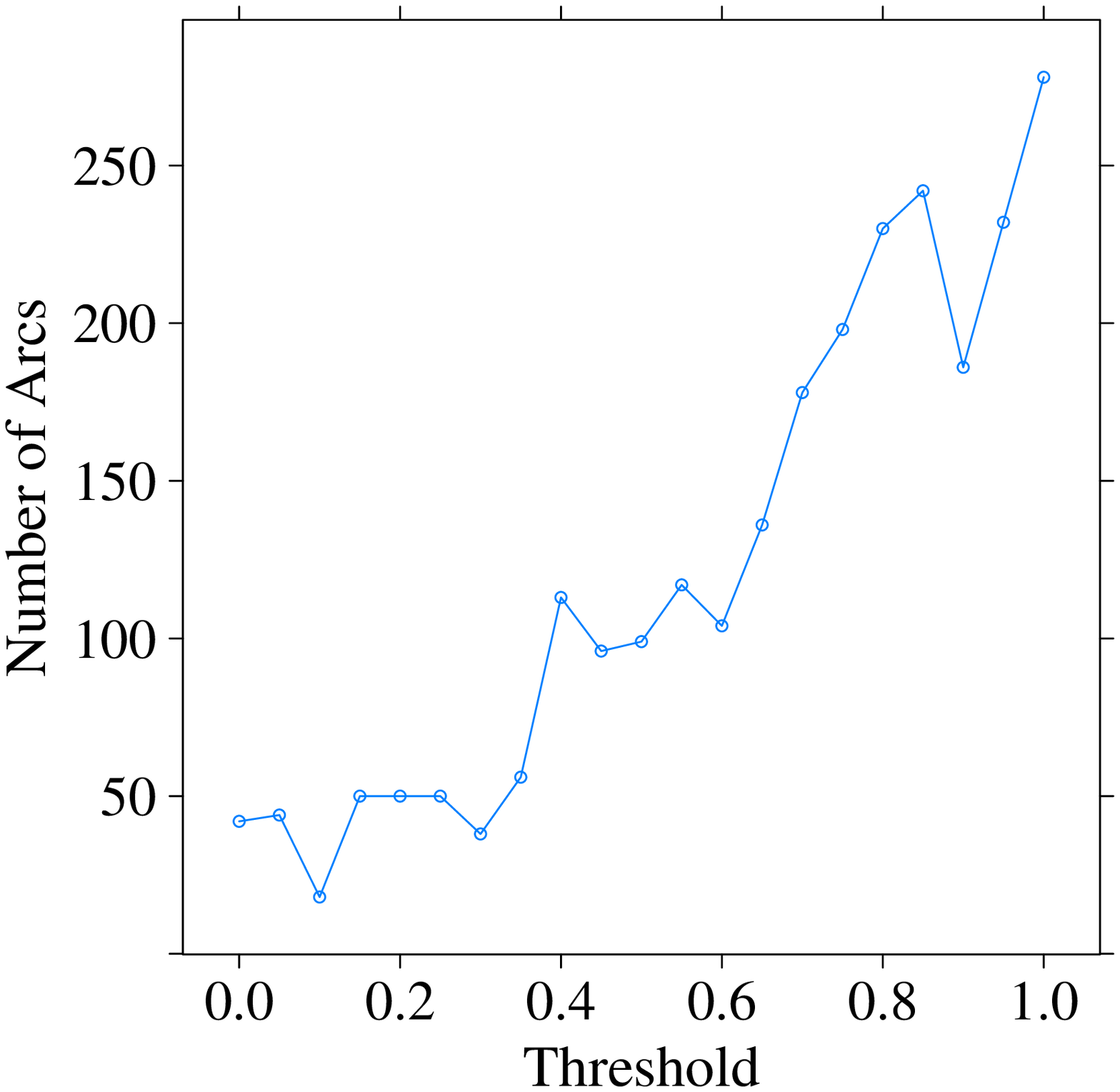}
			\caption{Number of Arcs}
			\label{fig:sepsis_complexity}
		\end{subfigure}
		\caption{Replay-fitness, precision and complexity based on the \texttt{Road Fines} log~\cite{https://doi.org/10.4121/uuid:270fd440-1057-4fb9-89a9-b699b47990f5} (\autoref{fig:road_fines_quality} and \autoref{fig:road_fines_complexity}) and the \texttt{Sepsis} log~\cite{https://doi.org/10.4121/uuid:915d2bfb-7e84-49ad-a286-dc35f063a460} (\autoref{fig:sepsis_quality} and \autoref{fig:sepsis_complexity}).}
		\label{fig:real_data}
	\end{figure}

	The results are presented in \autoref{fig:real_data}.
	In case of the \texttt{Road Fines} event log (figures on the left-hand side of \autoref{fig:real_data}) we observe that replay-fitness is around $0.46$ whereas precision is around $0.4$ for $\alpha$-values from $0$ to $0.5$.
	The number of arcs for the models of these $\alpha$-values remains constant (as well as the number of places and the number of transitions) suggesting that the models found are the same.
	After this the replay-fitness increases further to around $0.8$ and reaches $1$ for an $\alpha$-level of $1$.
	Interestingly, precision shows a little increase around $\alpha$-levels between $0.5$ and $0.75$ after which it drops slightly below its initial value.
	In this case, an $\alpha$-level in-between $0.5$ and $0.75$ seems most appropriate in terms of replay-fitness, precision and simplicity.
	
	In case of the \texttt{Sepsis} event log (figures on the left-hand side of \autoref{fig:real_data}) we observe that replay-fitness and precision are roughly behaving as each-other's inverse, i.e. replay-fitness increases whereas precision decreases for increasing $\alpha$-levels.
	We moreover observe that the number of arcs within the process models is steadily increasing for increasing $\alpha$-levels.
	In this case, an $\alpha$-level in-between $0.1$ and $0.4$ seems most appropriate in terms of replay-fitness, precision and simplicity.
	
	Finally, for each experiment we measured the associated computation time of solving all ILP problems.
	In case of the \texttt{Road Fines} event log, solving all ILP problems takes roughly 5 seconds.
	In case of the \texttt{Sepsis} event log, obtaining a model ILP problems takes less than 1 second.

	\section{Conclusion}
	\label{sec:conclusion}
	The work presented in this paper is motivated by the observation that existing region-based process discovery techniques are useful, as they are able to find non-local complex control-flow patterns.
	However, the techniques do not provide any structural guarantees w.r.t. the resulting process models, and, they are unable to cope with infrequent, exceptional behaviour in event logs.
	
	The approach presented in this paper extends techniques presented in~\cite{DBLP:journals/fuin/derWerfDHS09,DBLP:conf/apn/ZelstDA15,DBLP:conf/bpm/ZelstDA15a}.
	We have proven that our approach is able to discover relaxed sound workflow nets, i.e. we are now able to guarantee structural properties of the resulting process model.
	Additionally, we presented the sequence encoding filtering technique which enables us to apply filtering exceptional behaviour within the ILP-based process discovery algorithm.
	Our experiments confirm that the technique enables us to find Petri net structures in data consisting of exceptional behaviour, using ILP-based process discovery as an underlying technique.
	Sequence encoding filtering proves to be comparable to the IMi~\cite{DBLP:conf/bpm/LeemansFA13} approach, i.e. an integrated filter of the Inductive Miner~\cite{DBLP:conf/apn/LeemansFA13}, in terms of filtering behaviour.
	It is considerably faster than the general purpose filtering approach of~\cite{DBLP:journals/tkde/ConfortiRH17} and less sensible to variations in the filter threshold.
	
	\paragraph{Future Work}
	An interesting direction for future work concerns combining ILP-based process discovery techniques with other process discovery techniques.
	The Inductive Miner discovers sound workflow nets, however, these models are lack the ability to express complex control flow patterns such as a milestone pattern.
	Some of these patterns are however reconstructible using ILP-based process discovery.
	Hence, it is interesting to combine these approaches with possibly synergetic effects w.r.t. the process mining quality dimensions.
	
	Another interesting approach is the development of more advanced general purpose filtering techniques.
	Most discovery algorithms assume the input event logs to be free of noise, infrequent and/or exceptional behaviour.
	Real-life event logs however typically contain a lot of such behaviour.
	Surprisingly, little research is performed towards filtering techniques that greatly enhance process discovery results, independent of the discovery algorithm used.
	
	\bibliographystyle{spmpsci}

\end{document}